\documentclass[a4paper,11pt]{article}
\pdfoutput=1 
\usepackage{jheppub} 
\usepackage[T1]{fontenc} 
\usepackage{graphicx}
\newcommand{\be}{\begin{equation}}
\newcommand{\ee}{\end{equation}}
\newcommand{\ba}{\begin{eqnarray}}
\newcommand{\ea}{\end{eqnarray}}
\newcommand{\nn}{\nonumber\\}
\usepackage{color}

\title{\boldmath  Transport coefficients of  a hot QCD medium  and their relative significance in heavy-ion collisions}

\author[a,1]{Sukanya Mitra}
\author[a,2]{Vinod Chandra}

\affiliation[a]{Indian Institute of Technology Gandhinagar,\\ Palaj, Gandhinagar-382355, Gujarat, India}

\emailAdd{sukanyam@iitgn.ac.in}
\emailAdd{vinodc@iitgn.ac.in}
\abstract{
The main focus of this article is to obtain various transport coefficients for a hot QCD medium that is produced while colliding two heavy 
nuclei ultra-relativistically. As the hot QCD medium follows dissipative hydrodynamics while undergoing space-time evolution, the knowledge 
of the transport coefficients such as thermal conductivity, electrical conductivity, shear and bulk viscosities are essential to understand 
the underlying physics there. The approach adopted here is semi-classical transport theory. The determination of all these transport 
coefficients requires knowledge of the medium away from equilibrium. In this context, we setup the linearized transport equation 
employing the Chapman-Enskog technique from kinetic theory of many particle system with a collision term that includes the binary collisions of 
quarks/antiquarks and gluons. In order to include the effects of a strongly interacting, thermal medium, a quasi-particle description of 
realistic hot QCD equation of state has been employed through the equilibrium modeling of the momentum distributions of gluons and quarks with 
 non-trivial  dispersion relations while extending the model for finite but small quark chemical potential. The effective coupling for strong 
 interaction has  been redefined following the charge renormalization under the scheme of the quasiparticle model. The consolidated effects on transport coefficients are seen to have significant impact on their temperature 
dependence.  The relative significances of momentum and heat transfer as well as charge diffusion processes in hot QCD have been 
investigated by  studying the ratios of the respective  transport coefficients.
}

\begin{document} 
\maketitle
\flushbottom

\section{Introduction}
The sub-nucleonic world of partonic substructures (quarks and gluons)
 have been studied with greater precision in last few decades, by exploring a deconfined state of the 
nuclear matter at relativistically energetic heavy-ion collider experiments. The experimental 
facilities at Relativistic Heavy Ion Collider (RHIC), BNL and Large Hadron Collider 
(LHC), CERN, have provided a fortune of data, which helped in revealing the thermodynamic
and transport properties of the created medium after equilibration.  A closer 
inspection on the  experimental observables such as transverse momentum spectra and collective flows of 
charged hadrons or electromagnetic probes, reveals that their quantitative estimates 
should involve critical dependence upon the transport parameters of the system. This 
serves as a strong motivation for the  quantitative study of the  transport coefficients 
of this exotic medium (Quark-Gluon-Plasma (QGP))  that is created while colliding two heavy-ions such as Au-Au or Pb-Pb ultrarelativistically, along with a detailed study of 
their temperature dependences. The transport coefficients under investigation are, the shear and bulk viscosities ($\eta$ and $\zeta$) , electrical conductivity ($\sigma_{el}$) 
and thermal conductivity ($\lambda$) of the QGP medium.  Besides providing information about the 
dissipation and electromagnetic (EM) responses of the medium, these transport parameters give relevant insights about 
the nature of interaction and non-equilibrium dynamics of the system as well.
Earlier predictions of charged hadron elliptic flow from RHIC \cite{STAR} and their theoretical 
explanations using dissipative hydrodynamics \cite{Luzum} first provide the experimental
evidence of  existence of the transport processes in the QGP.  More recently,  a number  of ALICE results have 
re confirmed the relevance of transport processes \cite{ALICE1, ALICE2, ALICE3, ALICE4, ALICE5, ALICE-JHEP}.
In particular in the context of the signal properties of charged hadrons and thermally 
produced particles (photons and dileptons), electromagnetic responses of the QGP medium 
also observed to play vital role which have been explored in \cite {Hirano,Kharzeev1,Kharzeev2,Skokov,Toneev,Voloshin,Zahed},
in the due course of understanding the QGP medium.

The present article aims to  estimate the temperature dependence of $\eta$,  $\zeta$, $\lambda$ and $\sigma_{el}$
for a hot QCD medium/QGP created in heavy-ion collisions, including the effect of a finite quark chemical potential $\mu_q$.
To explore the relative importance of these transport parameters and associated physical transport processes, their ratios
in the form of known laws of known numbers in the literature have been investigated. The analysis has been done with semi-classical transport theory adopting Chapman-Enskog approach for many particle systems. 
The basic approach of determining the transport coefficients in kinetic theory is pursued by comparing the macroscopic
and microscopic definitions of thermodynamic flows, as a results of which the particle interactions
enter in the expressions of transport coefficients as dynamical inputs. Hence kinetic theory offers a unique  
scheme, that bridges between the microscopic events of particle interactions to its macroscopic effects
(transport phenomena) on the thermodynamic system. In order to initiate the analysis and setup 
the appropriate transport equation, the very first requirement is the knowledge of local equilibrium momentum distributions of the gluonic and quark degrees of 
freedom that constitute the QGP. To that end, the modeling of equilibrium momentum distributions of gluons and quarks/antiquarks at vanishing and 
non-vanishing quark chemical potentials is needed to be done in  a way that a realistic equation of state (EOS) for the QGP 
(such as lattice QCD EOS) could be 
mimicked. This has been done by adopting a recently introduced effective quasi-particle model by Chandra and 
Ravishankar~\cite{Chandra_quasi1,Chandra_quasi2} where 
the hot QCD medium effects, present in the equations of state (EOSs) have been mapped to the equilibrium momentum distributions of
quasiquarks and quasigluons containing temperature dependent quark and gluon effective fugacities. The modified thermodynamic quantities along with the non-trivial
dispersion relation and the effective coupling of the strong interaction within the scope of the
quasiparticle model, are observed to influence the temperature dependence of the transport parameters
and the ratios significantly.  It could be safely inferred that, the hot QCD medium effects, of a strongly
correlated QGP liquid, introduced through the quasiparticle model, are  being reflected in the 
temperature dependence of the estimated transport coefficients.

In order to quantify the energy-momentum dissipation during the 
space-time evolution of the system, shear and bulk viscous coefficients are needed to 
be estimated. The velocity gradient between the adjacent fluid layers results in the 
distortion of momentum distribution within the fluid elements, which gives rise to viscous forces. The viscous coefficients
provide a measure of how the microscopic interactions within the system restore back the
momentum distribution from skewed to isotropic. The thermal dissipation, occurring due to temperature
gradient over the spatial separations of fluid, is described in terms of thermal conductivity 
for a system with conserved baryon current density. Besides the dissipative properties, one  needs 
to investigate the electromagnetic (EM) responses in the QGP system, since a considerably
strong EM field ($eB\sim m_{\pi}^2$) is being generated in the early stages of heavy ion 
collisions. In order to quantify the impact of the fields on electromagnetically charged
QGP, the electrical conductivity plays quite useful role. It gives a measure of the electric current
being induced in the response of the early stage electric field. In the strongly correlated systems
like non-relativistic ultracold atomic Fermi gases or strongly coupled Bose fluids (in
particular liquid helium), and for the  QGP medium, the specific shear viscosity ($\eta/s$) 
is observed to have small values exhibiting near perfect fluidity of the system \cite{Schafer1,Schafer2}.
The value of shear viscosity has been constrained by its ratio over system's entropy density ($\eta/s$)
by a lower bound $1/4\pi$, following the uncertainty principle and substantiated using anti-de 
Sitter space/conformal-field-theory (AdS/CFT) correspondence \cite{ADSCFT}. The agreement of hydrodynamic 
description with the experimental data in \cite{Luzum} also confirms this small value ($\eta/s=2/4\pi$) 
of shear viscous coefficient, which appears to be consistent with the values extracted directly
from experiments \cite{Gavin} and lattice simulations \cite{Nakamura} as well.
The magnitude of bulk viscosity, $\zeta$ is found to be quite small as compared to the shear viscosity, $\eta$, 
due to which early viscous hydrodynamic simulations ignored bulk viscosity for simplicity \cite{Heinz}. 
Although it vanishes for a conformal fluid or massless QGP on the classical level,
quantum effects break the conformal symmetry of QCD and generate a nonzero
bulk viscosity even in the massless QGP phase, as recently shown by the lattice results
\cite{Meyer} in the $SU(3)$ pure gauge theory. Following the general argument that QCD to
hadron gas transition is a crossover, $\eta/s$ shows a minimum near $T_c$, the critical temperature, 
close to the lower bound \cite{Csernai,Lacey}, whereas the bulk viscosity to entropy density
ratio $\zeta/s$ shows large values around $T_c$ \cite{Karsch,Pratt}. Finally at FAIR energies and
in the low-energy runs at RHIC, where the baryon chemical potential will be significant,
thermal conductivity ($\lambda$) is expected to play important roles in the hydrodynamic evolution
of the system. In \cite{Kapusta}, the thermal conductivity is shown to diverge at the critical 
point and used to study the impact of hydrodynamic fluctuations on experimental observables. 
As a consequence of the strong electromagnetic field generated in the early stages of heavy ion collisions,
the produced matter, after thermalization involves a non negligible electrical conductivity $\sigma_{el}$. 
In \cite{Deng}, nontrivial time dependence of the electromagnetic fields is observed to be sensitive 
to this finite electrical conductivity. In \cite{Zakharov},  the electromagnetic responses in the 
plasma fireball is demonstrated in the presence of a realistic $\sigma_{el}$, demanding a finite 
value of electrical conductivity in the QGP system. The relative behavior of these transport
parameters leads to a comparative measure between different thermodynamic dissipations and
electromagnetic responses. The mutual ratios between $\eta$, $\lambda$ and $\sigma_{el}$
can reflect the competition between momentum transport, heat transport and charge transport
in the medium respectively, indicating different physical laws that will be elaborated in next section.

As already mentioned the physical laws connecting different transport coefficients provide a comparative
study between the various collective behavior of the system under consideration. We start with the Wiedemann-Franz 
law which  states that the thermal conductivity of a system is proportional to its electrical conductivity times the
bulk temperature ($T$) of the system, such that $\lambda/(\sigma_{el}T)$ is a constant of temperature.
The ratio is known as Lorentz number, which for most of the system including metals is independent
of temperature depending only on the fundamental constants. In Ref.  \cite{Crossno}, a break-down of Wiedemann-Franz law 
has been reported for electron-hole plasma in graphene, indicating the signature of a strongly coupled
Dirac fluid. In the same context, it is interesting to look into the behavior of this law in the strongly
interacting QGP medium as well. Next, we focus on the relative behavior of viscous and thermal dissipation.
From ADS/CFT studies of strongly coupled thermal gauge theories in the framework of the gauge-gravity duality,
a value of the ratio between shear viscosity and thermal conductivity has been reported \cite {Son1} providing an 
analogue of the Wiedemann-Franz law between momentum transport and heat transport. For a system with
finite chemical potential $\mu$, the ratio states $\frac{\lambda \mu^2}{\eta T_{H}}=8\pi^2$, where $T_{H}$ is the 
Hawking temperature. It is more customary to express the relative importance of kinematic viscosity or shear
viscosity and thermal conductivity, in a dimensionless ratio called Prandtl number ($Pr$), given by
$Pr=\eta c_{p}/\rho \lambda$, where $c_{p}$ is the specific heat at constant pressure of the system and
$\rho$ is mass density of the system. In non-relativistic conformal holographic fluid
this number is estimated to be $Pr=1$, from ADS/CFT computations \cite{Son2}. In \cite{Braby}
the Prandtl number is estimated to be $Pr=2/3$ for a dilute atomic Fermi gas, which agrees
with the classical gas result. Finally, we mention about the relative behavior between shear
viscosity and electrical conductivity which characterizes the relative importance of momentum diffusion
and charge diffusion in a electromagnetically charged system that undergoes dissipation. We can specify
this comparison by observing the ratio of two dimensionless quantities, $(\eta/s)/(\sigma/T)$.
Since the electromagnetic responses are mostly carried by the charged components of the system,
{\it i.e}, by the quarks in a strongly interacting QGP (although the diffusion flow of quarks and gluons 
are constrained to be coupled with each other, so that the gluon interaction rate in effect enters in the
expression of electrical conductivity), whereas  both quarks and gluons participate
in momentum transport, the shear viscosity should dominate over the electrical conductivity
as predicted by \cite{Greco}, for strongly interacting QGP system. These physical laws
and the associated ratios of transport parameters, providing useful informations
about the dynamics and relative responses about the system, is instructive to relook 
for the QGP system, which is one of the major motivation of this work.

In order to provide the spectrum of the theoretical estimations of these transport 
quantities, we need to review the state of the art developments in recent literature.
Turning out to be an useful signature of the phase transition occurring in the medium created
in heavy ion collisions, the estimations of shear and bulk viscous coefficients have emerged as  
celebrated topics for quite some time both below and above the QCD transition temperature $T_c$. 
Above $T_{c}$ in the QGP sector, there are a number of estimations of the viscosities employing 
the transport theory approach utilizing kinetic theory of a many particle system  
\cite{Danielewicz,Baym,Thoma,Heiselberg,Hosoya,AMY1,AMY2,Chen1,Chen2,Toneev-viscosity,Greiner1,Greiner2,Schaefer-viscosity,Jeon-Yaffe}.
Under the application of Kubo formalism the QGP viscosities have also been obtained by evaluating 
the correlation functions using linear response theory in 
\cite{Jeon,Hosoya-Sakagami,Carrington,Basagoiti,Moore}.
Describing the in medium constituent quark interactions under the scheme of Nambu-Jona-Lasinio(NJL)
model, both the shear and bulk viscous coefficients have been estimated in
\cite{Sasaki,Deb,Ghosh1,Ghosh2}. The quasiparticle approach, introduced in order to describe
the hot QCD medium, has been employed to estimate the viscosities as well \cite{Plumari,Chandra1,Chandra2}. 
The temperature dependence of $\eta$ and $\zeta$ have been constrained from hydrodynamic
simulations and by comparing with the experimental data in \cite{Denicol,Ryu,Hirano2}.
The molecular dynamics simulations have been employed in \cite{Shuryak} to extract
the shear viscosity to entropy density ratio for a strongly coupled QGP.
Below the transition temperature, i.e, in the confined hadronic regime also a number
of estimates of viscous coefficients are available 
\cite{Gavin-hadron,Prakash,NoronhaHostler,Itakura,Chen-hadron,Davesne,Dobado1,Dobado2,Dobado3,Dobado4,Dobado5,
Buballa,Albright,Lang,Wiranata,Moore-hadron,Moroz,Chakraborty,Pal,Mitra1,Mitra2,Mitra3}.
For quite a few times the viscous coefficients are being analyzed from holographic predictions as well.
These viscous parameters are studied in great detail in a number of recent ADS/CFT based literature
employing holographic QCD models \cite{Sachin,Policastro,Cremonini1,Cremonini2,Cremonini3,Cremonini4,Li,Buchel}.
In comparison to the estimations of viscosities, the study of thermal conductivity has received much less attention
in the current scenario. However a few estimated of $\lambda$ are available both in partonic and hadronic
areas \cite{Gavin-hadron,Prakash,Davesne,Greif,Marty,Nam,Dobado-conductivity,Nicola1,Nicola2,Mitra4}.
Electrical conductivity, turning out to be an effective signature of electromagnetic responses in strongly
interacting systems, has attained a lot of interest recently. In the strongly coupled QGP, the 
relativistic transport theory, dynamical quasiparticle model (DQPM) and the maximum entropy method (MEM), 
has found a number of applications to estimate the value and temperature dependence of $\sigma_{el}$
\cite{Greco,Greiner-econd,Greco-econd,Cassing1,Cassing2,Qin,Patra,Mitra-Chandra}. From
the soft photon spectrum in heavy-ion collisions $\sigma_{el}$ has been extracted in \cite{Yin}.
Quite a considerable number of estimations of $\sigma_{el}$ are available from Lattice QCD
computations as well \cite{Amato,Aarts1,Aarts2,Gupta1,Brandt1,Brandt2,Ding,Francis}.
In hadronic sector the contributions from \cite{Fraile,Denicol-econd} are prior
to mention. Finally a number of holographic estimations have been proposed for both thermal and electrical
conductivities in \cite{Finazzo,Huot,Sachin-cond1,Sachin-cond2,Sachin-cond3,Bu}. 
A detailed comparison of the various transport coefficients obtained in the present work to the above 
mentioned  existing interesting works will be presented in the later part of the manuscript.

The manuscript is organized as follows. Section 2, includes the formal developments
of the transport theory, containing the Quasiparticle description of hot QCD medium, the evaluation
and temperature behavior of thermal relaxation times of quarks and gluons within the medium and
the detailed estimations of the transport coefficients in different subsections. The physical
laws concerning the ratios of different transport coefficients have been discussed in Section 3.
The obtained results have been discussed in Section 4. Finally in Section 5, the article has been 
summarized with providing possible outlooks of the work.

\section{Formalism: Transport theory}
Determination of transport coefficients for a hot QCD system needs modeling of the system away from   
equilibrium. Their determination can be done within two equivalent approaches, {\it viz.}, the 
correlator technique in QCD using Green-Kubo formula, and  the semiclassical transport theory
(Chapman-Enskog or Grad's 14 method). The present analysis is done following the latter approach. 
To initiate the formalism, an appropriate modeling of the equilibrium, isotropic momentum distributions 
of gluons and quark-antiquarks in the hot QCD medium at vanishing or non-vanishing baryon density 
(whatever be the case), is required to be provided. This could be systematically  done by adopting 
an effective modeling of the hot QCD medium effects, encoded in the interacting QCD/QGP equations of 
state. To that end, the well accepted effective fugacity quasi-particle proposed by Chandra and Ravishankar 
~\cite{Chandra_quasi1,Chandra_quasi2,Chandra_quasi3} (EQPM) serves the current purpose which has been 
discussed below. The quasi-particle modeling of the system properties is followed by the estimations 
of the essential ingredients such as thermal relaxation times of interacting partons and other related 
quantities that are necessary while determining  various transport coefficients under consideration here. Finally 
the complete formalism for extracting the transport coefficients is presented with all the required
mathematical details below.
 
\subsection{Effective modeling of momentum distributions of gluons and matter sector}
The QCD medium at high temperature can conveniently be realized in terms of its effective quasi-particle 
degrees of freedom, {\it viz.}, the quasi-gluons and quasi-quarks/antiquarks with non-trivial dispersion 
relations. There have been several quasi-particle models, proposed over the last few decades, to describe 
the hot QCD equations of state in terms of non-interacting or weakly interacting effective gluons and effective 
quarks and anti-quarks. The  effective mass  models~\cite{Peshier1,Peshier2,Peshier3,Bannur1,Bannur2,Bannur3,Rebhan,Thaler,Szabo}, 
the effective mass models with Polyakov loop~\cite{DElia1,DElia2,Castorina1,Castorina2,Alba}, describe the 
medium effects in terms of effective thermal mass or effective coupling in the medium. In these models, 
thermodynamic consistency condition is needed to be handled carefully, sometimes by introducing a few additional 
temperature dependence parameters. Another set of these models include, the NJL (Nambu Jona Lasinio) and the 
PNJL (Polyakov loop extended Nambu Jona Lasinio) based effective models~\cite{Dumitru,Fukushima,SKGhosh,Abuki}. 
The EQPM which has been employed here, is described below in details. 

\subsubsection{The EQPM and its extension to finite quark chemical potential}
The EQPM models the hot QCD medium effects in terms of effective quasi-partons (quasi-gluons, quasi-quarks/antiquarks). 
The main idea is to map the hot QCD medium effects present in the hot QCD EOSs either computed within 
improved perturbative QCD (pQCD) or lattice QCD simulations, into the effective equilibrium distribution functions for 
the quasi-partons. The  EQPM for the QCD EOS at $O(g^5)$ (EOS1) and $O(g^6\ln(1/g)+\delta)$ (EOS2) along with a recent 
(2+1)-flavor lattice QCD EOS (LEOS) ~\cite{Cheng1} at physical quark masses, have been exploited in the present manuscript.
Note that, there are more recent lattice results with the improved hot QCD actions and refined lattices
\cite{Bazavov,Cheng2,Borsanyi1,Borsanyi2}, for which,  we need to re look the model with specific set of 
lattice data (specifically to define the effective gluonic degrees of freedom). Therefore, we will stick to the set of 
lattice data utilized in the model described in Ref.~\cite{Chandra_quasi2} and leave the issue for further investigations 
in near future.

In either of these EOSs,  form of the quasi-parton equilibrium distribution functions, 
 $ f_{eq}\equiv \lbrace f_{g}, f_{q, \bar{q}} \rbrace$  (describing the strong interaction effects in terms of effective fugacities $z_{g,q}$) can be written as,
\be
\label{eq1}
f_{g/q}= \frac{z_{g/q}\exp[-\beta E_p]}{\bigg(1\mp z_{g/q}\exp[-\beta E_p]\bigg)}
\ee
where $E_p=|\vec{p}|$ for the gluons and $\sqrt{|\vec{p}|^2+m_q^2}$ for the quarks/antiquarks  ($m_q$ denotes the mass of the quarks).
The parameter, $\beta=T^{-1}$ denotes inverse of the 
temperature, $\nu_g=2(N_c^2-1)$ denotes the gluonic degrees of freedom and 
$\nu_{q}=\nu_{\bar{q}}=2 N_c N_f$, are the quark-antiquark degrees of freedom for $SU(N_c)$ with $N_f$ number of flavors. 
Since the model is valid in the deconfined phase of QCD (beyond $T_c$), therefore, the mass of the light 
quarks can be neglected while comparing it with the temperature. Noteworthily, the EOS1 which is fully 
perturbative, is  proposed by Arnold and Zhai~\cite{Arnold-Zhai1,Arnold-Zhai2} and Zhai and Kastening
\cite{Zhai-Karstening} and the EOS2 which is at $O(g^6\ln(1/g)+\delta)$ is  determined by Kajantie 
{\it et al.}~\cite{Kajantie-Laine} while incorporating contributions from non-perturbative scales such as 
$g T$ and $g^2 T$.  In the case of vanishing baryon density, $f_q\equiv f_{\bar{q}}$.

It is important to note that these effective fugacities, $z_{g/q}$ are not merely temperature dependent 
parameters that encode the hot QCD medium effects; they lead to non-trivial dispersion relation both in 
the gluonic and quark sectors as,
\begin{equation}
\omega_{g/q}=E_p+T^2\partial_T ln(z_{g/q}),
\label{dispersion} 
\end{equation}

where $\omega_{g,q}$ denote the quasi-gluon and quasi-quark dispersions (single particle energy) respectively.
The second term in the right-hand side of Eq.(\ref{dispersion}), encodes the effects from collective excitations 
of the quasi-partons.

The extension of the model to finite baryon/quark chemical potential is quite straightforward. This could be done 
by introducing the quark-chemical potentials ($\mu_q$) in the momentum distributions in the matter sector as:
\be
f_{q/\bar{q}}= \frac{z_q \exp[-\beta (E_p\mp \mu_{q})]}{\bigg(1+ z_{q}\exp[-\beta (E_p\mp \mu_{q})]\bigg)}
\ee

It is important to note that the temperature dependence of effective fugacities, $z_g, z_q$ are set while 
implementing the EOS1, EOS2 and LEOS in terms of EQPM. In other words, while extending the EQPM for finite 
but small ($\mu_q/T<<1$) baryon densities, the same expressions for $z_g$ and $z_q$ have employed so that 
one can get the correct limit in case where $\mu_q=0$. The effective fugacities,  $z_g, z_q$ are not related 
with any conserved number current in the hot QCD medium. They have been merely introduced to encode the hot 
QCD medium effects in the EQPM. The physical interpretation of $z_g$ and $z_q$ emerges from the above mentioned 
non-trivial dispersion relations. The modified part of the energy dispersions in Eq. ({\ref{dispersion}) leads 
to the trace anomaly (interaction measure) in hot QCD and takes care of the thermodynamic consistency condition.  
It is straightforward to compute, gluon and quark number densities and all the thermodynamic quantities such as 
energy density, entropy, enthalpy {\it etc.} by realizing the hot QCD medium in terms of an effective Grand 
canonical system~\cite{Chandra_quasi1,Chandra_quasi2}. Furthermore, these effective fugacities lead to a very 
simple interpretation of hot QCD medium effects in terms of an effective Virial expansion. Note that $z_{g,q}$ 
scales with $T/T_c$, where $T_c$ is the QCD transition temperature. For the current analysis $T_c$ has been taken 
to be $170$ MeV. All the relevant thermodynamic quantities 
such as energy density, number density, pressure, entropy density, speed of sound {\it etc.} could 
straightforwardly be obtained in terms of $f_g$, $f_{q,\bar{q}}$ following their basic definitions. The 
detailed evaluation of these quantities at finite ($\mu_q$) are mentioned in the Appendix-B. The EQPM has 
recently been extended to non-extensive statistical systems keeping in view that the dense hadronic or QCD 
matter is generally produced in a nonextensive environment where the usual Boltzmann-Gibbs statistics is 
questionable and so the Tsallis statistics is being applied~\cite{Wilk1,Tsallis}.

\subsubsection{Charge renormalization and effective coupling at finite $T$ and $\mu_q$}
In contrast to the effective mass models where the effective mass is motivated from the mass renormalization  
in the hot QCD medium, the EQPM is based on the charge renormalization in high temperature QCD. This could be 
realized by computing the expression for the Debye mass in the medium following its definition that is derived 
in semi-classical transport theory~\cite{Kelly1,Kelly2,Litim,Blaizot} as, 
 \begin{eqnarray}
 \label{dm}
  m_D^2=4 \pi \alpha_{s}(T,\mu_q) \bigg(-2 N_c \int \frac{d^3 p}{(2 \pi)^3} \partial_p f_g (\vec{p})
  -  N_f  \int \frac{d^3 p}{(2 \pi)^3} \partial_p (f_q (\vec{p})+f_{\bar{q}}(\vec{p}))\bigg)
 \end{eqnarray}
 
 where, $\alpha_{s}(T)$ is the QCD running coupling constant at finite temperature and chemical potential
  
\begin{figure}[h]
\includegraphics[scale=0.55]{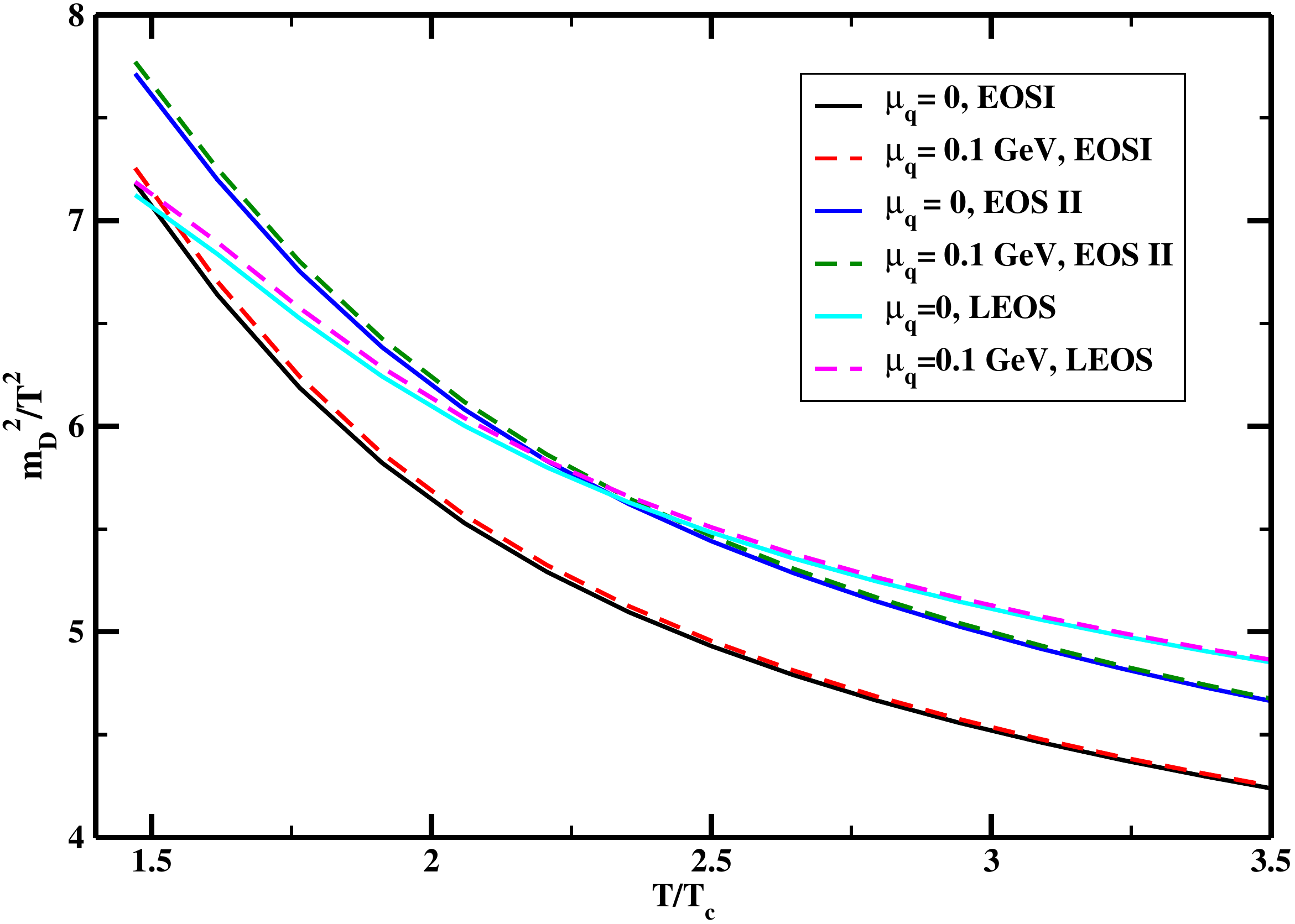}
\caption{(Color online)Effective coupling constant using various EOSs as a function of $T/T_c$.}
\label{alphas}
\end{figure} 

After performing the momentum integral and substituting the quasi-parton distribution function from 
Eq.(\ref{eq1}) to Eq.(\ref{dm}), we are left with the expression of Debye mass within the scheme of
EQPM model with finite quark chemical potential, upto the order $O(\tilde{\mu_q}^2)$ (since $\tilde{\mu_q}^2<<1$),
 \begin{eqnarray}
  m_D^2=4 \pi \alpha_{s}(T, \mu_q) T^2 \bigg\{ \bigg( \frac{2 N_c}{\pi^2} \textrm{PolyLog}[2,z_g]-\frac{2 N_f}{\pi^2} \textrm{PolyLog}[2,-z_q]\bigg) 
  +  {\tilde{\mu_q}^2} \frac{N_f}{\pi^2}\frac{z_q}{1+z_q}\bigg\}
 \label{dm1}
 \end{eqnarray}
 
 The Debye mass here reduces to the leading order HTL expression in the limit $z_{g,q}\rightarrow1$ 
 (ideal EoS: non-interacting ultra relativistic quarks and gluons),
 
 \begin{equation}
 \label{dm0}
 m_D^2(\textrm{HTL})=4 \pi \alpha_{s}(T,\mu_q) \ T^2 \bigg\{(\frac{N_c}{3}+\frac{N_f}{6})+ {\tilde{\mu_q}^2}\frac{N_f}{2\pi^2}\bigg\}~.
  \end{equation}

From Eq.(\ref{dm1}) and (\ref{dm0}), the effective coupling can be defined as:
\be
\alpha_{eff} (T, \mu_q)= \alpha_{s}(T, \mu_q)\times 
\frac{\big\{\frac{2 N_c}{\pi^2} \textrm{PolyLog}[2,z_g]-\frac{2 N_f}{\pi^2} \textrm{PolyLog}[2,-z_q]\big\}+
\tilde{\mu_q}^2\big\{\frac{N_f}{\pi^2} \frac{z_q}{1+z_q}\big\}}{\big\{\frac{N_c}{3}+\frac{N_f}{6}\big\} + \tilde{\mu_q}^2\big\{\frac{N_f}{2\pi^2} \big\}}~.
\ee   
The behavior of the ratio $m_D^2/T^2$ as a function of temperature ($T/T_c$) for various EOSs and finite $\mu_q$ is depicted in 
Fig. \ref{alphas}.  As expected, the finite but small $\mu_q$ effects are quite visible at lower temperatures, which are merging
with the zero quark chemical potential cases at higher temperatures. This is  seen to be valid for all the three EOSs considered 
here. The medium effects (thermal) manifested through the temperature dependent $z_{g,q}$, play the crucial role in modulating 
the quantity $m_D^2/T^2$ as a function of temperature.

There are only three free  functions  ($z_g$, $z_q$,  and $\tilde{\mu}_q$) in the EQPM employed here. The first two, depend on 
the chosen EOS. In the case of EOS1 and EOS2 employed in the present case, these functions are obtained in~\cite{Chandra_quasi1} 
and are continuous functions of $T/T_c$. On the other hand, for LEOS they are defined in terms of eight parameters obtained in 
Ref.~\cite{Chandra_quasi2} (See Table I of Ref.~\cite{Chandra_quasi2}). The quantity $\tilde{\mu_q}$ is chosen to be $0.0$ and 
$0.1$ GeV throughout our analysis. In addition, the effective coupling mentioned above depends on the QCD running coupling 
constant $g(T, \mu_q)=\sqrt{4\pi\alpha_s}$, that explicitly depends upon how we fix the QCD renormalization scale at finite temperature and $\mu_q$, 
and up to what order we define $g(T,\mu_q)$.  Henceforth, these are the only quantities that are needed to be supplied throughout the 
analysis here.

Notably, the EQPM employed here has been remarkably useful in understanding the bulk and the transport properties of the QGP in 
heavy-ion collisions~\cite{Chandra_eta1,Chandra_eta2,Chandra_dilep1,Chandra_dilep2,Chandra_hq1,Chandra_hq2,Chandra_hq3}.
Before, the formalism for estimations of thermal relaxation times of the constituent gluons and quarks are being discussed, 
it is important to highlight the utility of quasi-particle models in the context of understanding the bulk and transport properties 
of the hot QCD/QGP medium created out of the heavy-ion collisions. As already mentioned transport parameters of the QGP have 
estimated employing various quasiparticle models in~\cite{Plumari,Chandra1,Chandra2,Albright,Chandra_eta1,Chandra_eta2,Bluhm}. 
Note that Ref.~\cite{Bluhm} offered the estimation of $\eta$ and $\zeta$ for pure gluon plasma employing the effective mass 
quasi-particle model. On the other hand, Refs.~\cite{Chandra1,Chandra2,Chandra_eta1,Chandra_eta2}, reported their estimation 
for both gluonic as well as matter sector. Refs.~\cite{Chakraborty,Albright}, presented the quasi-particle estimations of 
$\eta$ and $\zeta$ in hadronic sector. The thermal conductivity has also been studied, in addition to the viscosity parameters
\cite{Albright}, within the effective mass model at finite baryon density. 

\subsection{Thermal Relaxation times}
As mentioned earlier, the microscopic interactions between the constituents
of the system, provide the dynamical inputs for different transport coefficients. Here, 
it is done by introducing the  thermal relaxation times of the partons, which in turn, 
introduce the transport cross sections  to the expressions of the transport coefficients.

In order to define the thermal relaxation times for quasi quarks/anti quarks and gluons, we start
with the relativistic transport equation of the momentum distribution functions of the constituent
partons in an out of equilibrium, multicomponent system that describes the binary elastic process 
$p_{k}+p_{l}\rightarrow p'_{k}+p'_{l}$,

\begin{equation}
 p_{k}\partial_{\mu}f_{k}=\sum_{l=1}^{N} C_{kl}[f_{k},f_{l}],~~~~~~~~~~~~~~~~[k=1,2,..........N]~.
\label{eq-R1}
\end{equation}

Here $f_{k}$ is the single particle distribution function for the $k^{th}$ species, that depends upon 
the particle 4-momentum $p_{k}$ and 4-space-time coordinates $x$. Here, the right hand side of Eq.(\ref{eq-R1}) denotes the collision term 
that quantifies the rate of change of $f_{k}$. For each $l$, $C_{kl}[f_{k},f_{l}]$ defines the collision contribution due to
the scattering of $k^{th}$ particle with $l^{th}$ one given in the following manner ~\cite{AMY1},

\begin{eqnarray}
 &&C_{kl}[f_{k},f_{l}]=\frac{1}{2} \frac{\nu_{l}}{2} \int d\Gamma_{p^{}_{l}} d\Gamma_{p'_{k}} d\Gamma_{p'_{l}} (2\pi)^4 
          \delta^{4}(p_{k}+p_{l}-p'_{k}-p'_{l}) \langle|M_{k+l\rightarrow k+l}|^{2}\rangle \nonumber\\ 
          &&[f_{k}(p'_{k}) f_{l}(p'_{l}) \{1\pm f_{k}(p_{k})\}\{1\pm f_{l}(p_{l})\}-
             f_{k}(p_{k})f_{l}(p_{l})\{1\pm f_{k}(p'_{k})\}\{1\pm f_{l}(p'_{l})\}]~.
 \label{eq-R2}        
\end{eqnarray}

The phase space factor is given by the notation $d\Gamma_{p_{i}}=\frac{d^3 \vec {p_{i}} }{(2\pi)^3 2\omega_i}$,
as $\omega_{k}$ is the energy of the scattered particle (of the $k^{th}$ species). 
The overall $\frac{1}{2}$ factor appears due to the symmetry in order to compensate for the double counting of
final states that occurs by interchanging $p'_{k}$ and $p'_{l}$. $\nu_{l}$ is the degeneracy of $2^{nd}$ particle 
that belongs to $l^{th}$ species. 

In the next section it will be shown that upto the next to leading order, the out of equilibrium
distribution function is constructed as follows,

\begin{eqnarray}
 f_{k}=f_{k}^0 +\delta f_{k}=f_{k}^0 + f_{k}^0 (1\pm f_{k}^0) \phi_{k}~,
 \label{eq-R3}
\end{eqnarray}
 
where the non-equilibrium part $\delta f_{k}$ of the distribution function is quantified by the deviation
function $\phi_k$. The distribution functions of the quasi partons at local thermal equilibrium is given by 
Eq.(\ref{eq1}).

In the next section, we will see that the simplest method of linearizing the transport equation (\ref{eq-R1})is 
to replace the collision term by the rate of change of the distribution function over the thermal relaxation time $\tau_k$
which is needed by the out of equilibrium distribution function to restore its equilibrium value, 
such that the transport equation becomes, 

\begin{equation}
 \frac{df_k}{dt}=-\frac{\delta f_{k}}{\tau_k}=-\frac{(f_{k}-f_{k}^{0})}{\tau_{k}}~. 
 \label{eq-R4}
\end{equation}

Consequently the collision term becomes,

\begin{equation}
C_{kl}[f_{k},f_{l}]=-\omega_k\frac{\delta f_{k}}{\tau_k}=-\omega_k \frac{f_{k}^0 (1\pm f_{k}^0) \phi_{k}}{\tau_k}~.
\label{eq-R5}
\end{equation}

Putting (\ref{eq-R3}) into the right hand side of (\ref{eq-R2}) by assuming the distribution functions of 
the particles other than the scattered one are very much close to equilibrium and comparing with (\ref{eq-R5}), 
the relaxation time finally becomes as the inverse of the reaction rate $\Gamma_k$ of the respective processes 
~\cite{Zhang},

\begin{eqnarray}
\tau_{k}^{-1}\equiv\Gamma_{k}=&&\frac{\nu_{l}}{2} \frac{1}{2\omega_{k}} \int d\Gamma_{p_{l}} d\Gamma_{p'_{k}} d\Gamma_{p'_{l}} (2\pi)^4
\delta^{4}(p_{k}+p_{l}-p'_{k}-p'_{l}) \nonumber\\
&&\langle|M_{k+l\rightarrow k+l}|^{2}\rangle 
\frac{f_{l}^{0} (1\pm f_{k}^{'0}) (1\pm f_{l}^{'0})}{(1\pm f_{k}^{0})}~.
\label{eq-R6}
\end{eqnarray}
Clearly the distribution function of final state particles are given by primed notation.

Simplifying $\tau_{k}$ utilizing the $\delta$-function we finally obtain its expression in the center of 
momentum frame of particle interaction as,

\begin{eqnarray}
 \tau_k^{-1}=\Gamma_{k}=
 \nu_{l}\int \frac{d^3 \vec{p_{l}}}{(2\pi)^3} d(\cos\theta) \frac{d\sigma}{d(\cos\theta)} 
 \frac{f_{l}^{0} (1\pm f_{k}^{'0}) (1\pm f_{l}^{'0})}{(1\pm f_{k}^{0})}~,
 \label{eq-R7}
\end{eqnarray}

where $\theta$ is the scattering angle in the center of momentum frame and $\sigma$ is the interaction
cross section for the respective scattering processes. Now in terms of the Mandelstam variables $s,t$ 
and $u$ the expression for $\tau_{k}$ can be reduced simply as,

\begin{equation}
 \tau_{k}^{-1}=\Gamma_{k}=\nu_{l}\int \frac{d^3 \vec{p_{l}}}{(2\pi)^3} dt \frac{d\sigma}{dt} 
 \frac{f_{l}^{0} (1\pm f_{k}^{'0}) (1\pm f_{l}^{'0})}{(1\pm f_{k}^{0})}~.
 \label{eq-R8}
\end{equation}

The differential cross section relates the scattering amplitudes as $\frac{d\sigma}{dt}=\frac{\langle|M|^2\rangle}{16\pi s^2}$.
The QCD scattering amplitudes for $2\rightarrow2$ binary, elastic processes are taken from ~\cite{Combridge}, 
that are averaged over the spin and color degrees of freedom of the initial states and summed over the  
final states. The inelastic processes like $q\overline{q}\rightarrow gg$, have been ignored in the present
case, because they do not have a forward peak in the differential cross section and thus their contributions 
will presumably be small compared to the elastic ones.
 
Now in order to take into account the small-angle scattering scenario that results into divergent contributions 
from $t$-channel diagrams of QCD interactions, a transport weight factor $(1-\cos\theta)=\frac{2tu}{s^2}$ 
has been introduced in the interaction rate ~\cite{Hosoya}. Furthermore considering the momentum transfer
$q=|\vec{p_k}-\vec{p'_{k}}|=|\vec{p_{l}}-\vec{p'_{l}}|$ is not too large we can make following assumptions,
$f_{k}^{0}\cong f_{k}^{'0}$ and $f_{l}^{0}\cong f_{l}^{'0}$ ~\cite{Thoma1} to finally obtain,

\begin{equation}
 \tau_k^{-1}=\Gamma_{k}=\nu_{l}\int \frac{d^3 \vec{p_{l}}}{(2\pi)^3} f_{l}^{0}(1\pm f_{l}^{0})
 \int dt \frac{d\sigma}{dt} \frac{2tu}{s^2}~.
 \label{eq-R9}
\end{equation}

This additional transport factor changes the infrared and ultraviolet behavior of the interaction rate 
quite significantly. Due to inclusion of this term all the higher order divergences reduce to simple
logarithmic singularities which can be simply handled by putting a small angle cut-off in the integration limit.
In the integration involving $t$-channel diagrams from where the infrared logarithmic singularity appears,
the limit of integration is restricted from $-s$ to $-k^2$ in order to avoid those divergent results 
using the cut-off $k^2=g^2 T^2$ as infrared regulator. Here $g^2=4\pi\alpha_s$ with $\alpha_s$ being 
the coupling constant of strong interaction as already mentioned in Section 1. 

Now in the QGP medium the quark and gluon interaction rates result from the following interactions respectively,
\begin{equation}
 \Gamma_g=\Gamma_{gg}+\Gamma_{gq}+\Gamma_{g\overline{q}}~,~~~~~~
 \Gamma_q=\Gamma_{qg}+\Gamma_{qq}+\Gamma_{q\overline{q}}~,~~~~~~
 \Gamma_{\overline{q}}=\Gamma_{\overline{q}g}+\Gamma_{\overline{q}q}+\Gamma_{\overline{q}\overline{q}}~
 \label{eq-R10}
\end{equation}
where $\Gamma_{kl}$, is the interaction rate of $k^{th}$ particle due to scattering with the $l^{th}$ one.

Finally after pursuing the angular integration in (\ref{eq-R9}) we are left with the thermal relaxation times of the 
quark, antiquark and gluon components in a QGP system in the following way,

\begin{eqnarray}
 \tau_g^{-1}=&&\{\nu_{g}\int \frac{d^3 \vec{p_{g}}}{(2\pi)^3} f_{g}^{0}(1+f_{g}^{0})\} 
              [\frac{9g^4}{16\pi\langle s \rangle_{gg}}\{ln\frac{\langle s \rangle_{gg}}{k^2}-1.267 \}] \nonumber\\
            +&&\{\nu_{q}\int \frac{d^3 \vec{p_{q}}}{(2\pi)^3} f_{q}^{0}(1-f_{q}^{0})\}
              [\frac{g^4}{4\pi\langle s \rangle_{gq}}\{ln\frac{\langle s \rangle_{gq}}{k^2}-1.287 \}] \nonumber\\
            +&&\{\nu_{\overline{q}}\int \frac{d^3 \vec{p_{\overline{q}}}}{(2\pi)^3} f_{\overline{q}}^{0}(1-f_{\overline{q}}^{0})\}
              [\frac{g^4}{4\pi\langle s \rangle_{g\overline{q}}}\{ln\frac{\langle s \rangle_{g\overline{q}}}{k^2}-1.287 \}] ,
 \label{eq-R11}
 \end{eqnarray}
 
 \begin{eqnarray}
 \tau_q^{-1}=&& \{\nu_{g}\int \frac{d^3 \vec{p_{g}}}{(2\pi)^3} f_{g}^{0}(1+f_{g}^{0})\}
               [\frac{g^4}{4\pi\langle s \rangle_{qg}}\{ln\frac{\langle s \rangle_{qg}}{k^2}-1.287 \}] \nonumber \\
            +&& \{\nu_{q}\int \frac{d^3 \vec{p_{q}}}{(2\pi)^3} f_{q}^{0}(1-f_{q}^{0})\}
               [\frac{g^4}{9\pi\langle s \rangle_{qq}}\{ln\frac{\langle s \rangle_{qq}}{k^2}-1.417 \}] \nonumber \\
            +&& \{\nu_{\overline{q}}\int \frac{d^3 \vec{p_{\overline{q}}}}{(2\pi)^3} f_{\overline{q}}^{0}(1-f_{\overline{q}}^{0})\}
               [\frac{g^4}{9\pi\langle s \rangle_{q\overline{q}}}\{ln\frac{\langle s \rangle_{q\overline{q}}}{k^2}-1.417 \}] ~   ,
 \label{eq-R12}
\end{eqnarray}

\begin{figure*}[h]
\label{relax}
\includegraphics[scale=0.32]{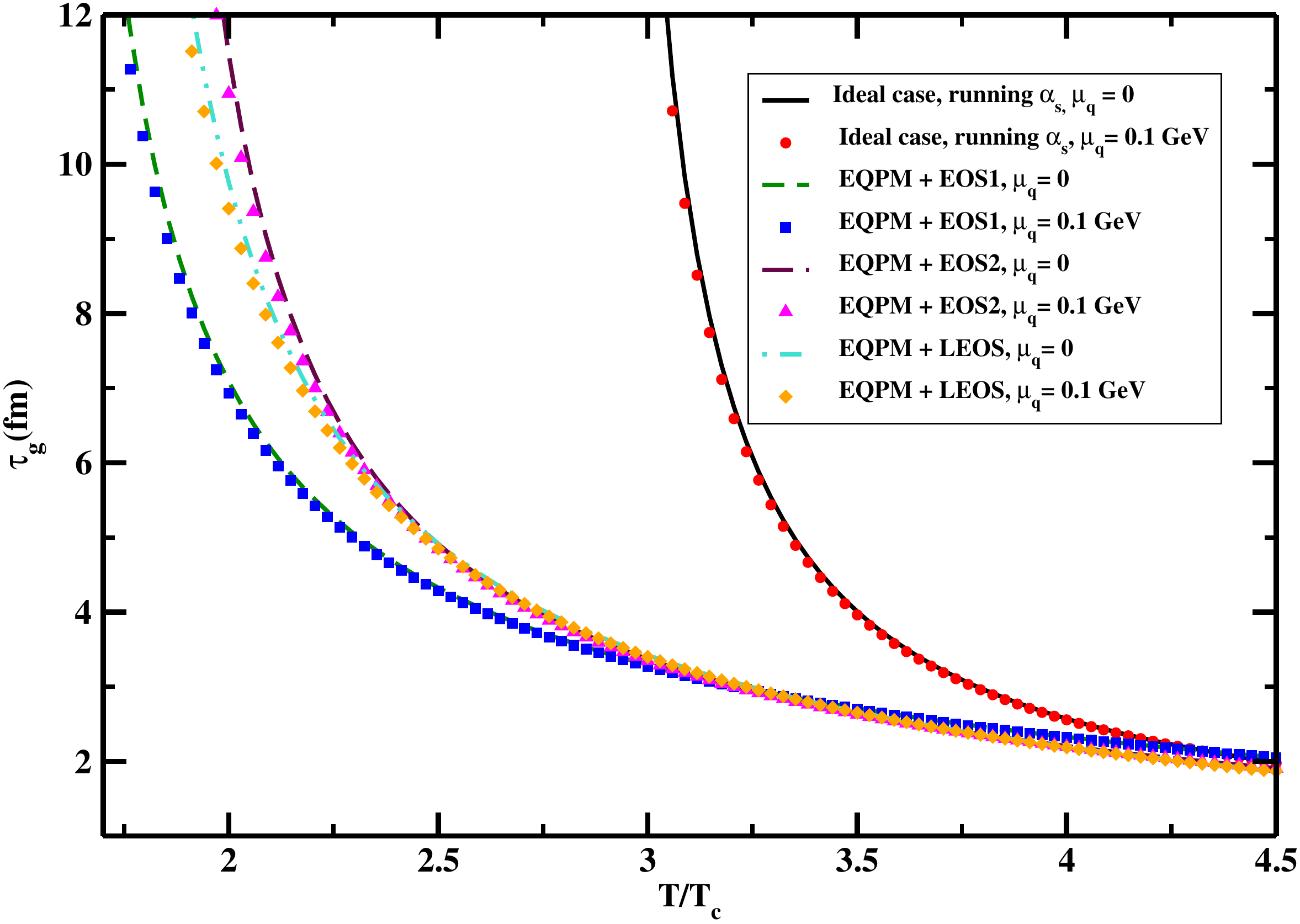}
\includegraphics[scale=0.32]{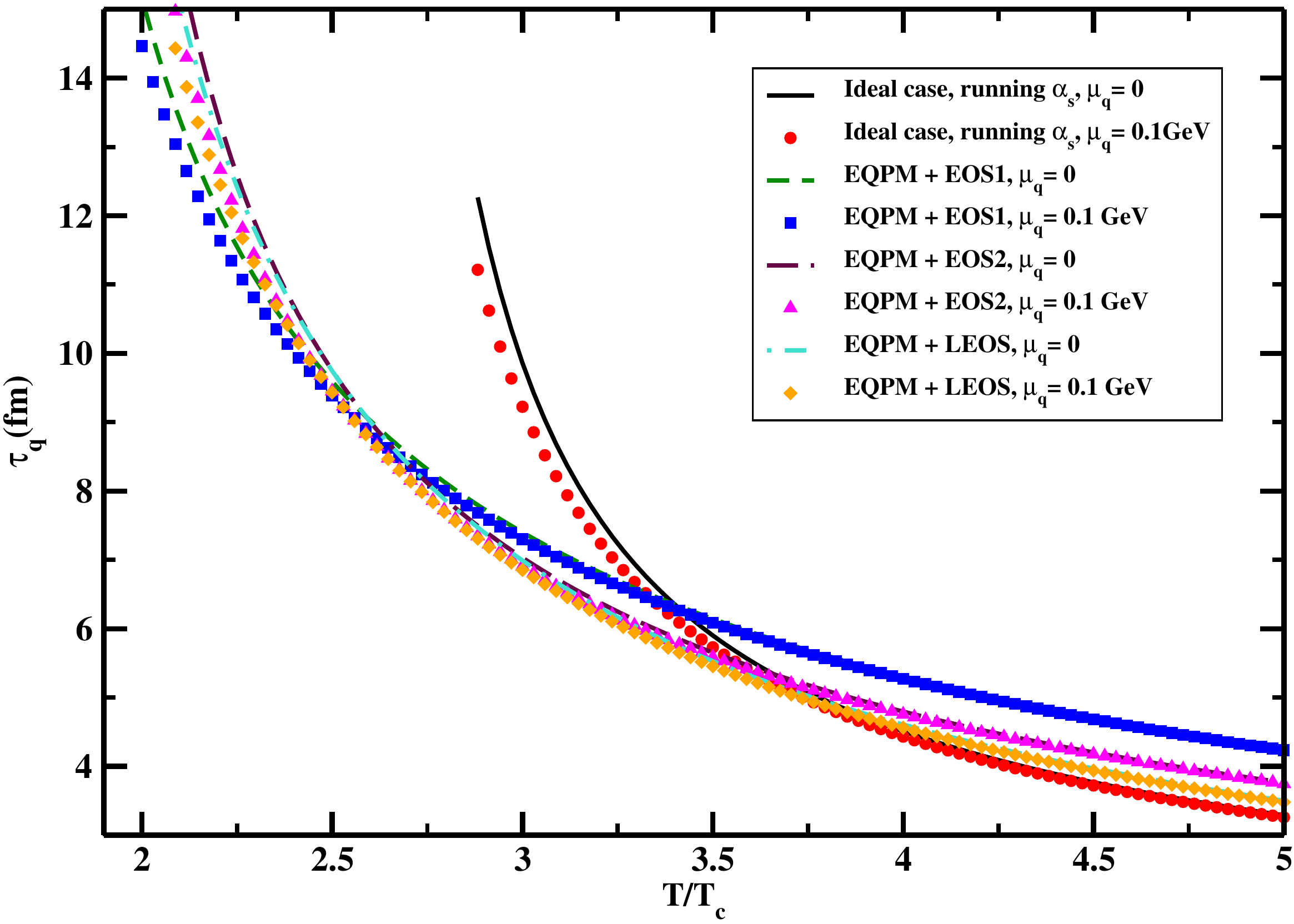}
\caption{Thermal relaxation times for gluons and quarks  using various EOSs as a function of $T/T_c$ at fixed $\tilde{\mu}_q$.}
\end{figure*} 

\begin{eqnarray}
 \tau_{\overline{q}}^{-1}=&& \{\nu_{g}\int \frac{d^3 \vec{p_{g}}}{(2\pi)^3} f_{g}^{0}(1+f_{g}^{0})\}
               [\frac{g^4}{4\pi\langle s \rangle_{\overline{q}g}}\{ln\frac{\langle s \rangle_{\overline{q}g}}{k^2}-1.287 \}] \nonumber \\
            +&& \{\nu_{q}\int \frac{d^3 \vec{p_{q}}}{(2\pi)^3} f_{q}^{0}(1-f_{q}^{0})\}
               [\frac{g^4}{9\pi\langle s \rangle_{\overline{q}q}}\{ln\frac{\langle s \rangle_{\overline{q}q}}{k^2}-1.417 \}] \nonumber \\
            +&& \{\nu_{\overline{q}}\int \frac{d^3 \vec{p_{\overline{q}}}}{(2\pi)^3} f_{\overline{q}}^{0}(1-f_{\overline{q}}^{0})\}
               [\frac{g^4}{9\pi\langle s \rangle_{\overline{q}\overline{q}}}\{ln\frac{\langle s \rangle_{\overline{q}\overline{q}}}{k^2}-1.417 \}] ~   ,
 \label{eq-R13}
\end{eqnarray}

where $\langle s \rangle_{kl}=2\langle p_{k} \rangle \langle p_{l} \rangle$ is the thermal average value of $s$ with
$\langle p_k \rangle =\frac{\int \frac{d^3 \vec{p_{k}}}{(2\pi)^3} |\vec{p_{k}}| f_{k}^{0}}{\int \frac{d^3 \vec{p_{k}}}{(2\pi)^3} f_{k}^{0}}$.
Clearly in order to account for a hot QCD medium the quasiparticle effects must be invoked in the expressions
of these thermal relaxation times obtained far. As discussed in Section 2.1, the distribution functions of quarks
and gluons and the coupling $g$ will carry the quasiparticle descriptions accordingly. Since the cut-off parameter
$k$ also depends upon $g$ and the thermal average of $s$ includes $f^{0}_{g,q,\overline{q}}$, they will reflect the hot QCD equation 
of state effect as well. Following the definition of equilibrium distribution function of quarks and gluons from Eq.(\ref{eq1}), 
within the quasiparticle framework, the thermal averages of gluon and quark momenta respectively  are obtained as,  
\begin{eqnarray}
 \langle p_g \rangle =&&3T\frac{\textrm{PolyLog}[4,z_{g}]}{\textrm{PolyLog}[3,z_{g}]}~,\\
 \langle p_q \rangle =&&3T\frac{\textrm{PolyLog}[4,-z_{q}]
              +\tilde{\mu_q}\textrm{PolyLog}[3,-z_{q}]
              +\frac{(\tilde{\mu_q})^2}{2}\textrm{PolyLog}[2,-z_{q}]}
              {\textrm{PolyLog}[3,-z_{q}]
              +\tilde{\mu_q}\textrm{PolyLog}[3,-z_{q}]
              -\frac{(\tilde{\mu_q})^2}{2}ln(1+z_q)}~,\\
 \langle p_{\overline{q}} \rangle =&&3T\frac{\textrm{PolyLog}[4,-z_{q}]
              -\tilde{\mu_q}\textrm{PolyLog}[3,-z_{q}]
              +\frac{(\tilde{\mu_q})^2}{2}\textrm{PolyLog}[2,-z_{q}]}
              {\textrm{PolyLog}[3,-z_{q}]
              -\tilde{\mu_q}\textrm{PolyLog}[3,-z_{q}]
              -\frac{(\tilde{\mu_q})^2}{2}ln(1+z_q)}~.              
\end{eqnarray}

The degeneracy factors used, are $\nu_g=2\times8=16$, $\nu_q=\nu_{\overline{q}}=2\times N_c\times N_f$, with
$N_f$ and $N_c$ are the quark number of flavors and colors respectively. From the above analysis, it turned out that the thermal relaxation times at a particular $\mu_q$ follows the 
form given as,
\be
\tau_{q/\overline{q},g}^{-1} \sim T\alpha_s^2 ln \bigg\{\frac{1}{\alpha_s} \bigg\}~.
\label{eq-R14}
\ee

In Fig.(\ref{relax}), the temperature dependence of the thermal relaxation times of quasi gluons and quarks, 
obtained from Eq.(\ref{eq-R11}) and (\ref{eq-R12}) respectively, have been plotted as a function of $T/T_c$. The 
temperature dependence of $\tau$ for both the gluonic and quark components, is observed to exhibit the obvious 
decreasing trend with increasing temperature, revealing that the enhanced interaction rates at higher temperatures 
make the thermal quarks and gluons restore down their equilibrium faster. We observe at a particular temperature, 
$\tau_q$ is quantitatively little greater than $\tau_g$, indicating stronger interaction rates of the gluonic part. 
The order of magnitude of the relaxation times and the fact that $\tau_q$ is larger than $\tau_g$, agree with the work 
given in \cite{Baym}.  

In the present case the $\tau_g$ and $\tau_q$ have been estimated for three different EOSs (EOS1, EOS2, LEOS)
within the scope of EQPM along with ideal EOS with running coupling and also for two different quark chemical 
potentials ($\mu_q=0, \ 0.1$GeV). As noticed earlier in the case of QCD coupling, here also the finite quark chemical 
potential effects are only significant at lower temperatures which almost diminishes at higher temperature
regions. The large values of $\tau_g$ and $\tau_q$ for the ideal case at lower temperature mostly result from 
the higher values of running $\alpha_s(T,\mu_q) (\sim0.4)$ compare to the $\alpha_{eff}(T,\mu_q) (\sim0.3)$, 
contributing through the logarithmic term. However at higher temperatures, the plots of $\tau$'s including the 
quasiparticle equation of state effects are merging with the ideal ones, as at those temperature regions the 
quasiparticle properties almost behave like that of the free particles. Three different set of plots with EQPM 
calculations are clearly showing the distinct effects of separate EOSs. In each set, the small but finite effects
of non-zero $\mu_q$ is observed at lower temperatures, which is more predominant in the plots of $\tau_q$ for 
obvious reasons. So we conclude that first the logarithmic term in Eq.(\ref{eq-R14}) is playing here the key role 
in determining the temperature behavior of $\tau$'s and secondly different EOSs describing the interacting medium 
through various models (pQCD or Lattice) and the non-zero $\mu_q$ is providing considerable effects on it.

\subsection{Estimation of transport coefficients in Chapman-Enskog method}

The basic scheme of determining the transport coefficients of a many particle system resides in comparing
the macroscopic and microscopic definition of thermodynamic flows. The description of irreversible phenomena 
taking place in non equilibrium systems is characterized by two kinds of concepts : the thermodynamic forces 
and thermodynamic flows. The first ones create spatial non-uniformities of the macroscopic thermodynamic 
state variables where the later tend to restore back the equilibration situation by wiping out these non-uniformities. 
Phenomenologically, one finds to a good approximation that these fluxes are linearly related to the thermodynamic forces
where the proportionality constants are termed as transport coefficients. As a consequence the irreversible part of the 
energy momentum tensor and the heat flow can be expressed in a linear law, directly proportional to the corresponding 
thermodynamic forces which is respectively the velocity gradient and temperature gradient of the system. From the second 
law of thermodynamics, it is known that the restoration of equilibrium  is achieved by the processes which involve 
increasing entropy. From these criteria the viscous pressure tensor and the irreversible heat flow of the system 
are expressed by the following equations respectively ~\cite{Degroot,Weinberg},
\begin{eqnarray}
\Pi^{\mu\nu}=&&2\eta\langle \partial^{\mu} u^{\nu} \rangle +\zeta\Delta^{\mu\nu}\partial\cdot u~,
\label{eq-T1A}\\
I^{\mu}=&&\lambda(\partial_\sigma T-TDu_\sigma)\Delta^{\mu\sigma}~,
\label{eq-T1B}
\end{eqnarray}
where the constant of proportionalities $\eta,\zeta$ and $\lambda$ are referred to as the transport coefficients.
The notation used are explained below. The  hydrodynamic velocity $u^{\mu}$ defined in a comoving frame as 
$u^{\mu}=(1,0,0,0)$. $\Delta^{\mu\nu}=g^{\mu\nu}-u^{\mu}u^{\nu}$ is the projection operator, with $g^{\mu\nu}=(1,-1,-1,-1)$
as the metric of the system. $\langle t^{\mu\nu}\rangle\equiv[\frac{1}{2}(\Delta^{\mu\alpha}\Delta^{\nu\beta}+
\Delta^{\nu\alpha}\Delta^{\mu\beta})-\frac{1}{3}\Delta^{\mu\nu}\Delta^{\alpha\beta}]
t_{\alpha\beta}$ indicates a space-like symmetric and traceless form of the tensor $t^{\mu\nu}$.

The alternative definition of thermodynamic fluxes at microscopic level involves an integral over the product of non-equilibrium 
or collisional part of the distribution function of particles and an irreducible tensor of the quantity which is 
being transported. Following this prescription the viscous pressure tensor and the irreversible heat flow can be 
given by the following integral equations,

\begin{eqnarray}
\Pi^{\mu\nu}=&&\sum_{k=1}^{N}\nu_{k}\int\frac{d^3\vec{p}_{k}}{(2\pi)^3 p_k^0}\Delta^{\mu}_{\sigma}\Delta^{\nu}_{\tau} p_{k}^\sigma p_{k}^\tau \delta f_{k},
\label{eq-T2A}\\
 I^{\mu}=&&\sum_{k=1}^{N}\nu_{k}\int\frac{d^3\vec{p}_{k}}{(2\pi)^3 p_k^0} (p_{k}.u-h_{k})p_{k}^{\sigma} \Delta^{\mu}_{\sigma} \delta f_{k}.
\label{eq-T2B}
\end{eqnarray}
Here $p_{k}$ and $h_k$ are the particle 4-momenta and enthalpy per particle respectively. So, comparing
the set of equations in (\ref{eq-T1A},\ref{eq-T1B}) and (\ref{eq-T2A},\ref{eq-T2B}), the values of $\eta$, $\zeta$ and $\lambda$ can be estimated
as a function of the particle distribution deviation $\delta f_{k}$.

For a system with electrically charged constituents, under the influence of an external electric field the induced current 
density relates with the field itself by a linear relation via electrical conductivity ($\sigma_{el}$) as,
 
\begin{equation}
J^{\mu}=\sigma_{el} E^{\mu}~. 
\label{eq-T3}
\end{equation}
In microscopic definition the current density of such a system is given by \cite{Greif-Thesis},
\begin{equation}
 J^{\mu}(x)=\sum_{k=1}^{N}q_{k}I_{k}^{\mu}=\sum_{k=1}^{N-1}(q_{k}-q_{N})I_{k}^{\mu}~,
 \label{eq-T4}
\end{equation}
where $q_{k}$ is the electric charge associated with the $k^{th}$ species.
The diffusion flow $I^{\mu}_a$ for a non equilibrium relativistic system including all reactive processes into
account, is given by,

\begin{eqnarray}
I_{a}^{\mu}=&&\sum_{k=1}^{N}q_{ak}I_{k},~~~~~~~~~~~~~~~~[a=1,2,..........N']\\
           =&&\sum_{k=1}^{N}q_{ak}\{N_{k}^{\mu}-x_{k}N^{\mu}\}~.
\label{eq-T5}
\end{eqnarray}
Here $a$ stands for the index of conserved quantum number and $q_{ak}$ is the $a^{th}$ conserved 
quantum number associated with $k^{th}$ component. $N_{k}^{\mu}(x)=\sum_{k=1}^{N}N_{k}^{\mu}(x)$ 
stands for the total particle 4-flow, where the particle 4-flow for the $k^{th}$ species in a 
multicomponent system is defined as, $N_{k}^{\mu}(x)=\int \frac{d^{3}\vec{p_{k}}}{(2\pi)^3 p_{k}^0} p_{k}^{\mu}f_{k}(x,p_{k}) $.
$x_k=\frac{n_k}{n}$ is defined as the particle fraction corresponding to $k^{th}$ species with $n_k=\int \frac{d^{3}\vec{p_{k}}}{(2\pi)^3 }f_{k}(x,p_{k})$~, 
and $n=\sum_{k=1}^{N}n_{k}(x)$, as the particle number density of $k^{th}$ species and total number density of the system respectively.

Putting Eq.(\ref{eq-T5}) into Eq.(\ref{eq-T4}), and comparing with Eq.(\ref{eq-T3}) we can obtain $\sigma_{el}$ 
again as a function of the particle distribution deviation $\delta f_{k}$.

Observing that the transport coefficients depend upon $\delta f_k=f^{0}_{k}(1\pm f^{0}_{k})\phi_{k}$,
we need to obtain a scheme to determine this quantity in an out of equilibrium thermodynamic system.
We proceed by solving the relativistic transport equation (\ref{eq-R1}), in a technique called Chapman-Enskog
method from the kinetic theory of a multicomponent, many particle system.
In Chapman-Enskog method the distribution function is
expanded in a series in terms of a parameter. This parameter must be a small, dimensionless quantity in order to make the series
asymptotic, such that leading order terms in the expansion must be significant as  compared to the next to leading order ones.
Before introducing a useful  quantity that can be used
as the expansion parameter,  let us investigate the transport equation (\ref{eq-R1}) again,

\begin{equation}
 p^{\mu}_{k}\partial_{\mu}f_{k}=\sum_{l=1}^{N} C_{kl}[f_{k},f_{l}],~~~~~~~~~~~~~~~~[k=1,2,..........N]~.
\label{eq-T6}
\end{equation}

The derivative on the left hand side of equation (\ref{eq-T6}) is decomposed into a time-like and a space-like part
as $\partial^{\mu}=u^{\mu}D+\nabla^{\mu}$, with the covariant time derivative $D=u^{\mu}\partial_{\mu}$ 
and the spatial gradient $\nabla_{\mu}=\Delta_{\mu\nu}\partial^{\nu}$.
The resulting equation is obtained as, 

\begin{equation}
 p_{k}^{\mu}u_{\mu}Df_{k}+p_{k}^{\mu}\nabla_{\mu}f_{k}=\sum_{l=1}^{N}C_{kl}[f_{k},f_{l}].
\label{eq-T7}
\end{equation}

We observe that the length scale associated with the collision term on the right hand side of the transport equation is the 
mean free path ($\lambda_c$) of the hydrodynamic system. The length scale associated with the terms on the 
left hand side of transport equation is the characteristic dimension for the spatial non-uniformities 
within the system, {\it i.e}, it is the typical length $L$ over which the macroscopic thermodynamic quantities 
within the system can vary appreciably. The dimensionless ratio $\lambda_c/L$ is called the Knudsen number
and let us denote it by $\epsilon=\lambda_c/L$.
The order of magnitude of the ratio of a typical term on the right hand side of transport equation to a 
typical term on the left hand side is the Knudsen number $\epsilon$ and due to this fact one can introduce 
$\epsilon$ (which must be small in the hydrodynamic regime where the deviation from equilibrium is small), 
as a dimensionless parameter in front of the left hand side of the transport 
equation depicted below to balance the magnitude of length scale of both sides of transport equation,
\begin{equation}
\epsilon\{p_{k}^{\mu}u_{\mu}Df_{k}+p_{k}^{\mu}\nabla_{\mu}f_{k}\}=\sum_{l=1}^{N}C_{kl}[f_{k},f_{l}].
\label{eq-T8}
\end{equation}

Next, we present the expansion of the particle distribution function in a  power series of $\epsilon$ in the following way,

\begin{equation}
 f_{k}=\sum_{n=0}^{\infty} \epsilon^{n} f_{k}^{(n)}~.
 \label{eq-T9}
\end{equation}
In covariant notation, the time derivative over distribution function is expanded as follows,
\begin{equation}
 Df_{k}=(Df_{k})^{0}+\epsilon (Df_{k})^{1}+\epsilon^{2} (Df_{k})^{2}+.............=\sum_{n=0}^{\infty} \epsilon^{n} (Df_{k})^{(n)} ~.
 \label{eq-T10}
\end{equation}
where the term $(Df_k)^{(n)}$ simply denotes $\partial f_{k}^{(n)}/\partial t$.
Equations (\ref{eq-T9}) and (\ref{eq-T10}) helps to expand transport equation in terms of the non-uniformity 
parameter $\epsilon$. Substituting them into the transport equation (\ref{eq-T8}) and equating the coefficients 
of equal power in $\epsilon$, we obtain the hierarchy of equations,
\begin{eqnarray}
&&0=\sum_{k=1}^{N}C_{kl}[f_{k}^{(0)},f_{l}^{(0)}]~,
\label{eq-T11}
\\
&&p_{k}^{\mu}u_{\mu}(Df_{k})^{r-1}+p_{k}^{\mu}\nabla_{\mu}f_{k}^{r-1}=\sum_{s=0}^{r}\sum_{l=1}^{N}C_{kl}[f_{k}^{(s)},f_{l}^{(r-s)}], ~~~~~~~~~~~r\geq1 ~.
\label{eq-T12}
\end{eqnarray}

Equation (\ref{eq-T11}) reveals nothing but the Boltzmann transport equation for a fluid in equilibrium 
where the collision term involving equilibrium distribution function vanishes.
Equation  (\ref{eq-T12}) provides a hierarchy of equations where in the left hand side of the transport 
equation the derivatives appear on the lower order of distribution function, and the next order
appear on the right hand side only under the collision term. If the $r$th order of distribution
function is expressed as $f_{k}^{(r)}=f_{k}^{(0)}(1\pm f_{k}^{(0)})\phi^{(r)}$, then employing the principle of detailed
balance $f_{k}^{(0)}(x,p_{k})f_{l}^{(0)}(x,p_{l})=f_{k}^{(0)}(x,p_{k}')f_{l}^{(0)}(x,p'_{l})$ we obtain,
\begin{equation}
 C[f_{k}^{(0)},f_{l}^{(r)}]+C[f_{k}^{(r)},f_{l}^{(0)}]=-{{\cal L}_{kl}}[\phi_{k}^{(r)},\phi_{l}^{(r)}]~.
 \label{Eq-T13}
\end{equation}

The unknown function $\phi_{k}^{(r)}$, which depends upon particle 4-momenta and fluid space-time co-ordinates,
is needed to be determined. Here,  we can see that the non-linear collision term is linearized under the 
function $\phi_{k}^{(r)}$ and the linearized collision operator is defined as,
\begin{eqnarray}
{\cal L}_{kl}[\phi_{k}^{(r)},\phi_{l}^{(r)}]=&&\frac{1}{2}\nu_{l}f_{k}^{(0)}(x,p_{k})\int d\Gamma_{p_l}\ d\Gamma_{p'_k}\ d\Gamma_{p'_l}f_{l}^{(0)}(x,p_l)
\{1+f_{k}^{(0)}(x,p'_{k})\}\{1+f_{l}^{(0)}(x,p'_{l})\}\nonumber\\
&&[\phi^{(r)}_{k}(x,p_{k})+\phi^{(r)}_{l}(x,p_{l})-\phi^{(r)}_{k}(x,p'_{k})-\phi^{(r)}_{l}(x,p'_{l})] W(p_{k},p_{l}|p'_{k},p'_{l})~.
\label{eq-T14} 
\end{eqnarray}
Here $W=\frac{1}{2}(2\pi)^4 \delta ^4 (p_{k}+p_{l}-p'_{k}-p'_{l})\langle|M_{k+l\rightarrow k+l}|^2\rangle$ is the interaction cross sections for the corresponding dynamical processes.
In this way the Chapman-Enskog hierarchy becomes,
\begin{equation}
 p_{k}^{\mu}u_{\mu}(Df_{k})^{r-1}+p_{k}^{\mu}\nabla_{\mu}f_{k}^{(r-1)}-\sum_{s=1}^{r-1}\sum_{l=1}^{N}C_{kl}[f_{k}^{(s)},f_{l}^{(r-s)}]=
 -\sum_{l=1}^{N}{{\cal L}_{kl}}[\phi_{k}^{(r)},\phi_{l}^{(r)}]~.
 \label{eq-T15}
\end{equation}

From the foregoing discussion it follows that the first Chapman-Enskog approximation is determined by
equation (\ref{eq-T15}) for r=1
\begin{equation}
 p_{k}^{\mu}u_{\mu}(Df_{k})^0+p_{k}^{\mu}\nabla_{\mu}f_{k}^{(0)}=-\sum_{l=1}^{N}{{\cal L}_{kl}}[\phi_{k}^{(1)},\phi_{l}^{(1)}]~.
 \label{eq-T16} 
\end{equation}

In equation (\ref{eq-T16}) the quantity $\phi_{k}^{(1)}$ is the measure of the deviation of the distribution 
function in the first approximation of Chapman-Enskog method from its equilibrium value and from here on 
we are restricting our estimations for $\phi_{k}^{(1)}$ only. Hence throughout this article this quantity
has been simply denoted by $\phi_{k}$. So the next to leading order correction in the leading order 
equilibrium distribution function is,
\begin{equation}
 \delta f_{k}=f_{k}^{(1)}=f_{k}^{(0)}(1\pm f_{k}^{(0)})\phi_{k} ~.
 \label{eq-T16A}
\end{equation}

From the above hierarchy of equations,  it can be well understood that the Chapman-Enskog technique is
an iterative method, where from the known lower order distribution function the unknown next order 
can be determined by successive approximation. 

Here the linearization of the collision integral, by turning it into a linear integral operator with 
symmetric kernel, in terms of the deviation function $\phi_{k}$, in the 
right hand side of Eq.(\ref{eq-T16}) is extremely essential. In general, the program of seeking solution 
of transport equation becomes non-trivial due to the non-linearity of the collision term. However if the 
state of the system is not considered to be too far from the equilibrium as in the present case, one may 
assume that a linearized form of the transport equation provides a reasonably accurate description of the 
non-linear phenomena. Now one of the conventional mathematical tool to treat the Integro-Differential 
equation containing the linearized collision term, is the variational approximation method, in which
the deviation function is expanded in a polynomial series to any desired degrees of accuracy. However
for a multicomponent system with $N$ number of independent particle species this polynomial
method leads to expressing the transport coefficients in terms of a $N\times N$ matrix whose elements 
include the 4-particle phase space integrals containing the explicit interaction cross sections. This
again becomes non-trivial to tackle along with the in medium corrections in the collective properties of a 
strongly interacting, thermal system. So in order to provide a solution without much loss in generality 
in the context of the situation concerned, we decide to proceed by treating the collision term in
relaxation time approximation (RTA). In this method the collision term is replaced by the rate of change of the
distribution function over the thermal relaxation times $\tau_{k}$ for particular species as discussed in 
section 2.2. Following the RTA scheme Eq.(\ref{eq-T16}) finally reduces to,

\begin{equation}
 p_{k}^{\mu}u_{\mu}(Df_{k})^0+p_{k}^{\mu}\nabla_{\mu}f_{k}^{0}=-\frac{\omega_k}{\tau_k}f_{k}^{0}(1\pm f_{k}^{0})\phi_{k}~.
 \label{eq-T17} 
\end{equation}

In the presence of an external electromagnetic force, the left hand side of the relativistic transport equation
includes also a covariant force term $q_{k}F^{\alpha\beta}p_{\beta}\frac{\partial f_{k}}{\partial p_{k}^{\alpha}}$,
where $q_{k}$ is the electronic charge of the $k^{th}$ species particle and $F^{\mu\nu}=\{-u^{\mu}E^{\nu}+u^{\nu}E^{\mu}\}$ 
is the electromagnetic field tensor with electric field $E^{\mu}$, in the absence of any magnetic field in the medium. 
After incorporating this force term in to Eq.(\ref{eq-T17}), we finally obtain the linearized transport equation under 
the Chapman-Enskog scheme as the following,

\begin{equation}
 p_{k}^{\mu}u_{\mu}(Df_{k})^0+p_{k}^{\mu}\nabla_{\mu}f_{k}^{0}
 +\frac{1}{T}f_{k}^0(1\pm f_{k}^0)q_{k}E_{\mu}p_{k}^{\mu}
 =-\frac{\omega_k}{\tau_k}f_{k}^{0}(1\pm f_{k}^{0})\phi_{k}~.
 \label{eq-T18} 
\end{equation}

Now applying the definition of equilibration momentum distribution function of quasi quarks/anti-quarks and quasi gluons 
from Eq.(\ref{eq1}) on the left hand side of Eq.(\ref{eq-T18}) and exploiting a number of thermodynamic identities which are 
nothing but equilibrium thermodynamic evolution equations of macroscopic state variables of the system, following from 
certain conservation laws (discussed in Appendix-A), we are left with a number of thermodynamic forces with different tensorial 
ranks,

\begin{equation}
Q_{k} X
+ \langle p_{k} ^{\nu}\rangle \{(p_{k}.u)-h_{k}\}X_{qk}
+ \langle p_{k}^{\nu} \rangle \sum_{a=1}^{N'-1}(q_{ak}-x_{a})X_{a\nu} -
 \langle p_{k}^{\mu} p_{k}^{\nu} \rangle \langle X_{\mu\nu}\rangle
 = -\frac{T \omega_{k}}{\tau_{k}}\phi_{k}~,
 \label{eq-T19}
\end{equation}
with,

\begin{eqnarray}
\label{eq-T20}
X=&&\partial \cdot u ~,\\
X_{q\mu}=&&[\frac{\partial_{\mu}T}{T}-\frac{\partial_{\mu}P}{nh}]+[-\frac{1}{h}\sum_{k=1}^{N}x_{k}q_{k}E_{\mu}]~,\\
\label{eq-T21}
X_{k\mu}=&&[(\partial_{\mu}\mu_{k})_{P,T}-\frac{h_{k}}{nh}\partial_{\mu}P]+
         [q_{k}-q_{N}-\frac{h_{k}-h_{N}}{h}\sum_{l=1}^{N}x_{l}q_{l}]E_{\mu}~,\\
\label{eq-T22}
\langle X_{\mu\nu}\rangle=&& \langle \partial _{\mu} u_{\nu} \rangle
=\frac{1}{2}\{\Delta_{\mu\alpha}\Delta_{\nu\beta}+\Delta_{\nu\alpha}\Delta_{\mu\beta}
-\frac{2}{3}\Delta_{\mu\nu}\Delta_{\alpha\beta}\}\partial^{\alpha}u^{\beta} ~.        
\label{eq-T23}
\end{eqnarray}

Here $Q_{k}=\frac{1}{3}\{|\vec{p_k}|^2-3\omega_{k}^2 c_s^2\}$, with $c_s$ is the velocity of sound
propagation within the medium and, $(\partial_{\mu}\mu_{a})_{P,T}=
\sum_{b=1}^{N'-1}\{\frac{\partial \mu_a}{\partial x_b}\}_{P,T,\{x_{a}\}} \partial_{\mu}x_b$, 
with $x_a$ and $\mu_a$ are the particle fraction and chemical potential associated with $a^{th}$ quantum number respectively.
Tensors of form $\langle p_{k}^{\mu} p_{k}^{\nu}\rangle=\frac{1}{2}\{\Delta^{\mu\alpha}\Delta^{\nu\beta}
+\Delta^{\nu\alpha}\Delta^{\mu\beta}-\frac{2}{3}\Delta^{\mu\nu}\Delta^{\alpha\beta}\}(p_k)_{\alpha}(p_k)_{\beta}$,
and $\langle p_{k} ^{\mu}\rangle=\Delta^{\mu\nu} (p_{k})_{\nu}$, are called irreducible tensor of rank 2 and 1 
respectively, where
the rank 0 is simply a scalar. These tensors are irreducible with respect to the transformation group, 
consisting of those Lorentz transformations $\Lambda$, which leaves the time-like vector $u^{\mu}$
invariant ($\Lambda^{\mu}_{\nu}u^{\nu}=u^{\mu}$). The natural occurrence of these
form of tensors in problems involving spherical symmetry and the fact that they can
form a complete set of tensors with minimum number of members, have made their application
quite convenient and advantageous in kinetic theory.

Now, we observe that different thermodynamic forces indicated by Eq.(\ref{eq-T20})-(\ref{eq-T23}) 
involves different transport processes. $X$, expressing the trace part of velocity gradient is 
known as bulk viscous force. The quantity, $X_{q\mu}$ is related to the temperature gradient
known as thermal driving force.
$X_{k\mu}$ includes the spatial gradient over chemical potential that can be translated
into the gradient over particle fraction (($\nabla^{\mu}\mu_{k})_{P,T}=
\frac{T}{x_k}\nabla^{\mu}x_k$), and thus known as diffusion driving force.
Finally $\langle X_{\mu\nu}\rangle$, containing the traceless part of velocity gradient
is known as the shear viscous force. The respective viscous forces give rise to
shear ($\eta$) and bulk ($\zeta$) viscous coefficient where as thermal driving force
gives rise to thermal conductivity$\lambda$. 
We notice that, apart of the spatial gradients over thermodynamic quantities, the thermal driving 
force and the diffusion driving forces include finite contributions purely from the $E^{\mu}$,
reflecting the response of the external electric field in the medium. So we can conclude  
that in the expressions of thermal and diffusion driving forces, terms proportional to
electric field give rise to electrical conductivity ($\sigma_{el}$).

Now in order to be a solution of Eq.(\ref{eq-T19}) the deviation function $\phi_{k}$ must be a 
linear combination of the thermodynamic forces in the following manner,

\begin{equation}
\phi_{k}=A_{k}X+B_{k}^{\mu}X_{q\mu}+\frac{1}{T}\sum_{a=1}^{N'-1}B_{ak}^{\mu}X_{a\mu}-C_{k}^{\mu\nu}\langle X_{\mu\nu}\rangle~,
\label{eq-T24}
\end{equation}
where $A,B$ and $C$ are the unknown coefficients with appropriate tensorial ranks consistent with the 
thermodynamic forces, such that $\phi_{k}$ becomes a scalar, needed to be estimated from the transport 
equation itself. In order to do so, we put Eq.(\ref{eq-T24}) on the right hand side of Eq.(\ref{eq-T19})
and by the virtue of the fact that thermodynamic forces are independent of each other we finally obtain,

\begin{eqnarray}
A_{k}=&&\frac{Q_k}{\{-\frac{T\omega_k}{\tau_k}\}}~,
\label{eq-T25}\\
B_{k}^{\mu}=&&\langle p_{k}^{\mu}\rangle\frac{\omega_{k}-h_{k}}{\{-\frac{T\omega_{k}}{\tau_k}\}}~,
\label{eq-T26}\\
B_{ak}^{\mu}=&&T\langle p_{k}^{\mu}\rangle\frac{q_{ak}-x_{a}}{\{-\frac{T\omega_{k}}{\tau_k }\}}~,
\label{eq-T27}\\
C_{k}^{\mu\nu}=&&\frac{\langle p_{k}^{\mu} p_{k}^{\nu} \rangle}{\{-\frac{T\omega_k}{\tau_k}\}}~.
\label{eq-T28}
\end{eqnarray}

Utilizing the expressions rom Eq.(\ref{eq-T25})-(\ref{eq-T28}) and putting the expression of $\phi_{k}$ 
from Eq.(\ref{eq-T24}) into Eq.(\ref{eq-T16A}), we finally obtain the full expression of the deviation
of the partonic distribution function $\delta f_{k}$. Now we are in a situation where by putting the 
expression of deviation of the distribution function into the microscopic definitions of
thermodynamic fluxes and comparing them with the macroscopic definitions of the same, the transport
coefficients can be estimated explicitly as discussed earlier. Here, we need to mention one crucial 
property of the irreducible tensor used so far. Due to isotropy and the relativistic invariance of the
of the collision operator, it can be observed that the thermodynamic flows and forces of different tensorial
rank do not couple, while they of equal rank do couple by via scalar coefficients. The inner product 
of two irreducible tensors of different ranks gives rise to zero, where with equal ranks they completely
contract giving rise to scalar transport coefficients. This statement is famous as Curie's principle
in the framework of relativistic kinetic theory. It beautifully takes care of the fact, that only the
relevant physical phenomena responsible for the deviation of particle's momentum distribution from equilibrium, 
will be contributed to the respective thermodynamic flows. The next four sub sections will be
contributed for the evaluation of the four different transport coefficients namely shear viscosity ($\eta$),
bulk viscosity ($\zeta$), thermal conductivity ($\lambda$) and electrical conductivity ($\sigma_{el}$).

\subsubsection{Shear viscosity}

As discussed in the previous section, in order to estimate the viscous coefficients we need to compare the 
expressions of viscous pressure tensor from Eq.(\ref{eq-T1A}) and Eq.(\ref{eq-T2A}). It is convenient to 
split $\Pi^{\mu\nu}$ into a traceless part and a remainder such as,
\begin{equation}
 \Pi^{\mu\nu}=\langle{\Pi}^{\mu\nu}\rangle +\Pi \Delta^{\mu\nu}~.
 \label{SV-1}
\end{equation}

The viscous pressure $\Pi$ is defined as one third of the trace of the viscous pressure tensor,                                                                                                                                                                          
\begin{equation}
\Pi=\sum_{k=1}^{N}\nu_{k}\frac{1}{3}\int\frac{d^3\vec{p}_{k}}{(2\pi)^3 p_{k}^0}\Delta_{\mu\nu}p_{k}^\mu p_{k}^\nu \delta f_{k}~.
\label{SV-2}
\end{equation}

So the trace less part of viscous pressure tensor comes out to be,
\begin{equation}
 \langle{\Pi}^{\mu\nu}\rangle=\Pi^{\mu\nu}-\Pi\Delta^{\mu\nu}=\sum_{k=1}^{N}\nu_{k}\int\frac{d^3\vec{p}}{(2\pi)^3 p^0}\langle p^\mu p^\nu \rangle \delta f_{k}~.
\label{SV-3}
\end{equation}

Clearly Eq.(\ref{SV-2}) will give rise to bulk viscosity while Eq.(\ref{SV-3}) gives rise to shear viscosity.
Putting the expression of $\delta f_{k}$, with the expression of $\phi_k$ from Eq.(\ref{eq-T24}), into Eq.(\ref{SV-3})
and comparing with Eq.(\ref{eq-T1A}) we obtain the expression of shear viscosity as,

\begin{equation}
 \eta=\sum_{k=1}^{N}\nu_{k}\frac{\tau_k}{15T}\int \frac{d^3 \vec{p}_{k}}{(2\pi)^3} \frac{|\vec{p_k}|^4}{\omega^2_k}f_{k}^{0}(1\pm f_{k}^{0})~.
 \label{SV-4}
\end{equation}
 
In the context of the current article, the shear viscosity for a strongly interacting QGP system is provided with the help 
of the thermal relaxation times of constituent partons from Eq.(\ref{eq-R11}), (\ref{eq-R12}) and (\ref{eq-R13}), and the 
quasiparticle equilibrium distribution functions of the same under EQPM scheme from Eq.(\ref{eq1}) as the following,

\begin{eqnarray}
 \eta=&&\nu_{q}\frac{\tau_q}{15T}\int \frac{d^3 \vec{p}_{q}}{(2\pi)^3} \frac{|\vec{p_q}|^4}{\omega^2_q}f_{q}^{0}(1 - f_{q}^{0})\nonumber\\
     +&&\nu_{\overline{q}}\frac{\tau_{\overline{q}}}{15T}\int \frac{d^3 \vec{p}_{\overline{q}}}{(2\pi)^3} 
      \frac{|\vec{p_{\overline{q}}}|^4}{\omega^2_{\overline{q}}}f_{\overline{q}}^{0}(1 - f_{\overline{q}}^{0})\nonumber\\
     +&&\nu_{g}\frac{\tau_g}{15T}\int \frac{d^3 \vec{p}_{g}}{(2\pi)^3} \frac{|\vec{p_g}|^4}{\omega^2_g}f_{g}^{0}(1 + f_{g}^{0})~.
     \label{SV-5}
\end{eqnarray}

Here the quasiparticle energy per partons under the EQPM model can be derived from the dispersion relation given in 
Eq.(\ref{dispersion}) as,

\begin{equation}
 \omega_k=|\vec{p_{k}}|[1+\{\frac{T}{|\vec{p_{k}}|}\}\{\frac{T}{T_c}\}\partial_{(\frac{T}{T_c})}\{ln z_k\}]~.
 \label{SV-6}
\end{equation}

We have estimated $\eta$ from Eq.(\ref{SV-5}) in two ways. First an exact estimation of $\eta$ from Eq.(\ref{SV-5}) has been
obtained using full numerical coding. Secondly we perform an analytical approximation of Eq.(\ref{SV-5}) in the following
manner by investigating its level of accuracy. By analyzing the temperature dependence of effective fugacity parameter $z_k$,
we have examined the second term on the right hand side of Eq.(\ref{SV-6}). For gluonic case, at $T/T_c=2.5$ we obtain from
Eq.(\ref{SV-6}), 

\begin{equation}
 \omega_g=|\vec{p_{g}}|[1+0.094]~.
 \label{SV-7}
\end{equation}

So the correction in $\omega_k$ due to the fugacity term is less than $10\%$. (Similar estimations can be shown for
quark degrees of freedom as well.) So Eq.(\ref{SV-7}) can be conveniently expanded in a binomial series keeping upto 
only $2^{nd}$ order term. Following this prescription the $\frac{1}{\omega_k^2}$ term in the expression of $\eta$, 
can be reduced to,

\begin{equation}
\frac{1}{\omega_k^2}=\frac{1}{|\vec{p_k}|^2}-\frac{2}{|\vec{p_k}|^3}[T(\frac{T}{T_c})\partial_{(\frac{T}{T_c})}\{ln z_k\}]~.
\label{SV-8}
\end{equation}

Following Eq.(\ref{SV-8}), the expression of $\eta$, with the analytical approximation performed, becomes

\begin{equation}
 \eta=\sum_{k=q,\overline{q},g}\nu_k \frac{\tau_k}{15T}
 [\int \frac{d^3 \vec{p}_{k}}{(2\pi)^3}|\vec{p_k}|^2 f_{k}^{0}(1\pm f_{k}^{0})
 -2\{T(\frac{T}{T_c})\partial_{(\frac{T}{T_c})}(ln z_k)\} \int \frac{d^3 \vec{p}_{k}}{(2\pi)^3}|\vec{p_k}| f_{k}^{0}(1\pm f_{k}^{0})]~.
 \label{SV-9}
\end{equation}

The momentum integrations over the equilibrium quasiparticle distribution functions are analytically computable giving compact
results in terms of PolyLog functions of the fugacity parameters of quasi-quarks and gluons,

\begin{eqnarray}
 \eta=&&\big(\frac{2T^4}{5\pi^2}\big)\nu_g \tau_g [2\textrm{Polylog}[4,z_g]
        -\{\big(\frac{T}{T_c}\big)\partial_{(\frac{T}{T_c})}(ln z_g)\}\textrm{Polylog}[3,z_g]]\nonumber\\
     +&&\big(\frac{2T^4}{5\pi^2}\big)\nu_q \tau_q [-2\{\textrm{Polylog}[4,-z_q]+\tilde{\mu_q}\textrm{Polylog}[3,-z_q]+\frac{{\tilde{\mu_q}}^2}{2}\textrm{Polylog}[2,-z_q]\}\nonumber\\
     +&&\{\big(\frac{T}{T_c}\big)\partial_{(\frac{T}{T_c})}(ln z_q)\}\{\textrm{Polylog}[3,-z_q]+\tilde{\mu_q}\textrm{Polylog}[2,-z_q]-\frac{{\tilde{\mu_q}}^2}{2} ln(1+z_q)\}]\nonumber\\
     +&&\big(\frac{2T^4}{5\pi^2}\big)\nu_{\overline{q}} \tau_{\overline{q}} [-2\{\textrm{Polylog}[4,-z_q]-\tilde{\mu_q}\textrm{Polylog}[3,-z_q]+\frac{{\tilde{\mu_q}}^2}{2}\textrm{Polylog}[2,-z_q]\}\nonumber\\
     +&&\{\big(\frac{T}{T_c}\big)\partial_{(\frac{T}{T_c})}(ln z_q)\}\{\textrm{Polylog}[3,-z_q]-\tilde{\mu_q}\textrm{Polylog}[2,-z_q]-\frac{{\tilde{\mu_q}}^2}{2} ln(1+z_q)\}]\nonumber\\
     \label{SV-10}
     \end{eqnarray}

\begin{figure*}[h]
\includegraphics[scale=0.32]{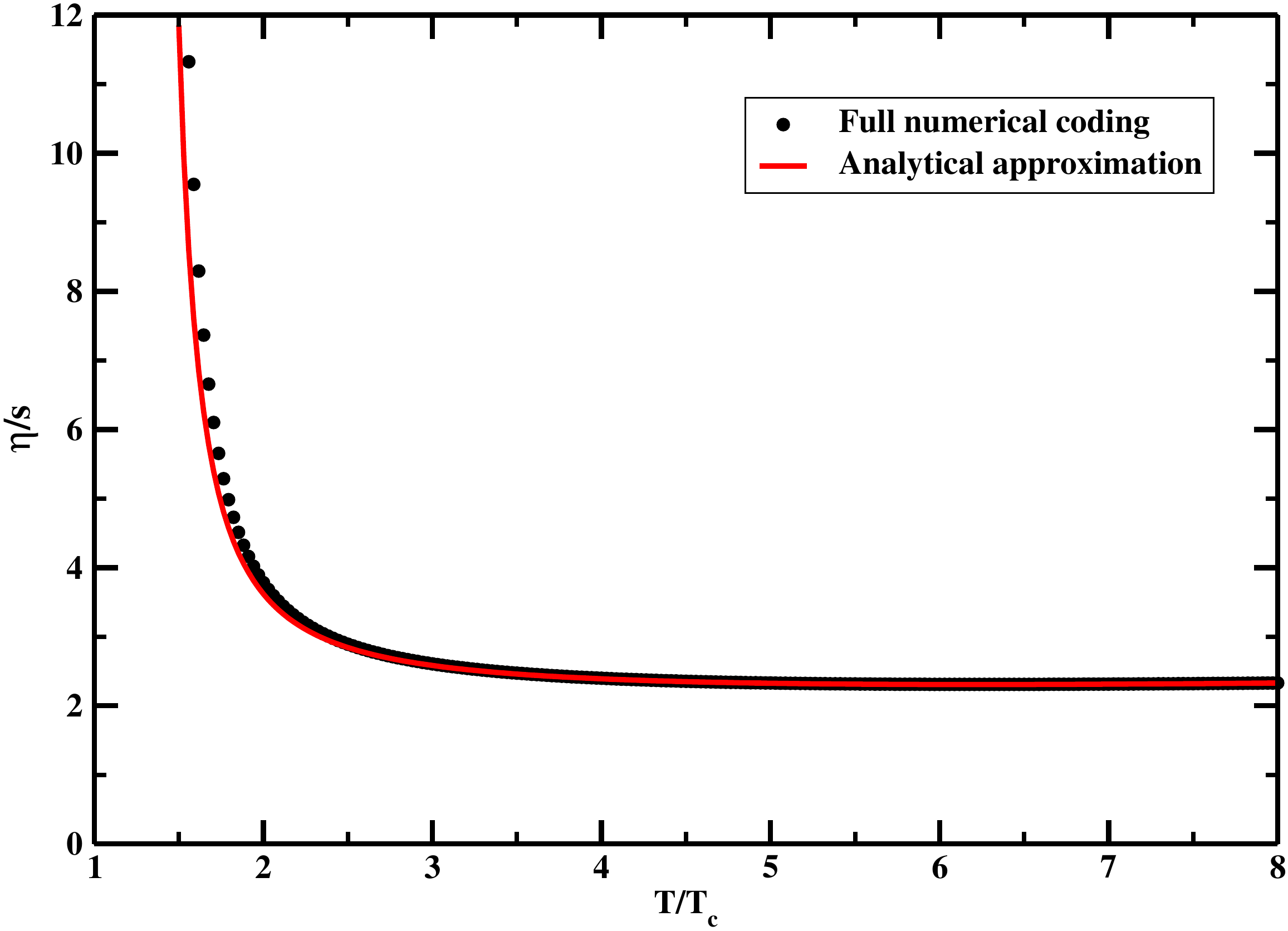}
\includegraphics[scale=0.32]{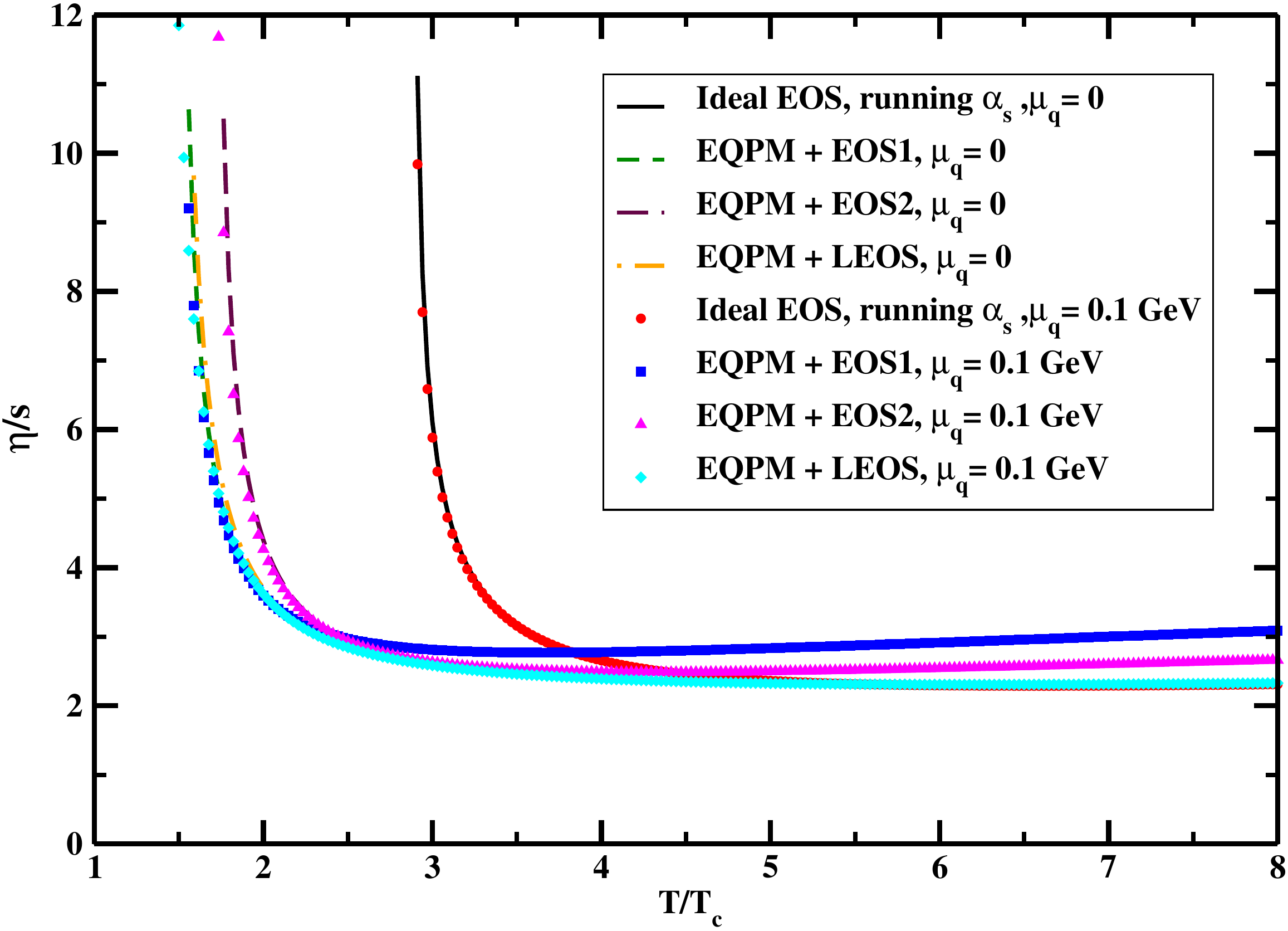}
\caption{Shear viscosity to entropy ratio  using various EOSs as a function of $T/T_c$ at fixed $\tilde{\mu}_q$.}
\label{shear_viscA}
\end{figure*} 

In Fig.(\ref{shear_viscA}) the obtained shear viscosity over entropy density ratio (derived in Appendix-B) has been
plotted as a function of $T/T_c$~. The first figure is showing the comparison between fully numerical estimation
directly from Eq.(\ref{SV-5}) and the approximated analytical estimation from Eq.(\ref{SV-10}) of $\eta/s$ for $N_f=3$,
with LEOS under EQPM scheme at $\mu_q=0.1$ GeV. The plot shows that the two curves are merely separable from 
each other above $T/T_c\sim 2~~ (T\sim 300$MeV). So it can be clearly inferred that the analytical approximation performed in the 
estimation of $\eta$ is quite reliable in the temperature range we are interested currently.
The right panel of the same figure exhibits the temperature dependence of $\eta/s$ estimated under the EQPM scheme using three separate
EOSs mentioned in section (2.1) for zero and non-zero $\mu_q$. As predicted by other pQCD estimates, the value of 
$\eta/s$ is observed to be greater than the experimental extractions and ADS/CFT predictions which is $\sim0.1$ (discussed
later in details). Here the leading log term in thermal relaxation time inputs are majorly responsible for the enhanced 
value of $\eta/s$. Upto $T/T_c\sim4$ the equation of state effects under EQPM are quite distinctly visible which is merging
with the ideal ones in high temperature ranges. The non-zero $\mu_q$ effects are only slightly visible in lower temperatures
which becomes negligible in high temperatures.

\subsubsection{Bulk viscosity}

The bulk viscous coefficients can be estimated in the same spirit as $\eta$ by comparing Eq.(\ref{SV-2}) and (\ref{eq-T1A}), 
and putting the expression of $\phi_k$ from Eq.(\ref{eq-T24}) into $\delta f_{k}$,

\begin{equation}
 \zeta=\sum_{k=q,\overline{q},g} \nu_{k}\frac{\tau_k}{9T}\int \frac{d^3 \vec{p}_{k}}{(2\pi)^3} \frac{1}{\omega^2_k}
       \{p_k^2-3 \omega_k^2 c_s^2\}^2 f_{k}^{0}(1\pm f_{k}^{0})~.
 \label{BV-1}
\end{equation}

Under the analytical approximation mentioned in the earlier section Eq.(\ref{BV-1}) becomes,

\begin{eqnarray}
\zeta&&=(1-3c_s^2)^2\sum_{k=q,\overline{q},g} \nu_{k}\frac{\tau_k}{9T}\int \frac{d^3 \vec{p}_{k}}{(2\pi)^3} |\vec{p_k}|^2 f_{k}^{0}(1\pm f_{k}^{0})\nonumber\\
     &&-2(1-9c_s^4) \sum_{k=q,\overline{q},g} \nu_{k}\frac{\tau_k}{9T}\big\{T\big(\frac{T}{T_c}\big)\partial_{(T/T_c)}\{lnz_{k}\}\big\}\int \frac{d^3 \vec{p}_{k}}{(2\pi)^3} |\vec{p_k}| f_{k}^{0}(1\pm f_{k}^{0})\nonumber\\
     &&+(1+3c_s^2)^2\sum_{k=q,\overline{q},g} \nu_{k}\frac{\tau_k}{9T}\big\{T\big(\frac{T}{T_c}\big)\partial_{(T/T_c)}\{lnz_{k}\}\big\}^2\int \frac{d^3 \vec{p}_{k}}{(2\pi)^3} f_{k}^{0}(1\pm f_{k}^{0})~. 
 \label{BV-2}
\end{eqnarray}

\begin{figure*}[h]
\includegraphics[scale=0.32]{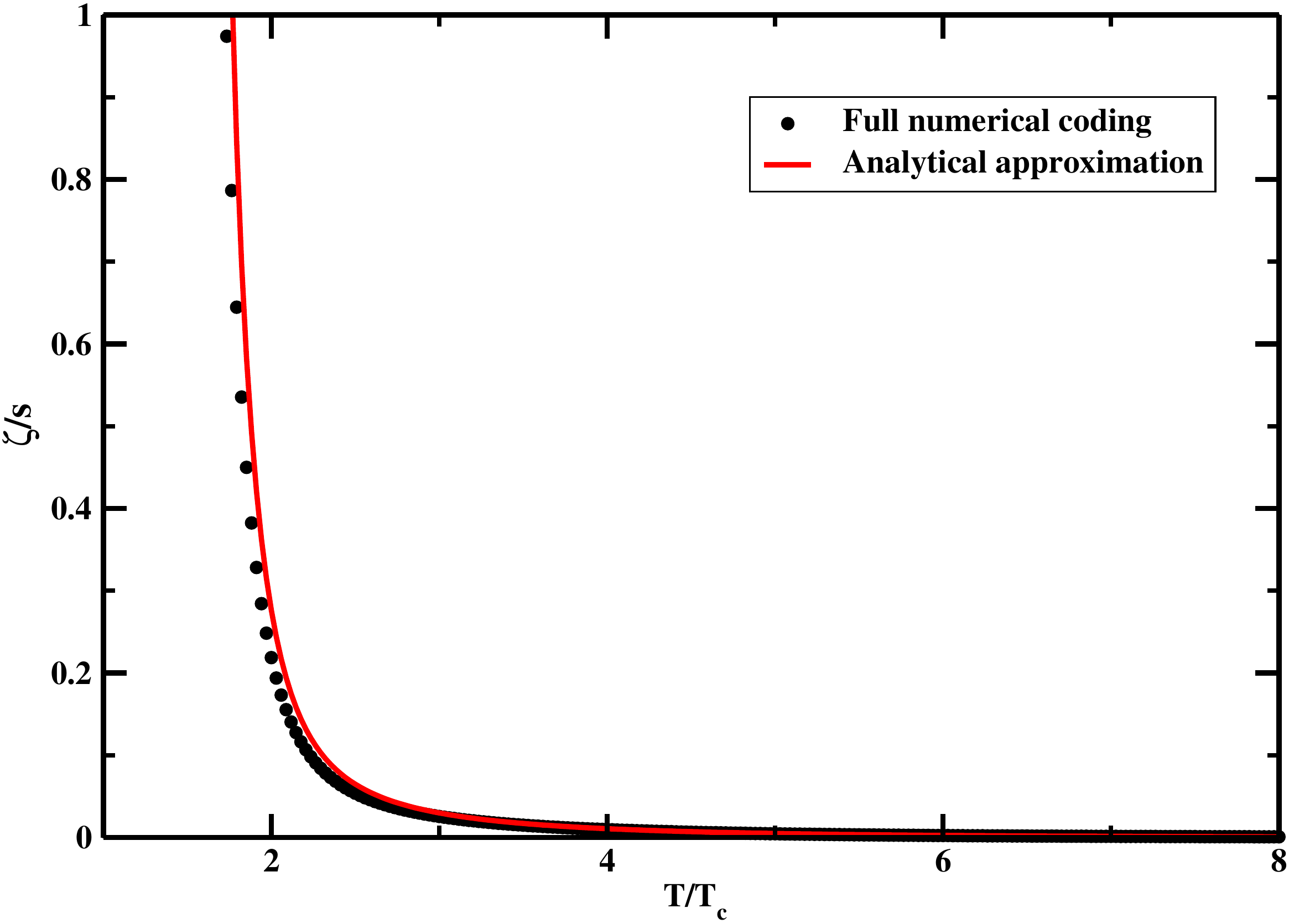}
\includegraphics[scale=0.32]{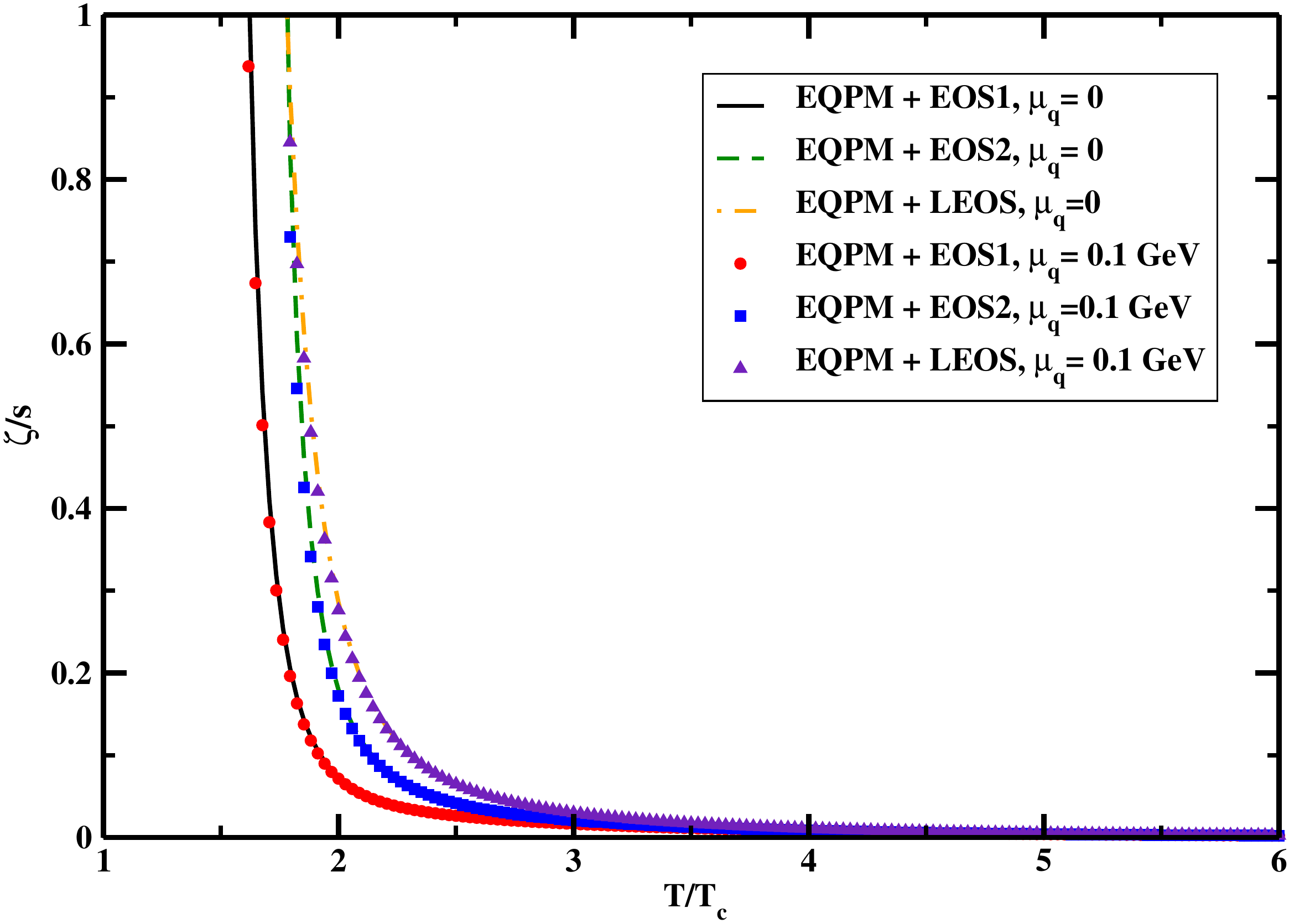}
\caption{Bulk viscosity to entropy ratio  using various EOSs as a function of $T/T_c$ at fixed $\tilde{\mu}_q$.}
\label{bulk}
\end{figure*}

After performing the momentum integrals over separate partonic degrees of freedom, we obtain the consolidated expression 
of $\zeta$ in terms of the Polylog function over fugacity parameters in the following way,

\begin{eqnarray}
\zeta&&=(1-3c_s^2)^2 \tau_g \nu_g (\frac{4T^4}{3\pi^2})\textrm{PolyLog}[4,z_g]\nonumber\\
     &&-(1-9c_s^4)   \tau_g \nu_g (\frac{2T^4}{3\pi^2})[(\frac{T}{T_c})\partial_{(\frac{T}{T_c})}(ln z_g)]\textrm{PolyLog}[3,z_g]\nonumber\\
     &&+(1+3c_s^2)^2 \tau_g \nu_g (\frac{T^4}{9\pi^2}) [(\frac{T}{T_c})\partial_{(\frac{T}{T_c})}(ln z_g)]^2 \textrm{PolyLog}[2,z_g]\nonumber\\
     &&+(1-3c_s^2)^2 \tau_q \nu_q (\frac{4T^4}{3\pi^2})[-\{\textrm{PolyLog}[4,-z_q]+\tilde{\mu_q}\textrm{Polylog}[3,-z_q]
     +\frac{{\tilde{\mu_q}}^2}{2}\textrm{Polylog}[2,-z_q]\}]\nonumber\\
     &&-(1-9c_s^4)   \tau_q \nu_q (\frac{2T^4}{3\pi^2}) [(\frac{T}{T_c})\partial_{(\frac{T}{T_c})}(ln z_q)][-\{\textrm{PolyLog}[3,-z_q]
     +\tilde{\mu_q}\textrm{Polylog}[2,-z_q]-\frac{{\tilde{\mu_q}}^2}{2}ln(1+z_q)\}]\nonumber\\
     &&+(1+3c_s^2)^2 \tau_q \nu_q (\frac{T^4}{9\pi^2}) [(\frac{T}{T_c})\partial_{(\frac{T}{T_c})}(ln z_q)]^2
     [-\{\textrm{PolyLog}[2,-z_q]-\tilde{\mu_q}ln(1+z_q)-\frac{{\tilde{\mu_q}}^2}{2}\frac{z_q}{1+z_q}\}]\nonumber\\
     &&+(1-3c_s^2)^2 \tau_{\overline{q}} \nu_{\overline{q}} (\frac{4T^4}{3\pi^2})[-\{\textrm{PolyLog}[4,-z_q]-\tilde{\mu_q}\textrm{Polylog}[3,-z_q]
     +\frac{{\tilde{\mu_q}}^2}{2}\textrm{Polylog}[2,-z_q]\}]\nonumber\\
     &&-(1-9c_s^4)   \tau_{\overline{q}} \nu_{\overline{q}} (\frac{2T^4}{3\pi^2}) [(\frac{T}{T_c})\partial_{(\frac{T}{T_c})}(ln z_q)][-\{\textrm{PolyLog}[3,-z_q]
     -\tilde{\mu_q}\textrm{Polylog}[2,-z_q]-\frac{{\tilde{\mu_q}}^2}{2}ln(1+z_q)\}]\nonumber\\
     &&+(1+3c_s^2)^2 \tau_{\overline{q}} \nu_{\overline{q}} (\frac{T^4}{9\pi^2}) [(\frac{T}{T_c})\partial_{(\frac{T}{T_c})}(ln z_q)]^2
     [-\{\textrm{PolyLog}[2,-z_q]+\tilde{\mu_q}ln(1+z_q)-\frac{{\tilde{\mu_q}}^2}{2}\frac{z_q}{1+z_q}\}]\nonumber\\
     \label{BV-3}
     \end{eqnarray}

In Fig.(\ref{bulk}) the temperature dependence of the estimated $\zeta/s$ has been depicted. The first figure
again proves the authenticity of the analytical approximation (Eq.(\ref{BV-3})), as it agrees sensibly with the 
full numerical coding (Eq.(\ref{BV-2})). The temperature dependence of $\zeta/s$ is displaying the conventional
decreasing trend with increasing temperature above $T_c$, and away from ($T/T_c\sim2$) its magnitude appears to be
quite small as expected, indicating the diverging nature of $\zeta$ only around $T_c$. We note from Eq.(\ref{BV-1}),
that the ideal EOS will result in vanishing contribution to $\zeta$ for massless QGP. The different EOSs 
under EQPM are showing distinct temperature behavior of $\zeta/s$ around $T/T_c\sim2$ which are merging together
into extremely small values at higher temperatures. However due to small order of magnitude of $\zeta$ even around 
$T_c$ (in comparison with other transport coefficients), the nonzero quark chemical potential effects are barely 
visible in this case. 

After obtaining the expressions of $\eta$ and $\zeta$, their ratios have been plotted while  scaled with \{$2(1/3-c_s^2)$\} 
(scaling 1) and \{$15(1/3-c_s^2)^2$\} (scaling 2) including three different EOS effects and with $\mu_q=0.1$GeV in 
Fig.(\ref{zetabyeta}). These scaling factors have been widely used to illustrate the interplay
between bulk and shear viscous coefficients in a number of literature based on pQCD, ADS/CFT and experimental extractions 
of transport parameters (details mentioned in discussion section). However in our case, the second one is offering a
better scaling at least at higher temperature regions for all three EOSs, whereas the first one fails to prove a sensible 
scaling of $\zeta/\eta$ ratio.

\begin{figure*}[h]
\includegraphics[scale=0.32]{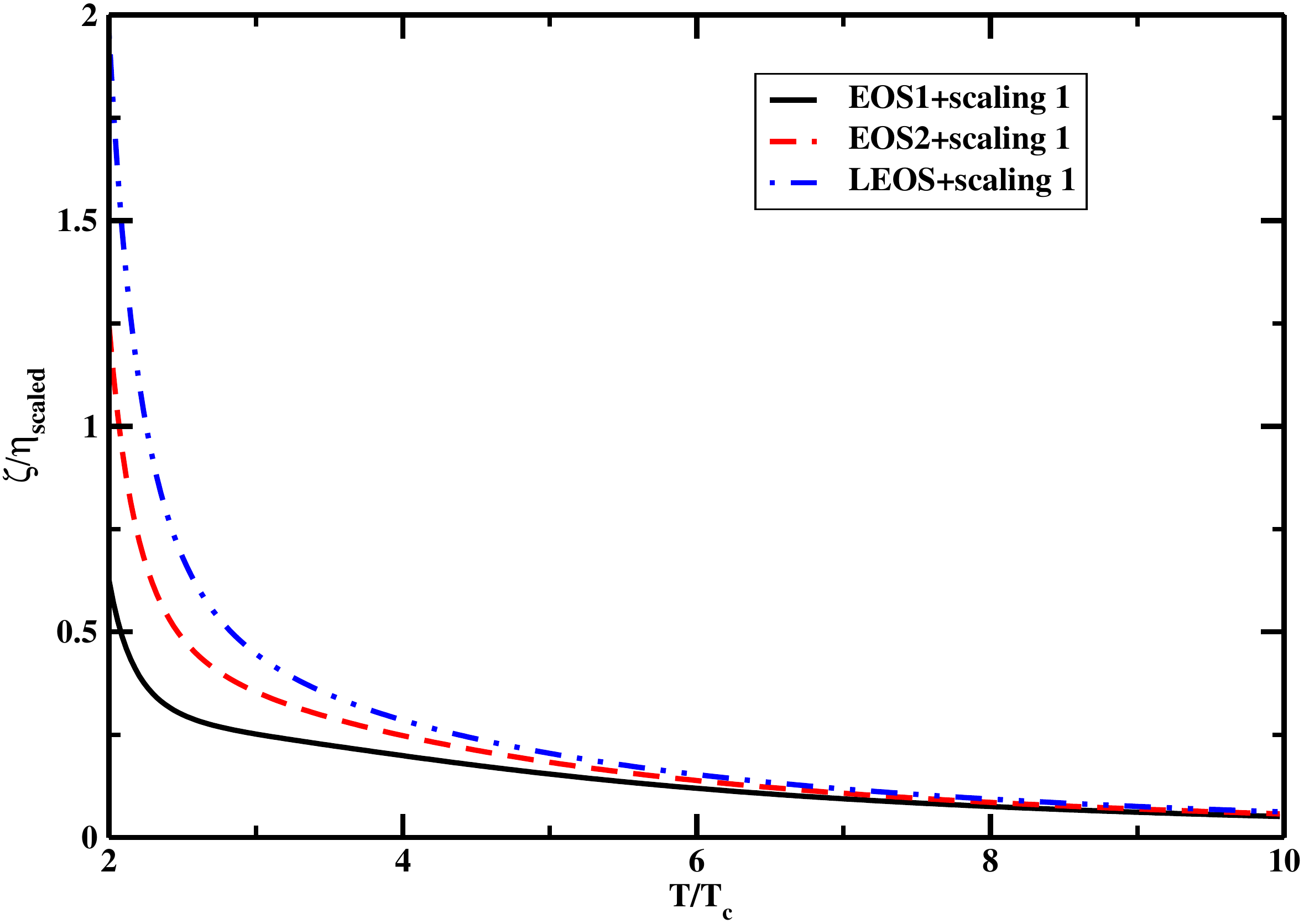}
\includegraphics[scale=0.32]{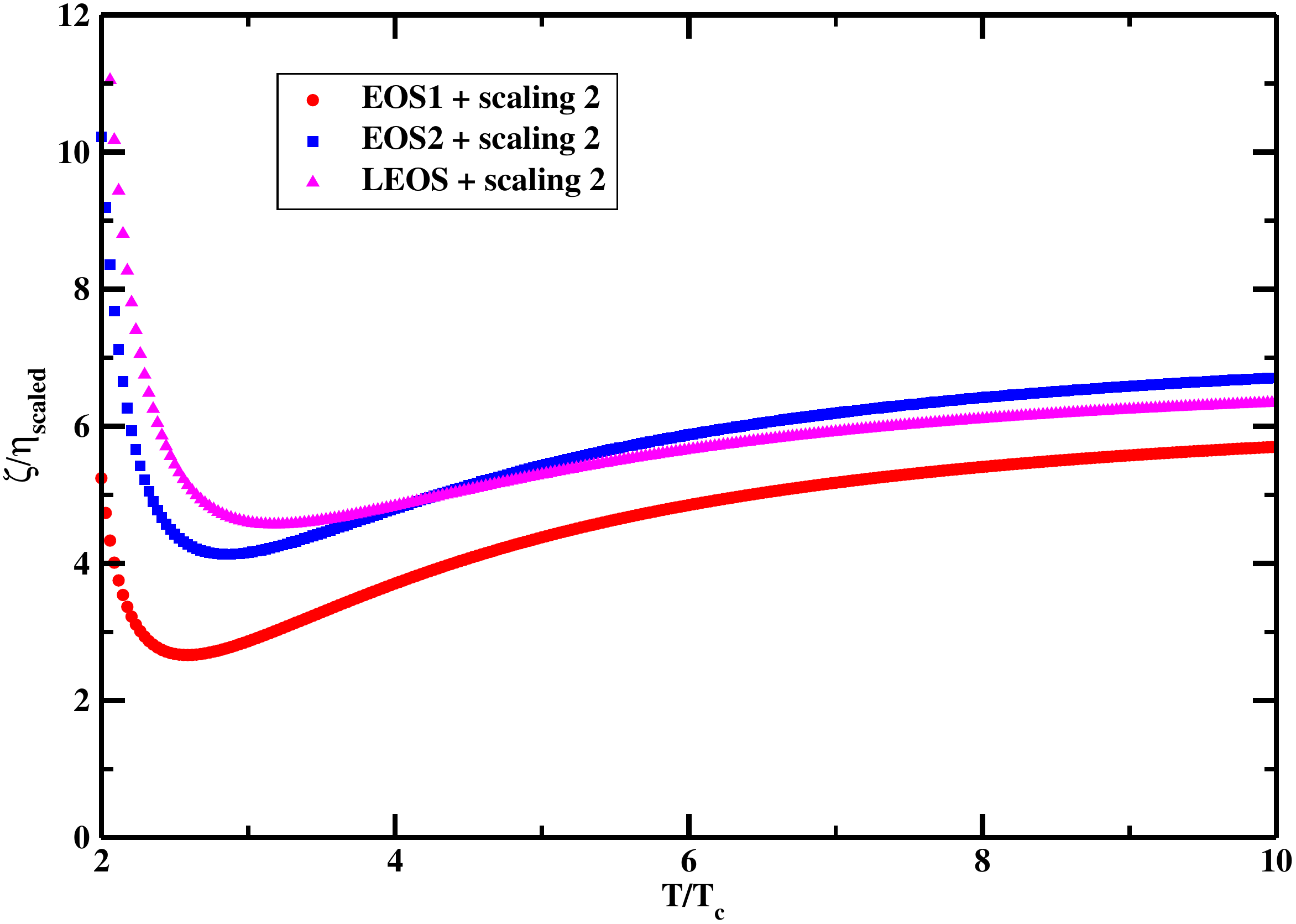}
\caption{Bulk viscosity to shear viscosity ratio  for various EOSs as a function of $T/T_c$ at fixed $\tilde{\mu}_q$
using different scalings.}
\label{zetabyeta}
\end{figure*} 

\subsubsection{Thermal conductivity}

The analytical expression of thermal conductivity can be obtained by comparing Eq.(\ref{eq-T1B}) and (\ref{eq-T2B}),
and replacing $\phi_k$ in $\delta f_k$ from Eq.(\ref{eq-T24}) in the following form,

\begin{equation}
 \lambda=\sum_{k=q,\overline{q},g}\nu_{k}\frac{\tau_k}{3T^2}
 \int \frac{d^3 \vec{p}_{k}}{(2\pi)^3} \frac{|\vec{p_k}|^2}{\omega^2_k}(\omega_k - h_k)^2f_{k}^{0}(1\pm f_{k}^{0})~.
 \label{TH-1}
\end{equation}

In analytical approximation $\lambda$ comes out to be,

\begin{eqnarray}
 \lambda=&&(\frac{2T^3}{9\pi^2})\nu_g \tau_g [2\textrm{Polylog}[4,z_g]
        -\{(\frac{T}{T_c})\partial_{(\frac{T}{T_c})}(ln z_g)\}\textrm{Polylog}[3,z_g]]\nonumber\\
     +&&(\frac{2T^3}{9\pi^2})\nu_q \tau_q [-2\{\textrm{Polylog}[4,-z_q]+\tilde{\mu_q}\textrm{Polylog}[3,-z_q]+\frac{{\tilde{\mu_q}}^2}{2}\textrm{Polylog}[2,-z_q]\}\nonumber\\
     +&&\{(\frac{T}{T_c})\partial_{(\frac{T}{T_c})}(ln z_q)\}\{\textrm{Polylog}[3,-z_q]+\tilde{\mu_q}\textrm{Polylog}[2,-z_q]-\frac{{\tilde{\mu_q}}^2}{2} ln(1+z_q)\}]\nonumber\\
     +&&(\frac{2T^3}{9\pi^2})\nu_{\overline{q}} \tau_{\overline{q}} [-2\{\textrm{Polylog}[4,-z_q]-\tilde{\mu_q}\textrm{Polylog}[3,-z_q]+\frac{{\tilde{\mu_q}}^2}{2}\textrm{Polylog}[2,-z_q]\}\nonumber\\
     +&&\{(\frac{T}{T_c})\partial_{(\frac{T}{T_c})}(ln z_q)\}\{\textrm{Polylog}[3,-z_q]-\tilde{\mu_q}\textrm{Polylog}[2,-z_q]-\frac{{\tilde{\mu_q}}^2}{2} ln(1+z_q)\}]\nonumber\\
\label{TH-2}
\end{eqnarray}

\begin{figure*}[h]
\includegraphics[scale=0.32]{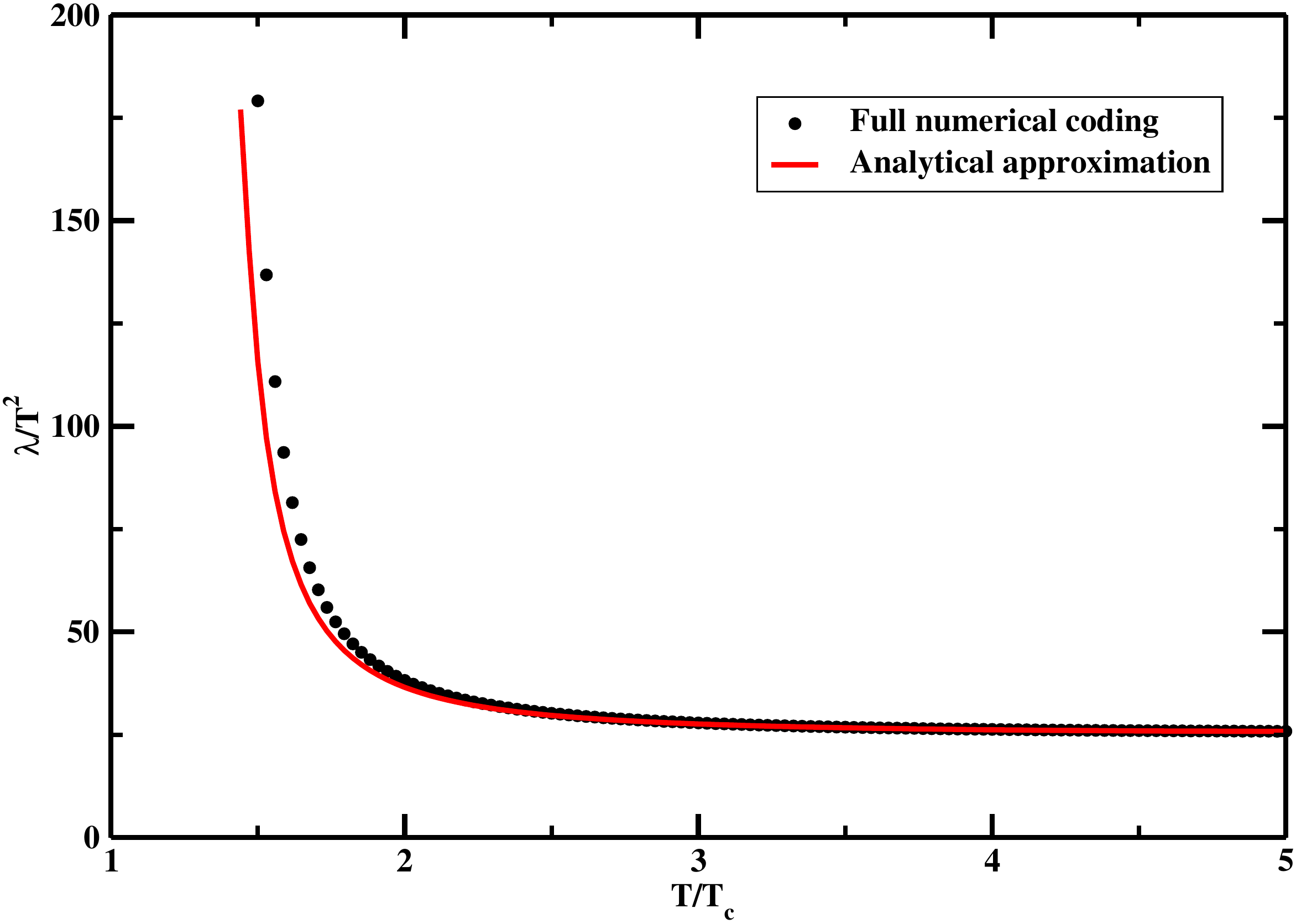}
\includegraphics[scale=0.32]{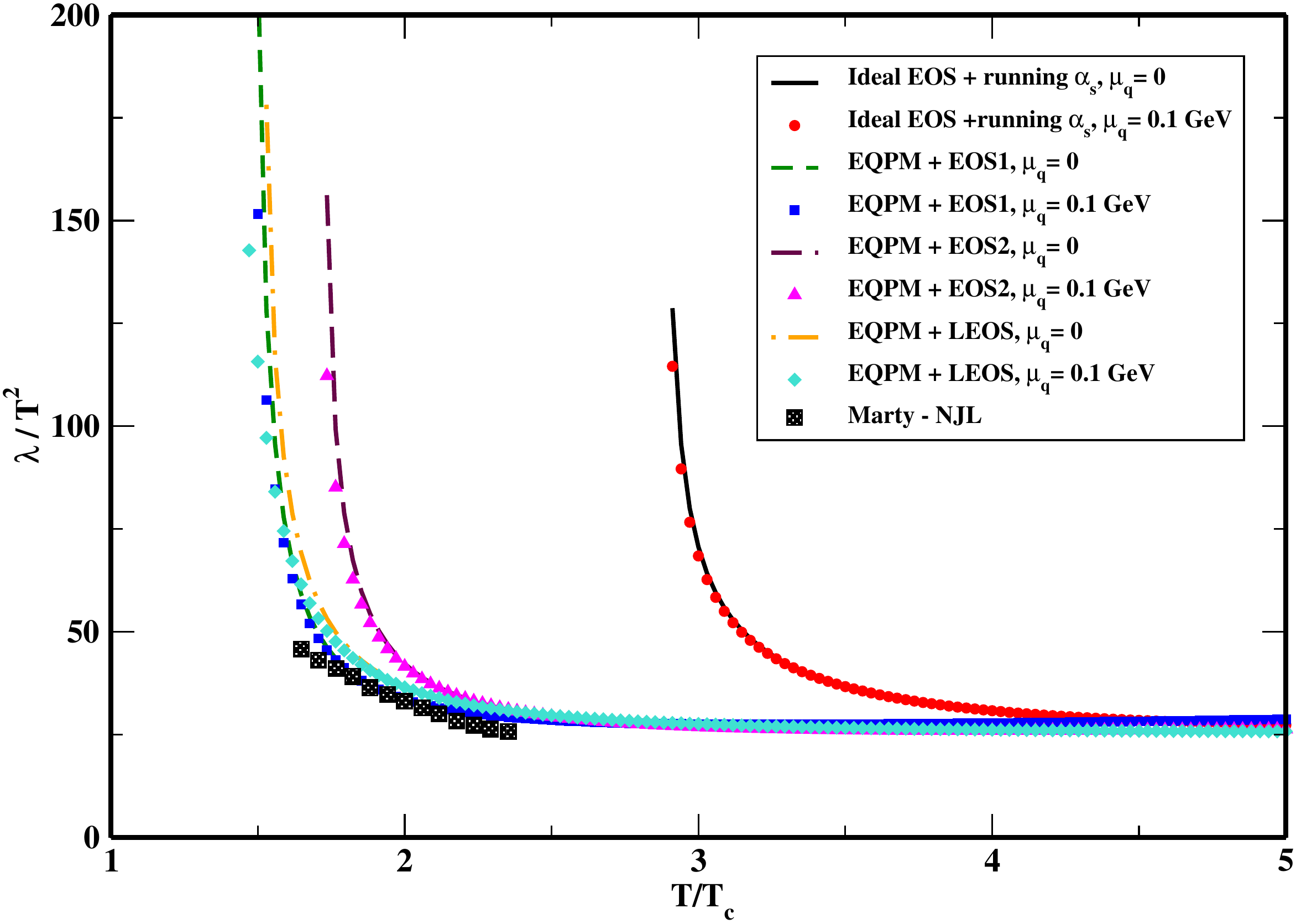}
\caption{Thermal conductivity  using various EOSs as a function of $T/T_c$.}
\label{Tcond}
\end{figure*} 

The results of thermal conductivity displayed as a function of temperature in Fig.(\ref{Tcond}). Like other two
previous cases, here too the analytical approximation works wonderfully well, showing convincing agreement with full
numerical coding. We have plotted the dimensionless quantity $\lambda/T^2$ as a function of $T/T_c$ in the 
second plot, for all possible EOSs and both zero and nonzero quark chemical potentials. As before, the different 
EOSs are providing recognizably different effects at lower temperatures which are fusing with the ideal one at
higher temperatures. The nonzero $\mu_q$ effects are only visible at quite low temperatures. Around $T/T_c\sim2$,
the LEOS results with $\mu_q=0.1$GeV is in good agreement with the NJL estimation of thermal conductivity by
Marty \cite{Marty}.

\subsubsection{Electrical conductivity}
In order to estimate $\sigma_{el}$, we start with the expression of diffusion flow given in Eq.(\ref{eq-T5}).
We clearly observe that at leading order with equilibrium distribution function $f_k^0$ in the 
definition of $N_{k}^{\mu}$, $N^{\mu}$ and $x_k$, the diffusion flow vanishes, while in the next
to leading order the correction term $\delta f_{k}=f_{k}^0(1\pm f_k^0)$, gives finite contribution
to the diffusion flow as follows,

\begin{equation}
I_{a}^{\mu}=\sum_{k=1}^{N}(q_{ak}-x_{a})\int \frac{d^{3}\vec{p_{k}}}{(2\pi)^3 p_{k}^0} p_{k}^{\mu} f_k^0(1\pm f_k^0)\phi_k~. 
\label{EC-1}
\end{equation}
Putting the value of $\phi_{k}$ from (\ref{eq-T24}) with the help of Eq. (\ref{eq-T26}) and (\ref{eq-T27})
we get the linear law obeyed by the diffusion flow,

\begin{equation}
I_{a}^{\mu}=l_{aq}X_{q}^{\mu}+\sum_{b=1}^{N'-1}l_{ab}X_{b}^{\mu}~, a=1,....,(N'-1)~,
\label{EC-2}
\end{equation}
where the coefficients associated with thermal diffusion and particle concentration diffusion are given respectively as,

\begin{eqnarray}
l_{aq}=&&\sum_{k=1}^{N}(q_{ak}-x_{a}) \frac{1}{T}\int \frac{d^{3}\vec{p_{k}}}{(2\pi)^3}f_k^0(1\pm f_k^0)\tau_{k}\frac{|\vec{p}_k|^2}{\omega_k^2}(\omega_k-h_k)~,\\
\label{EC-3}
l_{ab}=&&\sum_{k=1}^{N}(q_{ak}-x_{a})(q_{bk}-x_{b})\frac{1}{T}\int \frac{d^{3}\vec{p_{k}}}{(2\pi)^3}f_k^0(1\pm f_k^0)\tau_{k}\frac{|\vec{p}_k|^2}{\omega_k^2}.  
\label{EC-4}
\end{eqnarray}

Substituting the expression of diffusion flow from Eq.(\ref{EC-2}) into the microscopic definition of current density 
in Eq.(\ref{eq-T4}), and pertaining the terms proportional to electric field only we finally obtain the expression for the 
electric current density as,

\begin{eqnarray}
 J^{\mu}=\sum_{k=1}^{N-1}(q_k-q_N)[\sum_{l=1}^{N-1}l_{kl}\{q_l-q_N-\frac{h_l-h_N}{h}\sum_{n=1}^{N}x_{n}q_{n}\}-
 \frac{l_{kq}}{h}\sum_{n=1}^{N}x_{n}q_{n}]E^{\mu}~.
 \label{EC-5}
\end{eqnarray}

Finally comparing the Eq.(\ref{EC-5}) with the macroscopic definition of induced current density from Eq.(\ref{eq-T3})
we get the expression for electrical conductivity as the following,

\begin{eqnarray}
 \sigma_{el}=\sum_{k=1}^{N-1}(q_k-q_N) [\sum_{l=1}^{N-1}l_{kl}\{q_l-q_N-  
             \frac{h_l-h_N}{h}\sum_{n=1}^{N}x_{n}q_{n}\}- \frac{l_{kq}}{h}\sum_{n=1}^{N}x_{n}q_{n}]~.
 \label{EC-6}
\end{eqnarray}

For a QGP system with quarks, antiquarks and gluons as the degrees of freedom, the expression of $\sigma_{el}$
turns out to be,

\begin{eqnarray}
 \sigma_{el}=q^2_{q}[&&(l_{11}+l_{21})\{1-\frac{h_{q}-h_{g}}{h}(x_{q}+x_{\overline{q}})\}
                    +(l_{12}+l_{22})\{1-\frac{h_{\overline{q}}-h_{g}}{h}(x_{q}+x_{\overline{q}})\}\nonumber\\
                    -&&(l_{1q}+l_{2q})\frac{(x_{q}+x_{\overline{q}})}{h}]  ~,       
\label{EC-7}
\end{eqnarray}

with,

\begin{eqnarray}
l_{1q}=
&&(1-x_{q})\frac{\tau_{q}}{T}\int\frac{d^{3}\vec{p_{q}}}{(2\pi)^3}f^{0}_{q}(1-f^{0}_{q})\frac{|\vec{p_{q}}|^{2}}{\omega_{q}^2}(\omega_{q}-h_{q})\nonumber\\
&&+(-x_{q})\frac{\tau_{\overline{q}}}{T}\int\frac{d^{3}\vec{p_{\overline{q}}}}{(2\pi)^3}f^{0}_{\overline{q}}(1-f^{0}_{\overline{q}})\frac{|\vec{p_{\overline{q}}}|^{2}}{\omega_{\overline{q}}^2}(\omega_{\overline{q}}-h_{\overline{q}})\nonumber\\
&&+(-x_{q})\frac{\tau_{g}}{T}\int\frac{d^{3}\vec{p_{g}}}{(2\pi)^3}f^{0}_{g}(1+f^{0}_{g})\frac{|\vec{p_{g}}|^{2}}{\omega_{g}^2}(\omega_{g}-h_{g})~,
\label{EC-8}
\end{eqnarray}

\begin{eqnarray}
l_{2q}=
&&(-x_{\overline{q}})\frac{\tau_{q}}{T}\int\frac{d^{3}\vec{p_{q}}}{(2\pi)^3}f^{0}_{q}(1-f^{0}_{q})\frac{|\vec{p_{q}}|^{2}}{\omega_{q}^2}(\omega_{q}-h_{q})\nonumber\\
&&+(1-x_{\overline{q}})\frac{\tau_{\overline{q}}}{T}\int\frac{d^{3}\vec{p_{\overline{q}}}}{(2\pi)^3}f^{0}_{\overline{q}}(1-f^{0}_{\overline{q}})\frac{|\vec{p_{\overline{q}}}|^{2}}{\omega_{\overline{q}}^2}(\omega_{\overline{q}}-h_{\overline{q}})\nonumber\\
&&+(-x_{\overline{q}})\frac{\tau_{g}}{T}\int\frac{d^{3}\vec{p_{g}}}{(2\pi)^3}f^{0}_{g}(1+f^{0}_{g})\frac{|\vec{p_{g}}|^{2}}{\omega_{g}^2}(\omega_{g}-h_{g})~,
\label{EC-9}
\end{eqnarray}

\begin{eqnarray}
l_{11}=
(1-x_{q})^{2}\frac{\tau_{q}}{T}\int\frac{d^{3}\vec{p_{q}}}{(2\pi)^3}f^{0}_{q}(1-f^{0}_{q})\frac{|\vec{p_{q}}|^{2}}{\omega_{q}^2}\nonumber\\
+x_{q}^2\frac{\tau_{\overline{q}}}{T}\int\frac{d^{3}\vec{p_{\overline{q}}}}{(2\pi)^3}f^{0}_{\overline{q}}(1-f^{0}_{\overline{q}})\frac{|\vec{p_{\overline{q}}}|^{2}}{\omega_{\overline{q}}^2}\nonumber\\
+x_{q}^2\frac{\tau_{g}}{T}\int\frac{d^{3}\vec{p_{g}}}{(2\pi)^3}f^{0}_{g}(1+f^{0}_{g})\frac{|\vec{p_{g}}|^{2}}{\omega_{g}^2}~,
\label{EC-10}
\end{eqnarray}

\begin{eqnarray}
l_{22}=
&&x_{\overline{q}}^{2}\frac{\tau_{q}}{T}\int\frac{d^{3}\vec{p_{q}}}{(2\pi)^3}f^{0}_{q}(1-f^{0}_{q})\frac{|\vec{p_{q}}|^{2}}{\omega_{q}^2}\nonumber\\
&&+(1-x_{\overline{q}})^2\frac{\tau_{\overline{q}}}{T}\int\frac{d^{3}\vec{p_{\overline{q}}}}{(2\pi)^3}f^{0}_{\overline{q}}(1-f^{0}_{\overline{q}})\frac{|\vec{p_{\overline{q}}}|^{2}}{\omega_{\overline{q}}^2}\nonumber\\
&&+x_{\overline{q}}^2\frac{\tau_{g}}{T}\int\frac{d^{3}\vec{p_{g}}}{(2\pi)^3}f^{0}_{g}(1+f^{0}_{g})\frac{|\vec{p_{g}}|^{2}}{\omega_{g}^2}~,
\label{EC-11}
\end{eqnarray}
and
\begin{eqnarray}
l_{12}=l_{21}=
&&x_{\overline{q}}(x_{q}-1)\frac{\tau_{q}}{T}\int\frac{d^{3}\vec{p_{q}}}{(2\pi)^3}f^{0}_{q}(1-f^{0}_{q})\frac{|\vec{p_{q}}|^{2}}{\omega_{q}^2}\nonumber\\
&&+x_{q}(x_{\overline{q}}-1)\frac{\tau_{\overline{q}}}{T}\int\frac{d^{3}\vec{p_{\overline{q}}}}{(2\pi)^3}f^{0}_{\overline{q}}(1-f^{0}_{\overline{q}})\frac{|\vec{p_{\overline{q}}}|^{2}}{\omega_{\overline{q}}^2}\nonumber\\
&&+x_{q}x_{\overline{q}}\frac{\tau_{g}}{T}\int\frac{d^{3}\vec{p_{g}}}{(2\pi)^3}f^{0}_{g}(1+f^{0}_{g})\frac{|\vec{p_{g}}|^{2}}{\omega_{g}^2}~.
\label{EC-12}
\end{eqnarray}

The $q_q^2=\sum_k \nu_k q_{qk}^2$ is simply the square of the fractional quark charges taking sum over quark degeneracy.
For up, down and strange quarks the fractions quark charges are taken to be $2/3$, $-1/3$ and $-1/3$ respectively.

Apart of the full numerical coding, we have done the analytical approximation as well in estimating the value of $\sigma_{el}$.
For this purpose the two relevant integrals present in Eq.(\ref{EC-8})-(\ref{EC-12}), indicated by $I_{1}$ and $I_{2}$ as
following,

\begin{eqnarray}
 \{I_{1}\}_{k}=&&\int\frac{d^{3}\vec{p_{k}}}{(2\pi)^3}f^{0}_{k}(1\pm f^{0}_{k})\frac{|\vec{p_{k}}|^{2}}{\omega_{k}^2}(\omega_{k}-h_{k})~,
 \label{EC-13}\\
 \{I_{2}\}_{k}=&&\int\frac{d^{3}\vec{p_{k}}}{(2\pi)^3}f^{0}_{k}(1\pm f^{0}_{k})\frac{|\vec{p_{k}}|^{2}}{\omega_{k}^2}~,
 \label{EC-14}
\end{eqnarray}
are needed to be computed analytically as indicated earlier. The estimated values of the integrals for different partonic 
degrees of freedom in terms of the fugacity parameters and its derivatives are given below

\begin{eqnarray}
 \{I_{1}\}_{g}=&&-\frac{T^4}{\pi^2}[\textrm{PolyLog}[3,z_g]-\frac{2}{3}\{(\frac{T}{T_c})\partial_{(\frac{T}{T_c})}(ln z_g)\}\textrm{PolyLog}[2,z_g]]~,
 \label{EC-15}\\
 \{I_{1}\}_{q}=&&\frac{T^4}{\pi^2}[\textrm{PolyLog}[3,-z_q]+\tilde{\mu_q}\textrm{PolyLog}[2,-z_q]-\frac{{\tilde{\mu_q}}^2}{2}ln(1+z_q)]\nonumber\\
               &&-\frac{2T^4}{3\pi^2}\{(\frac{T}{T_c})\partial_{(\frac{T}{T_c})}(ln z_q)\}
               [\textrm{PolyLog}[2,-z_q]-\tilde{\mu_q}ln(1+z_q)-\frac{{\tilde{\mu_q}}^2}{2}\frac{z_q}{1+z_q}]~,
 \label{EC-16}\\
 \{I_{1}\}_{\overline{q}}=&&\frac{T^4}{\pi^2}[\textrm{PolyLog}[3,-z_q]-\tilde{\mu_q}\textrm{PolyLog}[2,-z_q]-
                     \frac{{\tilde{\mu_q}}^2}{2}ln(1+z_q)]\nonumber\\
               &&-\frac{2T^4}{3\pi^2}\{(\frac{T}{T_c})\partial_{(\frac{T}{T_c})}(ln z_q)\}
               [\textrm{PolyLog}[2,-z_q]+\tilde{\mu_q}ln(1+z_q)-\frac{{\tilde{\mu_q}}^2}{2}\frac{z_q}{1+z_q}]~,
 \label{EC-17}\\
 \{I_{2}\}_{g}=&&\frac{T^3}{\pi^2}\textrm{PolyLog}[2,z_g]+\frac{T^3}{\pi^2}\{(\frac{T}{T_c})\partial_{(\frac{T}{T_c})}(ln z_g)\}ln(1-z_g)~,
 \label{EC-18}\\
 \{I_{2}\}_{q}=&&\frac{-T^3}{\pi^2}[\textrm{PolyLog}[2,-z_q]-\tilde{\mu_q}ln(1+z_q)-\frac{{\tilde{\mu_q}}^2}{2}\frac{z_q}{1+z_q}]\nonumber\\
               &&-\frac{T^3}{\pi^2}\{(\frac{T}{T_c})\partial_{(\frac{T}{T_c})}(ln z_q)\}
               [ln(1+z_q)+\tilde{\mu_q}\frac{z_q}{1+z_q}+\frac{{\tilde{\mu_q}}^2}{2}\frac{z_q}{(1+z_q)^2}]~,
 \label{EC-19}\\
 \{I_{2}\}_{\overline{q}}=&&\frac{-T^3}{\pi^2}[\textrm{PolyLog}[2,-z_q]+\tilde{\mu_q}ln(1+z_q)-\frac{{\tilde{\mu_q}}^2}{2}\frac{z_q}{1+z_q}]\nonumber\\
               &&-\frac{T^3}{\pi^2}\{(\frac{T}{T_c})\partial_{(\frac{T}{T_c})}(ln z_q)\}
               [ln(1+z_q)-\tilde{\mu_q}\frac{z_q}{1+z_q}+\frac{{\tilde{\mu_q}}^2}{2}\frac{z_q}{(1+z_q)^2}]~.
 \label{EC-20}
\end{eqnarray}

\begin{figure*}[h]
\includegraphics[scale=0.32]{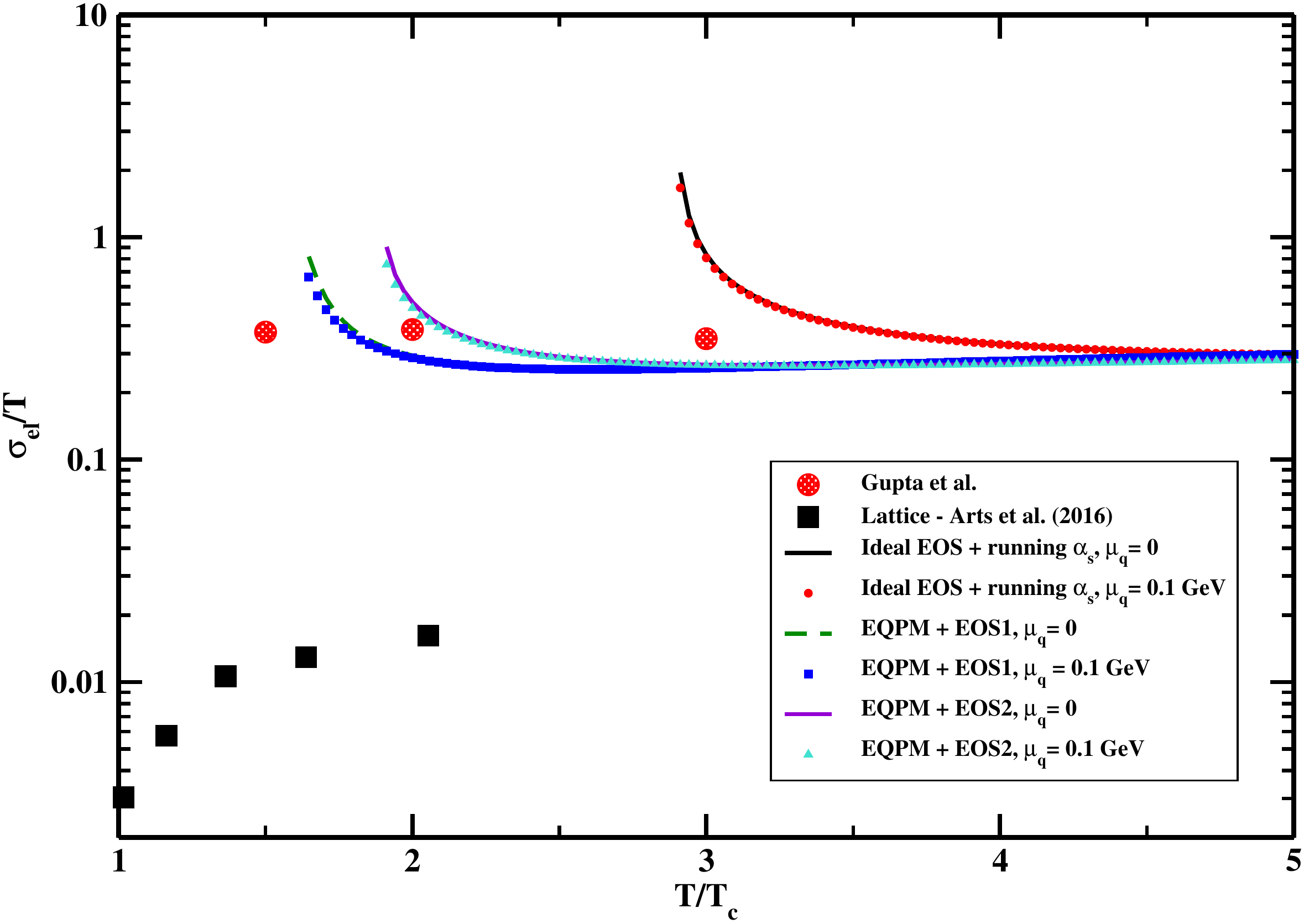}
\includegraphics[scale=0.32]{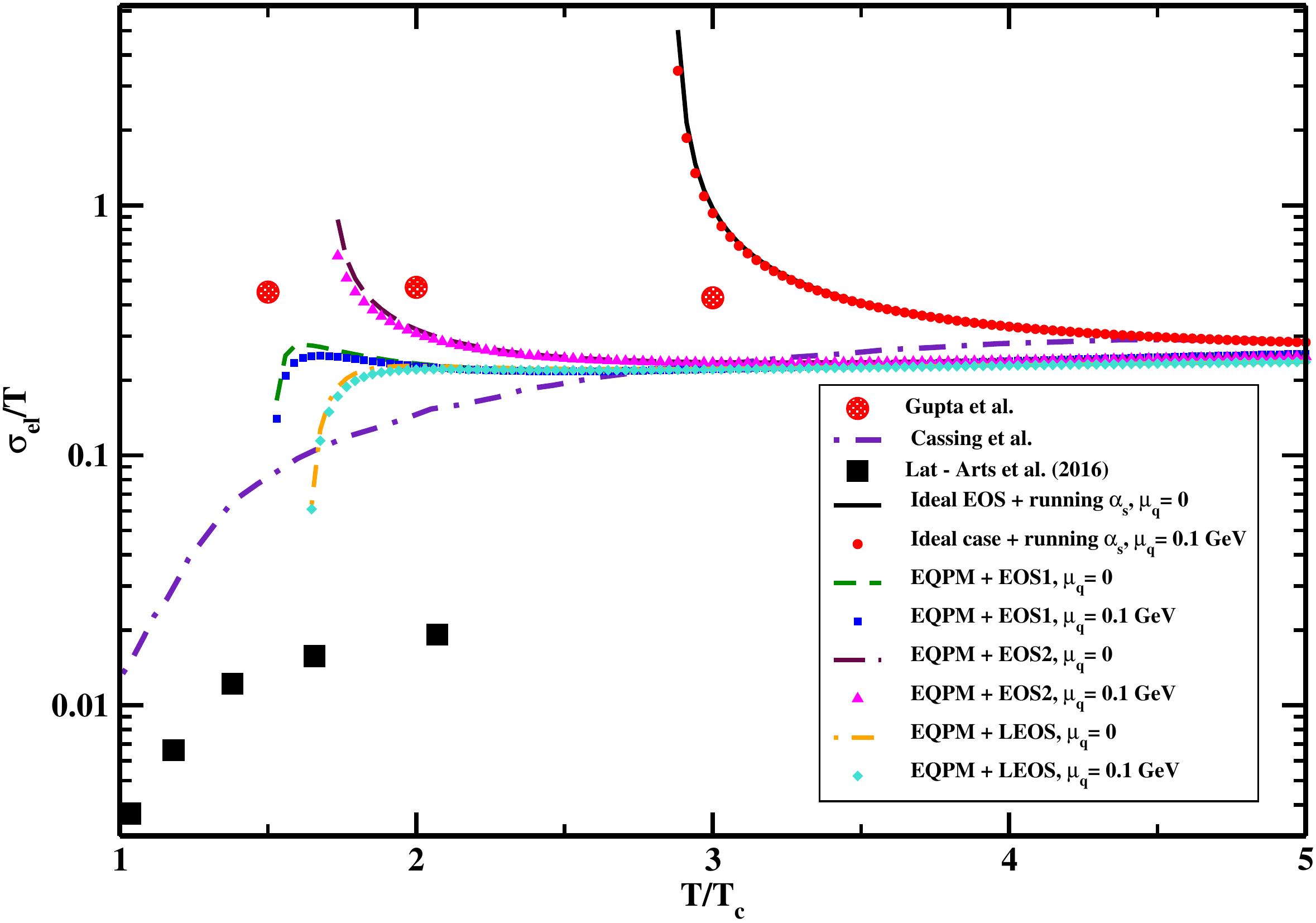}
\caption{Electrical conductivity  using various EOSs as a function of $T/T_c$.}
\label{Econd}
\end{figure*} 

We end this section by giving the results of electrical conductivity in Fig.(\ref{Econd}). The dimensionless ratio $\sigma_{el}/T$
has been plotted against $T/T_c$ for $N_f=2$ and $N_f=3$ employing different EOSs in the EQPM. The results with $\mu_q=0.1$ GeV
differ from the same with $\mu_q=0$ as given in \cite{Mitra-Chandra}, below $T/T_c\sim2$, in a small but quantitative amount.
The 3-flavor case appears to be slightly greater than the 2-flavor ones, since the quark charge $q_{q}^2$ in Eq. (\ref{EC-7}) 
includes the fractional quark charge of strange quark also. At lower temperature the lattice data from \cite{Aarts1} is 
observed to under predict the current results, however the quenched lattice estimations of electrical conductivity from 
Gupta et al. \cite{Gupta1} agrees with the current estimation of $\sigma_{el}$ quite sensibly. For 3-flavor case beyond 
$T/T_c\sim3$, the estimations of $\sigma_{el}$ is matching with the trend given in Cassing  {\it et al}. \cite{Cassing1} and agrees 
with their statement that above $T\sim5T_c$ the dimensionless ratio $\sigma_{el}/T$ becomes approximately constant ($\approx0.3$). 
In the estimations of $\sigma_{el}$ throughout, the electronic charges are explicitly given by the relation 
$\frac{e^2}{4\pi}=\frac{1}{137}$.

\section{Ratios of  transport coefficients and related  physical laws}
This section deals with the highlights on relative important of various transport parameters computed in the previous sections. 
In a nutshell, the relative importance of the charge diffusion, the momentum diffusion and the heat diffusion in  a hot QCD medium are 
being explored by studying the ratios of various transport coefficients, explicated below.

\subsection{Thermal diffusion {\it vs} charge diffusion: The ratio $\lambda/{\sigma_{el} T}$}
The relation between electrical conductivity, $\sigma_{el}$ and thermal conductivity, $\lambda$ for any substance can be understood  in terms 
of  the Wiedemann-Franz law. The basic mathematical statement of the law is,
\be
\label{lr}
\frac{\lambda}{\sigma_{el} T}= {\mathbf L}  \  \  [{\mathbf L}: \text{Lorenz number}]
\ee
\begin{figure*}[h]
\begin{center}
\includegraphics[scale=0.50]{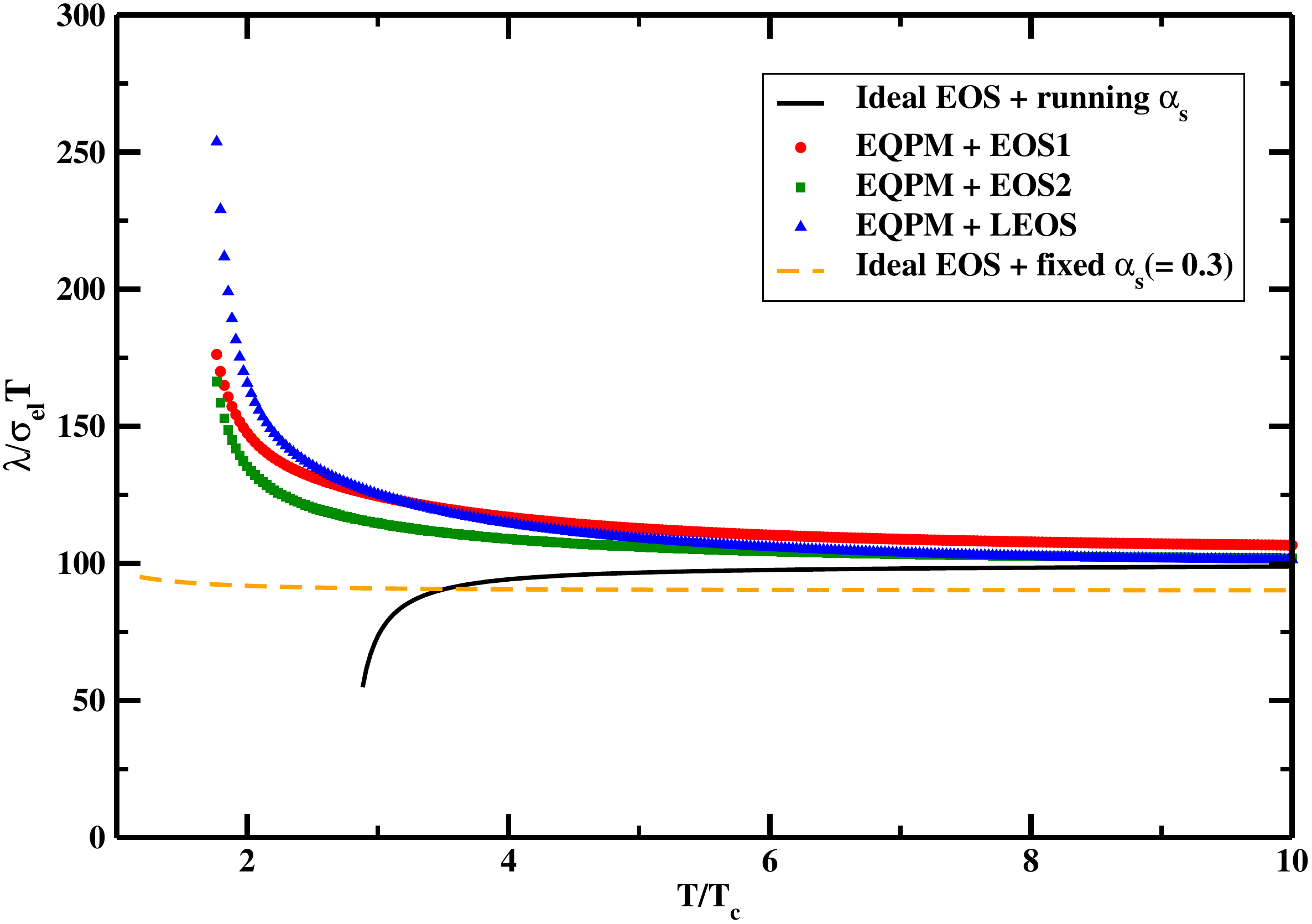}
\caption{ Lorentz number using various EOSs as a function of $T/T_c$.}
\label{econd1}
\end{center}
\end{figure*} 
For instance, in the case of metals, ${\mathbf L}$ is a constant  and quantifies the fact that metals are good electrical as well thermal conductors. In general cases, where the system under consideration, 
does not possess a high symmetry (for {\it e.g.} an anisotropic crystal), both $\lambda$ and $\sigma_{el}$ are tensorial quantities (rank-2 tensors), in effect, ${\mathbf L}$ will be a forth rank tensor. In the case of hot QCD medium, it is possible to 
compute both of these transport coefficients independently either  within  field theory approach or transport theory with Chapman-Enskog technique (or any  other equivalent method). As shown earlier, here, the latter approach has been 
utilized. The main aim  here is to investigate, whether the hot QCD medium/QGP also follow the above mentioned law. If there are deviations, what interesting aspects could emerge while understanding the quantum aspects of its  liquidity. The temperature dependence of the Lorentz number for the QGP is 
depicted in Fig. \ref{econd1}. For the temperature range $2-10 \ T_c$,  ${\mathbf L}$ varies between $250-100$ for various realistic QGP EOSs.  For $T\geq 4 T_c$ the number saturates closer to a value $100$ which is also the Stefan-Boltzmann (SB) limit (QGP as a ultra-relativistic gas of gluons and quarks). 
Clearly, the violation is quite apparent for the 
temperatures which are smaller than $4T_c$ (in fact the violation becomes quite prominent while moving towards $2 T_c$). The law in 
the case of the QGP mainly depends on the effective coupling, the EOS chosen and the contributions that are included while computing the thermal relaxation times.
To make any sensible argument about the violation and its connection with the other Wiedemann-Franz law violating quantum fluids such as graphene~\cite{Crossno}, a more through and deeper analysis is needed (inclusion of higher order QCD processes and appropriate collision and source terms in the transport equation and 
also effects from momentum anisotropy). Nevertheless, the observation from our study perhaps indicates towards much more complex nature of the QGP as a strongly interacting quantum fluid for the temperatures that not very large as compared to the QCD transition temperature, $T_c$.
 Noticeably, such deviations have also been observed in holographic anisotropic models that are dual to spatially anisotropic, ${\cal N}=4$ Super Yang-Mills theory at finite chemical potential~\cite{XianGe} which further violates the KSS bound of $\eta/S$ in the longitudinal direction.
Further, within the holographic set up while considering the charged black holes,  Jain~\cite{Sachin-cond1,Sachin-cond2} has been able to show that 
thermal and electrical conductivities show universal properties and so their universal ratio  in higher dimension.  The number obtained in Fig. \ref{econd1} are slightly higher as compared to  those from holographic models~\cite{Son1}, 
by holographic estimates.

\subsection{Momentum diffusion  {\it vs}  charge diffusion}
The relative significance of the momentum diffusion and the charge diffusion in a medium,
could be understood in terms of a dimensionless ratio, 
\be
\label{etsig}
{\mathcal R}_{\eta/\sigma_{el}}=\frac{\eta/S}{\sigma_{el}/T}.
\ee
\begin{figure}[h]
\begin{center}
\includegraphics[scale=0.50]{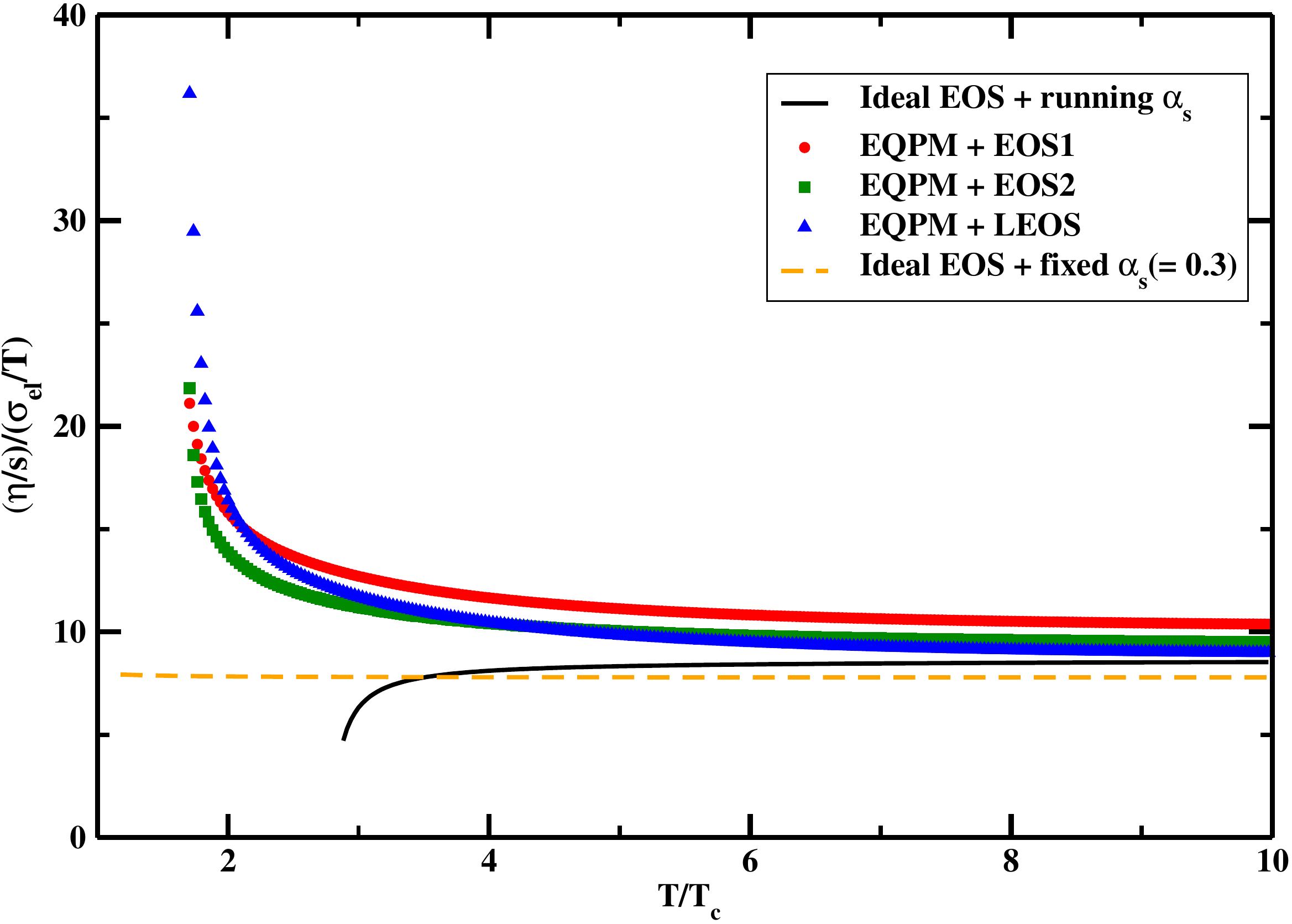}
\caption{Shear viscosity to entropy ratio vs Electrical conductivity as a function of $T/T_c$. }
\label{econd2}
\end{center}
\end{figure} 
In a hot QCD medium, unlike gluons, a quark (antiquark)  carries EM charge. Therefore, it is reasonable to expect that their contribution to the 
$\sigma_{el}$ would be predominant as gluons enter only through the interactions ($qg$ and $gg$ scattering  contributions).  Since the gluonic scattering rates are larger as compared to that 
for the quarks and antiquarks, hence the quark contribution to the shear  viscosity  is also expected to be dominant. The ratio,  ${\mathcal R}_{\eta/\sigma_{el}}$
could be an indicator of relative significance of the gluonic and matter sector as far as the relative importance of the momentum and the charge transports in the QGP medium are concerned. This point has been 
realized in some degree of detail in~\cite{Greco} where a scaling in terms of gluon and quark relaxation times was seen. In contrast, in the present analysis, such a scaling highlighting the relative importance of gluonic and quark contributions is not expected due to more systematic treatment 
of the  scattering cross-sections and computation of relaxation times and inclusion of  all the relevant effects from gluonic sector. The ratio decreases with increasing temperature and subsequently saturating towards  the SB limit (the black line in Fig. \ref{econd2}). The ratio is always greater than unity for the whole range of temperature. It can be inferred that 
the momentum transfer has dominant impact over the charge diffusion.

\subsection{Momentum diffusion  {\it vs}  thermal diffusion: The Prandtl number for the QGP}
The relative magnitude of the momentum and  the thermal  diffusions is quantified in terms of {\it Prandtl number},  $Pr$ (the ratio of momentum diffusibility  by thermal diffusibility):
\be
\label{pr}
Pr= \frac{\eta\ c_p}{\rho\  \lambda}, 
\ee
where $c_p$ is the specific heat at constant pressure.  This number signifies the relative importance of the  shear viscosity and the thermal conductivity in the sound attenuation in a liquid medium. Before, describing it for a hot QCD/QGP medium, let us have an idea about its magnitude for other 
strongly coupled systems. For liquid Helium the {\it Prandtl number} is around $2.5$~\cite{Schafer2}, for  weakly  interacting  unitary Fermi gas at high temperature, it is $2/3$~\cite{Schafer2,schafer}. On the other hand for conformal non-relativistic theories, the number is of the order of $1$~\cite{Son2}.

To define {\it Prandtl number} for the QGP, apart from $\eta$ and $\sigma$, and  $c_p$,  one  requires
 to know the mass density, $\rho$. The only mass scale in high temperature QCD is the thermal mass of a
 dressed parton (gluon or quark) that is obtained in terms of the QCD effective coupling constant at high temperature and 
the temperature scale of the system. In our case, the mass density for the QGP can be  defined as,
\be
\rho = m_g\  n_g + m_q\  (n_q + n_{\bar{q}}), 
\ee
where, $m_g$ and $m_q$ are the thermal(medium)  masses of the gluons and  quarks respectively and $n_g$, $n_{q,\bar{q}}$ are their 
respective number densities obtained from the momentum distributions, following the basic thermodynamic definitions.

\begin{figure}[h]
\begin{center}
\includegraphics[scale=0.50]{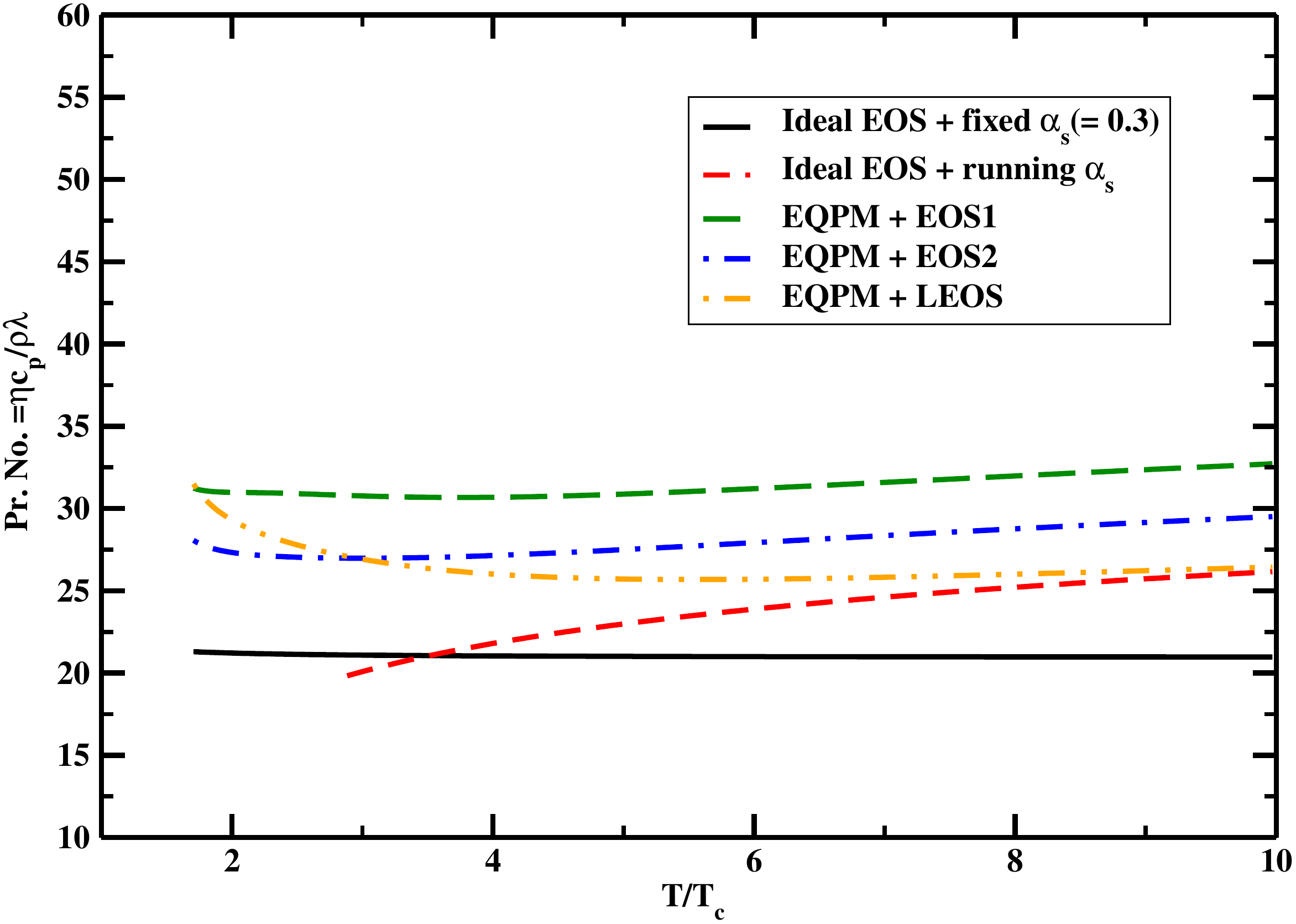}
\caption{{\it{Pr}.No.}  using various EOSs as a function of $T/T_c$.}
\label{econd3}
\end{center}
\end{figure} 

The  gluon medium mass $m_g$  and quark medium mass, $m_q$ for the QCD are obtained in terms of gluon and quark/antiquark distributions 
functions as~\cite{greiner_dm}\\
\ba
 m_g^2(= m_D^2)\equiv &&4 \pi \alpha_{s}(T,\mu_q) \bigg\lbrace -2 N_c \int \frac{d^3 \vec{p}}{(2 \pi)^3} \partial_p f_g (\vec{p})
  -  N_f  \int \frac{d^3 p}{(2 \pi)^3} \partial_p (f_q (\vec{p})+f_{\bar{q}}(\vec{p}))\bigg\rbrace\nn
  m_q^2&=& \bigg(\frac{N_c^2-1}{2 N_c}\bigg) 4 \pi \alpha_s(T, \mu_q) \int  \frac{d^3 \vec{p}}{(2 \pi)^3|\vec{p}|} \bigg \lbrace f_g (\vec{p}) 
  +\bigg(\frac{f_q(\vec{p})+f_{\bar{q}} (\vec{p})}{2}\bigg)\bigg\rbrace.
   \ea
   The number density and specific heat computations are presented in the appendix. The number is plotted as a function of  $T/T_c$  in Fig.  \ref{econd3} for various EOSs within their EQPM descriptions. The number is in the range $20-35$ for the all cases considered here.  
   All the other liquids/systems mentioned previously possess much smaller numbers as compared to the  QGP,  therefore sound attenuation is mostly governed by the momentum diffusion here. In this context, we may perhaps, ignore the effects that are coming out from thermal diffusion unlike holographic models where both these effects are at an equal footing.

\section{Results  and discussions}
The temperature dependences of the viscosities, conductivities and their mutual ratios obtained so far 
leads to a situation where our results are needed to  be analyzed in the light of the estimations of these 
quantities employing various discrete techniques, available in existing literature. In the following, 
the results regarding different transport parameters and the corresponding physical laws are discussed
subsequently. 

\begin{itemize}
 \item As earlier experimental predictions of viscosity values, \cite{Luzum} introduces $\eta/s=0.08$ and $0.16$ 
 in their 2+1 dimensional viscous hydrodynamic code using Glauber and CGC initial conditions respectively, for 
 explaining the elliptic flow data of charged hadron from STAR and PHOBOS. Ref. \cite{Heinz}, in their work
 has concluded a robust upper limit $(\eta/s)_{QGP}<\frac{5}{4\pi}$ by analyzing the phenomenological 
 aspects to be compatible with the experimental $v_2$ data.  Ref. \cite{Song} has obtained  a range $\frac{1}{4\pi}<(\eta/s)_{QGP}<\frac{2.5}{4\pi}$ 
 in their hybrid code ((VISH2+1)+URQMD), in order to describe the charged hadron multiplicity density obtained 
 from 200  GeV,  Au+Au RHIC data. In \cite{Denicol},  the reported value is,  $\eta/s\simeq0.5$, which when implemented 
 in hydrodynamic  simulation approves the multiplicity and collective flow data of charged hadrons from PHOBOS.
 \cite{Rischke} shows a temperature dependence of $\eta/s$ in explaining the transverse momentum spectra and elliptic 
 flow of hadrons in RHIC.  ALICE predictions \cite {ALICE1,ALICE2,ALICE2,ALICE4,ALICE5,ALICE-JHEP} mostly reports
 $\eta/s\simeq0.2$. The analysis of BAMPS data in \cite{Greiner2} gives $\eta/s=0.15$ at $\alpha_s=0.3$. The lattice 
 prediction by \cite{Nakamura} for pure glue leads to $\eta/s\simeq0.08-0.75$ around $T_c$.
 In comparison to all these experimental extractions and lattice simulations, the values of $\eta/s$ obtained from all the 
 pQCD based studies \cite {AMY1,AMY2,Chen1,Chen2,Greiner1,Greco,Hirano2} are at least a factor of 5-10 times larger.
 Mostly leading log results with bare particles results $\eta/s\sim1$, where in \cite{AMY2} the medium properties
 folded within the hard thermal loops of exchanged propagator leads to $\eta/s\sim2.7$ at $\alpha_s\simeq0.1$.
 So hence we conclude that the EQPM model, which is basically dressing up the bare particles by defining their
 collective properties in a thermal medium, and those properties are being reflected through the logarithmic
 term in the viscosity, is responsible for the large values ($\sim 2$) of $\eta/s$. Nevertheless, these are comparable with 
 other estimates based on effective field theory estimations at high temperature.
 
 \item In all the estimations of bulk viscous coefficient the value of $\zeta/s$ has been reported extremely
 smaller except near $T_c$. In \cite{Ryu}, the 3D hybrid simulation agrees with data from ALICE and CMS
 for $\zeta/s\approx0.3$ around $T_c$. The lattice Monte Carlo calculation for $SU(3)$ gluodynamics in \cite{Meyer}
 reports $\zeta/s<0.15$ near $T_c$ ($T/T_c=1.65$) which becomes negligibly small $\zeta/s<0.015$ away from $T_c$. The 
 leading order pQCD estimation from \cite{ADY} also gives small value of bulk viscosity, $\zeta=0.2\alpha_s^2 T^3$
 for reasonable perturbative value of $0.2\lesssim\alpha_s\lesssim0.3$. Our estimation of $\zeta$ in the current
 work mostly consistent with these results giving considerable magnitude only below $T/T_c \sim 2$. Regarding
 the scaling law obeyed by the ratio $\zeta/\eta$, most of the relativistic hydrodynamic calculations \cite{Weinberg},
 pQCD estimations \cite{ADY} and the flow harmonics studies from experimental data \cite{Rose} follow the scaling-2 
 ($\sim \{\frac{1}{3}-c_s^2\}^2$), whereas ADS/CFT based measurements mostly report the scaling-1 ($\sim \{\frac{1}{3}-c_s^2\}$)
 \cite{Benincasa,Gubser}. In our study we observed that $\eta/\zeta$ ratio is more consistent with the scaling-2 at higher temperatures. However, at 
 lower temperature ranges non of the scaling works indicating the peculiar nature of the QGP as a strongly coupled quantum fluid.

 \item In comparison to the viscosities, the investigations on thermal conductivity are really quite 
 limited in  the  existing literature. In \cite{Greif},  the $\lambda$ of a massless Boltzmann gas using partonic cascade model 
 has been listed as a temperature independent quantity for a number of isotropic cross sections. In \cite{Marty},
 the dimensionless quantity,  $\lambda/T^2$ has been plotted against $T/T_c$ from for both NJL model and dynamical 
 quasiparticle particle model (DQPM), which displays two completely different trends. However, at lower temperature 
 ranges ($T/T_c\sim1.5-2.5$),  our estimations appear to be quite consistent at the quantitative level with the NJL
 one.
 
 \item As a prime signature of electromagnetic responses in the heavy ion collision, the estimation of 
$\sigma_{el}$ has achieved a considerable amount of attention now a days in the QGP related fields. 
 The lattice data from \cite{Amato,Aarts1,Aarts2,Brandt1,Brandt2,Ding} provide an estimation for  $\sigma_{el}/T$ that is  below $0.05$,  upto temperature $350$ MeV, 
 which quite under predict our results but the quenched lattice
 estimation from \cite{Gupta1} offers quite sensible agreement. The DQPM results from \cite{Cassing1} also
 exhibits compatible trend with our result. The maximum entropy method (MEM) by \cite{Qin} gives $\sigma_{el}/T\sim0.4$
 at $T/T_c=3$ which is also close to our result. In \cite{Kharzeev2}, the sensitivity of early stage strong
 magnetic field in non central heavy ion collisions have been tested on the azimuthal distributions and correlations
 of the produced charged hadrons. In their analysis they have chosen the electric conductivity $\sigma_{el}=0.023 fm^{-1}$
 in order to obtain the directed flow ($v_{1}$) of charged pions at LHC energies which is not away from our
  estimations around a temperature region $T/T_c\sim2.5$.  

 \item
Interplay of the thermal diffusion and the electrical diffusion could be understood in terms of the Wiedemann-Franz law--the 
 dimensionless number (Lorentz number). For a large class of metals this is constant and depicts the common origin of both the 
 transport processes. In other words, the metals are good thermal and electrical conductors at the same time. The number for the hot QCD system in the current 
 works converges to a number slightly higher than $100$ at higher temperatures. For the temperatures that are lower than $3 T_c$, there is significant rise as we decrease the temperature. This indicates, towards the violation 
 of   the above mentioned law. In some known systems, such as Graphene this violation leads to a strongly interacting quantum fluid also termed as Dirac fluid \cite{Crossno}. In this present case, the violation is mainly due to the 
 $1/\alpha_s$ term and the strongly interacting EOS. To make any such concrete   connection with the other interesting quantum fluids is quite early as it will require more refined computation of the Lorenz number while including higher order hot QCD effects to the current analysis. 
 This will be one direction where the future investigation will focus on. 
 In order to explore the relative importance of the  momentum diffusion and the charge diffusion in the hot QCD medium, ratio of $\eta/s$ to $\sigma_{el}/T$ is studied
 as a function of  temperature. The ratio for the QGP is turned out to be much greater than unity for the whole range of the temperature considered here indicating the more prominent role of the momentum diffusion in agreement to the 
 prediction of ~\cite{Greco}. Finally the relative significance of the thermal and the momentum diffusions has been quantified
  in terms of {\it Prandtl number}. For the hot QCD system here, this number came out to be much greater than unity signifying the dominance of the momentum diffusion over thermal one. 
  In other words, sound attenuation in the hot QCD/QGP system will mainly be governed by the  shear viscous effects which is in contrast to 
  the observations for dilute fermi-gases \cite{Schafer2} or the holographic systems~\cite{Son2}. 
  For {\it e.g.} in liquid Helium, the number is $2.5$~\cite{Schafer2} which is but  an order of magnitude smaller.
  
 \end{itemize}

\section{Conclusions and outlook}
The current article concerns about the temperature behavior of various transport coefficients that measures the 
dissipative and electromagnetic responses in a strongly interacting QCD system at finite temperature with non zero
quark chemical potential. The most important feature of this work is to highlight the concerning physical laws
expressing the relative importance of different transport phenomena, by obtaining the temperature dependences of
their mutual ratios. The detail Chapman-Enskog technique for a multi-component fluid, adopted from the 
kinetic theory of many particle systems has been discussed which gives the mathematical expressions of shear and bulk viscosities,
 thermal conductivity and electrical conductivity in terms of the medium interactions.
The interaction cross sections are provided through the thermal relaxation times of constituent quarks, antiquarks
and gluon by leading order QCD estimations. The effects of a strongly couped thermal medium has been
introduced in the evaluation of these transport parameters through the EQPM model, which describes the collective
properties of quarks and gluons by considering them as quasi particles rather than bare ones. The finite temperature
effects have been folded through this EQPM scheme by introducing the pQCD and Lattice QCD based equation of state effects
in particle momentum distribution and effective couplings. Finally they are applied to the current formalism of 
estimating transport coefficients and studying the related physical laws. So we conclude by saying that we have 
investigated the transport properties and electromagnetic responses along with the associated physical laws in a 
strongly interacting hot QCD medium quite throughly and reasonably, presenting a sensible realistic scenario created 
out of the relativistic heavy ion collisions. The results obtained in our approach are seen to be consistent with other parallel or 
distinct approaches.

The current work opens a horizon of possible extensions and applications in the related areas in near future.  A few interesting ones
 are listed below which could be a matter of immediate future investigations.
\begin{itemize}
 \item All the above mentioned transport coefficients have wide spread applications in signal extractions by affecting 
 the quantitative estimates of the signals for QGP from heavy ion collisions, particularly, where hydrodynamic simulations
 are involved. In \cite{Dusling1,Dusling2,Dusling3,Bhatt-Mishra,Dion-Gale,Sarkar-Alam,Mitra-photon,Denicol-2017}, the 
 effect of viscosities have been tested on transverse momentum spectra and collective flows of charged hadron and 
 electromagnetic probes rigorously, revealing the necessity of incorporating them in the out of equilibrium particle 
 distribution functions and dissipative hydrodynamics. They are observed to modify the collective behavior of plasma
 as well, such as in \cite{Strickland} viscous corrections are shown to affect Debye screening and Landau damping
 mass scales and in \cite{Torrieri} leads to clusterization and early freeze out of the QGP. In \cite{Bhatt-cavitation,Rajagopal}
 it is discussed that the viscous effects lead to fluid cavitation predicting the breakdown of hydrodynamic calculations.
 In \cite{Skokov-th} the effect of $\lambda$ is studied in the dynamics of first-order phase transitions. In 
 \cite{Yin,Ding,Huot} the soft photon and dilepton emission rates are shown to depend up on $\sigma_{el}$ so that their
 transverse momentum spectra and elliptic flow are also sensitive on the temperature dependence of $\sigma_{el}$. 
 In the light of the above analysis, implementation of the obtained temperature dependence of transport parameters, in the 
 particle emission rates and in the hydrodynamic evolution codes as well as in the collective mode measurements, 
 is aimed to be a matter of immediate future investigation.
 
 \item Realizing the high event statistics and high precision knowledge offered by relativistic heavy ion collisions
 where the application of the usual Boltzmann-Gibbs statistics is questionable, the inclusion of a nonextensive statistical 
 model \cite{Wilk1,Tsallis,Biro} is necessary via Tsallis distribution. So all the definitions of thermodynamic and transport
 quantities used in this article are to be revisited for a dynamical, non-equilibrium system with strong intrinsic fluctuations 
 and long-range correlations.

 \item The collision term is needed to be re looked by keeping the instantaneous charge and energy-momentum conservation
 consistent \cite{BGK} and including appropriate Vlasov term and source contributions in order to study the production 
 and evolution of QGP. There may be possibilities to  exclude the relaxation time technique as a whole and treat the 
 collision term by explicit polynomial expansion using variational method \cite{Degroot}.

 \item The viscosities controlling the magnitude of hydrodynamic fluctuations in the fluid can be extracted directly from the 
 correlation observables in heavy ion collisions \cite{Kapusta-fluctuation}. This can offer a means to predict the viscosity, 
 independent from the traditional collective flow analysis and can shed some light regarding the dissimilarities 
 in the viscosity values extracted by them from the pQCD measurements.
 
 \item Finally, the estimations of all the above mentioned transport coefficients by including higher order thermal QCD effects 
 with an appropriate collision, Vlasov and source terms in the transport equation also including the anisotropic aspects of the QGP
 in heavy-ion collisions could be another interesting direction where the future explorations could focus on.

\end{itemize}

\appendix

\section{Evolution equation of thermodynamic quantities and conservation laws}
The conservation laws and the consequent time evolution equations for the thermodynamic 
macroscopic quantities which define the state of the system have been enlisted below. In order to determine the transport coefficients 
in Chapman-Enskog technique, these equations of motions have been used as the thermodynamic identities in 
Section 2.3.

In order to do this we start with the relativistic transport equation in presence of the external
electric field,

\begin{equation}
 p_{k}^{\mu}\partial_{\mu}f_{k}^{0}
 +\frac{1}{T}f_{k}^0(1\pm f_{k}^0)q_{k}E_{\mu}p_{k}^{\mu}
 =\sum_{l=1}^{N} C_{kl}[f_{k},f_{l}],~~~~~~~~~~~~~~~~[k=1,2,..........N]~.
 \label{Ap-1} 
\end{equation}

Integrating both sides of Eq.(\ref{Ap-1}) over $\int \frac{d^3p}{(2\pi)^3 p_k^0}$ and summing over $\sum_{k=1}^{N}$,
the right hand side of Eq.(\ref{Ap-1}) vanishes, since the first moment of collision term is zero by the virtue
of summational invariants. Finally what we left with is the conservation of particle number as the following,

\begin{equation}
 \partial_{\mu}N^{\mu}=0~.
 \label{Ap-2}
\end{equation}

In a system where the number of particles of each components are conserved separately, like the current system
where only elastic collisions are being considered, we obtain,

\begin{equation}
 \partial_{\mu}N_{k}^{\mu}=0~~~~~~~~[k=1,....,N].
 \label{Ap-3}
\end{equation}

From Eq.(\ref{Ap-3}) the continuity equation, i.e, the evolution equation of the particle number
density for each species can be obtained in the following manner,

\begin{equation}
 Dn_{k}=-n_{k}\partial\cdot u ~.
 \label{Ap-4}
\end{equation}

In an analogous way, multiplying both sides of Eq.(\ref{Ap-1}) with $p_{k}^{\nu}$, integrating
over $\int \frac{d^3p}{(2\pi)^3 p_k^0}$ and summing over $\sum_{k=1}^{N}$,
the right hand side of Eq.(\ref{Ap-1}) again vanishes, since the second moment of collision term is zero also.
Finally contracting the resulting equation with $u_{\nu}$ from left we get,

\begin{equation}
 u_{\nu}\partial_{\mu}T^{\mu\nu}=0~,
 \label{Ap-5}
\end{equation}
where $T^{\mu\nu}$ is the energy-momentum stress tensor.
From Eq.(\ref{Ap-5}) the equation of energy evolution of the system can be easily traced out as,

\begin{equation}
 \sum_{k=1}^{N}x_{k}D\omega_{k}=-\frac{\sum_{k=1}^N P_{k}}{\sum_{k=1}^N n_{k}}\partial \cdot u ~.
\end{equation}

In the second case a contraction with $\Delta^{\sigma}_{\nu}$ from left gives,

\begin{equation}
 \Delta^{\sigma}_{\nu}\partial_{\mu}T^{\mu\nu}+E^{\sigma}\sum_{k=1}^{N}q_k n_k=0~,
\end{equation}
from which in a multicomponent system in the presence of an electric field the equation of motion becomes,

\begin{equation}
 Du^{\mu}=\frac{\nabla^{\mu} P}{\sum_{k=1}^{N} n_{k} h_{k}}+\frac{\sum_{k=1}^N q_{k}n_{k}}{\sum_{k=1}^N h_{k}n_{k}}E^{\mu}~.
 \label{velid}
 \end{equation}
Clearly even the pressure gradient is zero, the Lorentz force acting on the electrically charged particle produces non-zero acceleration.

\section{Thermodynamical quantities at finite chemical potential,  $\mu_q$}
Here, the expressions for the thermodynamic quantities are presented following their  fundamental definitions, within EQPM at finite  $\mu_q$.  
These quantities have been used throughout the article during the 
formalism development. The quantities under consideration are , {\it viz.},  particle number density,  pressure, energy density, enthalpy density, 
entropy density, specific heat at constant pressure and the velocity of sound.
To compute these quantities, we need their definitions in terms of EQPM degrees of freedom. Following the corresponding  definitions of the particle number density, pressure and energy density, we can straightforwardly derive all the 
quantities mentioned here.

The  particle number density, $n$ is obtained as, 
\begin{eqnarray}
n=&&\sum_{k=g,q,\overline{q}}\nu_{k}\int\frac{d^{3}\vec{p_{k}}}{(2\pi)^3}f^{0}_{k}\\
     =&&\frac{T^3}{\pi^2}[\nu_g \textrm{PolyLog}[3,z_g]-\nu_q\{ 2\textrm{PolyLog}[3,-z_q]-{\tilde{\mu_q}}^2 ln(1+z_q)\}]~.
\end{eqnarray}

On the other hand, Pressure, $P$ and the energy density, $\epsilon$, following their respective fundamental definitions can be obtained as, 
\begin{eqnarray}
P= &&\sum_{k=g,q,\overline{q}}\frac{1}{3}\nu_{k}\int\frac{d^{3}\vec{p_{k}}}{(2\pi)^3p_{k}^0}|\vec{p_k}|^2f^{0}_{k}\\
     =&&\frac{T^4}{\pi^2}[\nu_g \textrm{PolyLog}[4,z_g]-\nu_q\{ 2\textrm{PolyLog}[4,-z_q]+{\tilde{\mu_q}}^2\textrm{PolyLog}[2,-z_q]\}]~.
\end{eqnarray}
\begin{eqnarray}
\epsilon= &&\sum_{k=g,q,\overline{q}}\nu_{k}\bigg[\int\frac{d^{3}\vec{p_{k}}}{(2\pi)^3p_{k}^0} {(p_k^0)}^2f^{0}_{k}
+T\big\{\frac{T}{T_c}\big\}\{\partial_{(\frac{T}{T_c})}(ln z_k)\} \int\frac{d^{3}\vec{p_{k}}}{(2\pi)^3}f^{0}_{k}\bigg]\\
     =&&\frac{3T^4}{\pi^2}[\nu_g \textrm{PolyLog}[4,z_g]-\nu_q\{2\textrm{PolyLog}[4,-z_q]+{\tilde{\mu_q}}^2\textrm{PolyLog}[2,-z_q]\}]\nonumber\\
     &&+T\{\frac{T}{T_c}\}\{\partial_{(\frac{T}{T_c})}(ln z_g)\}[\frac{T^3}{\pi^2}\nu_g \textrm{PolyLog}[3,z_g]]\nonumber\\
     &&-T\{\frac{T}{T_c}\}\{\partial_{(\frac{T}{T_c})}(ln z_q)\}[\frac{T^3}{\pi^2}\nu_q \{ 2\textrm{PolyLog}[3,-z_q]-{\tilde{\mu_q}}^2 ln(1+z_q)\}]~.
\end{eqnarray}

The enthalpy density, $H$, entropy density, $s$ and specific heat at constant pressure, $c_p$, could be obtained in terms of 
$P$ and $\epsilon$ as, 
\begin{eqnarray}
H=&&\epsilon + P\\
  &&\frac{4T^4}{\pi^2}[\nu_g \textrm{PolyLog}[4,z_g]-\nu_q\{2\textrm{PolyLog}[4,-z_q]+{\tilde{\mu_q}}^2\textrm{PolyLog}[2,-z_q]\}]\nonumber\\
     &&+T\{\frac{T}{T_c}\}\{\partial_{(\frac{T}{T_c})}(ln z_g)\}[\frac{T^3}{\pi^2}\nu_g \textrm{PolyLog}[3,z_g]]\nonumber\\
     &&-T\{\frac{T}{T_c}\}\{\partial_{(\frac{T}{T_c})}(ln z_q)\}[\frac{T^3}{\pi^2}\nu_q \{ 2\textrm{PolyLog}[3,-z_q]-{\tilde{\mu_q}}^2 ln(1+z_q)\}]~.
\end{eqnarray}
\begin{eqnarray}
 s=&&\frac{\epsilon +P}{T}-\frac{n\mu}{T}\\
 &&\frac{4T^3}{\pi^2}\nu_g \textrm{PolyLog}[4,z_g]-\frac{8T^3}{\pi^2}\nu_q \textrm{PolyLog}[4,-z_q]
 -\frac{2T^3}{\pi^2}{\tilde{\mu_q}}^2\nu_q\textrm{PolyLog}[2,-z_q]\nonumber\\
     &&+\{\frac{T}{T_c}\}\{\partial_{(\frac{T}{T_c})}(ln z_g)\}[\frac{T^3}{\pi^2}\nu_g \textrm{PolyLog}[3,z_g]]\nonumber\\
     &&-\{\frac{T}{T_c}\}\{\partial_{(\frac{T}{T_c})}(ln z_q)\}[\frac{T^3}{\pi^2}\nu_q \{ 2\textrm{PolyLog}[3,-z_q]-
     {\tilde{\mu_q}}^2 ln(1+z_q)\}]~.    
\end{eqnarray}
\begin{eqnarray}
 c_p=&&\frac{\partial H}{\partial T}+\frac{\partial H}{\partial \mu_q}\tilde{\mu_q}\\
 =&&\frac{16T^3}{\pi^2}\nu_g\textrm{PolyLog}[4,z_g]
 +\frac{9T^3}{\pi^2}\{\frac{T}{T_c}\}\{\partial_{(\frac{T}{T_c})}(ln z_g)\}\nu_g\textrm{PolyLog}[3,z_g]\nonumber\\
 +&&\frac{T^3}{\pi^2}\{\frac{T}{T_c}\}^2\{\partial_{(\frac{T}{T_c})}(ln z_g)\}^2\nu_g\textrm{PolyLog}[2,z_g]
 +\frac{T^3}{\pi^2}\{\frac{T}{T_c}\}^2\{\partial^2_{(\frac{T}{T_c})}(ln z_g)\} \nu_g\textrm{PolyLog}[3,z_g]\nonumber\\
 -&&\frac{16T^3}{\pi^2}\nu_q[2\textrm{PolyLog}[4,-z_q]+\tilde{\mu_q}^2 \textrm{PolyLog}[2,-z_q]]\nonumber\\
 -&&\frac{9T^3}{\pi^2}\{\frac{T}{T_c}\}\{\partial_{(\frac{T}{T_c})}(ln z_q)\}\nu_q[2\textrm{PolyLog}[3,-z_q]-\tilde{\mu_q}^2ln(1+z_q)]\nonumber\\
 -&&\frac{T^3}{\pi^2}\{\frac{T}{T_c}\}^2\{\partial_{(\frac{T}{T_c})}(ln z_q)\}^2\nu_q[2\textrm{PolyLog}[2,-z_q]-\tilde{\mu_q}^2\frac{z_q}{1+z_q}]\nonumber\\
 -&&\frac{T^3}{\pi^2}\{\frac{T}{T_c}\}^2\{\partial^2_{(\frac{T}{T_c})}(ln z_q)\} \nu_q[2\textrm{PolyLog}[3,-z_q]-\tilde{\mu_q}^2ln(1+z_q)]\nonumber\\
 +&&\tilde{\mu_q}^2 [-\frac{8T^4}{\pi^2}\nu_q \textrm{PolyLog}[3,-z_q]-
 \{\frac{T}{T_c}\}\{\partial_{(\frac{T}{T_c})}(ln z_q)\}\frac{2T^4}{\pi^2}\nu_q \textrm{PolyLog}[2,-z_q]]~.
 \end{eqnarray}

Finally,  the velocity of sound could be obtained in terms of the first order differentials of the $P$ and $\epsilon$, leading to the following expression,
\begin{eqnarray}
c_s^2=&&\frac{\frac{\partial P}{\partial T}+\frac{\partial P}{\partial \mu_q}\tilde{\mu_q}}
           {\frac{\partial \epsilon}{\partial T}+\frac{\partial \epsilon}{\partial \mu_q}\tilde{\mu_q}}\\
=&&\frac{12T^3}{\pi^2}\nu_g\textrm{PolyLog}[4,z_g]
 +\frac{8T^3}{\pi^2}\{\frac{T}{T_c}\}\{\partial_{(\frac{T}{T_c})}(ln z_g)\}\nu_g\textrm{PolyLog}[3,z_g]\nonumber\\
 +&&\frac{T^3}{\pi^2}\{\frac{T}{T_c}\}^2\{\partial_{(\frac{T}{T_c})}(ln z_g)\}^2\nu_g\textrm{PolyLog}[2,z_g]
 +\frac{T^3}{\pi^2}\{\frac{T}{T_c}\}^2\{\partial^2_{(\frac{T}{T_c})}(ln z_g)\} \nu_g\textrm{PolyLog}[3,z_g]\nonumber\\
 -&&\frac{12T^3}{\pi^2}\nu_q[2\textrm{PolyLog}[4,-z_q]+\tilde{\mu_q}^2 \textrm{PolyLog}[2,-z_q]]\nonumber\\
 -&&\frac{8T^3}{\pi^2}\{\frac{T}{T_c}\}\{\partial_{(\frac{T}{T_c})}(ln z_q)\}\nu_q[2\textrm{PolyLog}[3,-z_q]-\tilde{\mu_q}^2ln(1+z_q)]\nonumber\\
 -&&\frac{T^3}{\pi^2}\{\frac{T}{T_c}\}^2\{\partial_{(\frac{T}{T_c})}(ln z_q)\}^2\nu_q[2\textrm{PolyLog}[2,-z_q]-
 \tilde{\mu_q}^2\frac{z_q}{1+z_q}]\nonumber\\
 -&&\frac{T^3}{\pi^2}\{\frac{T}{T_c}\}^2\{\partial^2_{(\frac{T}{T_c})}(ln z_q)\} \nu_q[2\textrm{PolyLog}[3,-z_q]-
 \tilde{\mu_q}^2ln(1+z_q)]]~.
\end{eqnarray}

\section{Some useful  identities related to $\textrm{PolyLog}[a,z]$ function}
Below, we enlist the  $PolyLog$ identities that are required for our calculations.
\begin{equation}
 \partial_{T}\{\textrm{PolyLog}[a,z_g]\}=\textrm{PolyLog}[(a-1),z_g]\{\partial_{T}(lnz_g)\}
\end{equation}

\begin{equation}
 \partial_{T}\{\textrm{PolyLog}[a,-z_q e^{\pm \tilde{\mu_q}}]\}
 =\textrm{PolyLog}[(a-1),-z_q e^{\pm \tilde{\mu_q}}]\{\partial_{T}(lnz_q)\}
\end{equation}

\begin{equation}
 \partial_{\tilde{\mu_q}}\{\textrm{PolyLog}[a,-z_q e^{\pm \tilde{\mu_q}}]\}
 =\pm\textrm{PolyLog}[(a-1),-z_q e^{\pm \tilde{\mu_q}}]
\end{equation}

\begin{equation}
 \partial^2_{\tilde{\mu_q}}\{\textrm{PolyLog}[a,-z_q e^{\pm \tilde{\mu_q}}]\}
 =\textrm{PolyLog}[(a-2),-z_q e^{\pm \tilde{\mu_q}}]
\end{equation}

\begin{equation}
 \textrm{PolyLog}[a,-z_q e^{\pm \tilde{\mu_q}}]
 =\textrm{PolyLog}[a,-z_q] \pm\tilde{\mu_q} \textrm{PolyLog}[(a-1),-z_q]
 +\frac{\tilde{\mu_q}^2}{2}\textrm{PolyLog}[(a-2),-z_q]
\end{equation}

\acknowledgments
VC would like to acknowledge Department of Science and Technology  (DST), Govt. of India for INSPIRE Faculty Fellowship (IFA-13/PH-55) and 
Science and Engineering Research Board (SERB), DST for granting  funds under Early Career Research Award  (ECRA).
 SM acknowledges Indian Institute of Technology Gandhinagar for the Institute Postdoctoral 
Fellowship. We record our sincere gratitude to the people of India for their generous  help  
for the research in basic sciences.

\end{document}